%% file: B2G-24-016_temp.tex
\documentclass[11pt,twoside,a4paper,cmspaper,final,collab]{cms-tdr}
\def\svnVersion{10dbc00-D}\def\svnDate{2026/10/02}\def\cmsCernNoTag{CERN-EP-2026-248}\def\cmsCernDate{\today}\def\cmsMessage{Submitted to Physical Review D}
\begin{document}\cmsNoteHeader{B2G-24-016}

\providecommand{\cmsTable}[1]{\resizebox{\textwidth}{!}{#1}}
\newlength\cmsTabSkip\setlength{\cmsTabSkip}{1ex}
\newcommand{\cPGf}{\PGf}

\cmsNoteHeader{B2G-24-016}
\title{Search for pair-produced vector-like top quarks decaying into Lorentz-boosted top quarks and scalar bosons in proton-proton collisions at \texorpdfstring{$\sqrt{s}=13$\TeV}{sqrt(s) = 13 TeV}}

\date{\today}

\abstract{
A search for pair-produced vector-like top quarks, \PQT, decaying into a new scalar boson \cPGf and a standard model (SM) top quark is presented. The search targets events containing a photon pair originating from the decay of one of the scalar bosons, accompanied by all-hadronic decays of the SM top quark pair. The analysis is optimized for boosted top quark signatures in which the decay products of the top quarks are reconstructed as large-radius jets. The search is based on proton-proton collision data at $\sqrt{s}=13\TeV$, collected by the CMS experiment at the CERN LHC during 2016--2018, corresponding to an integrated luminosity of 138\fbinv. The search is performed for an array of \PQT and \cPGf masses. No significant deviation from the predicted background is observed.  The largest local excess has a significance of 2.7 standard deviations, reduced to 1.5 standard deviations after accounting for the look-elsewhere effect.  Upper limits at 95\% confidence level are set on the product of $\PQT\PAQT$ production cross section and the branching fraction.}

\hypersetup{%
pdfauthor={CMS Collaboration},%
pdftitle={Search for pair-produced vector-like top quarks decaying into Lorentz-boosted top quarks and scalar bosons in proton-proton collisions at sqrt(s) = 13 TeV},%
pdfsubject={CMS},%
pdfkeywords={LHC, CMS, vector-like quarks, pair-production, new scalar, diphotons with boosted jets}}

\maketitle 

\section{Introduction}\label{sec:introduction}

The observation of the Higgs boson~\cite{ATLAS:2012yve, CMS:2012qbp, CMS:2022dwd} confirmed key aspects of the standard model (SM).  However, the SM does not contain a mechanism to control the quadratic divergence of the quantum corrections to the Higgs boson mass at large mass scales; this issue is referred to as the hierarchy problem~\cite{Barbieri:1987fn, Peskin:2025hp}.

Vector-like quarks (VLQs) are hypothetical particles proposed in many models beyond the SM~\cite{Guo:2018uxx,Giudice:2007fh}, including the ones that address the hierarchy problem.  They are heavy partners of the SM quarks and, like the latter, transform as triplets under SU(3) color symmetry. Their left- and right-handed components have identical electroweak quantum numbers, allowing gauge-invariant mass terms. In such models, vector-like top quark partners can potentially resolve the hierarchy problem by offsetting large quantum corrections to the Higgs boson mass resulting from top quark loops, thereby stabilizing the electroweak scale. 

In composite Higgs models~\cite{Lodone:2008yy}, VLQs emerge as composite states that interact strongly with the Higgs field.  These models extend the Higgs sector beyond the single scalar field of the SM, introducing additional symmetries and particles, such as charged Higgs bosons and pseudoscalar Higgs bosons~\cite{Benbrik:2019zdp}.

This paper presents a search probing a composite Higgs model~\cite{Benbrik:2019zdp}, which predicts the pair production of vector-like top quark partners, denoted as \PQT, via the strong interaction, followed by the decay of each \PQT quark into an SM top quark and a new neutral scalar boson \cPGf.  The search utilizes 138\fbinv of proton-proton collision data collected at a center-of-mass energy of 13\TeV by the CMS experiment at the CERN LHC during  2016--2018.  While the \PQT quark has four primary decay channels, $\PQT \to \cPGf\PQt$, $\PQT \to \PZ\PQt$, $\PQT \to \PH\PQt$, and $\PQT \to \PW\PQb$, the first channel dominates in the considered model.  For \PQT quark masses around 800\GeV, the branching fraction for the decay $\PQT \to \cPGf\PQt$ is approximately 90\%, reaching nearly 100\% at around 1.0\TeV and above.  We search for \PQT quark masses above 800\GeV and assume a branching fraction of 100\% for $\PQT \to \cPGf\PQt$.  Existing LHC searches constrain vector-like top quarks with masses up to about 1.5\TeV for the conventional $\PQT \to \PZ\PQt, \PH\PQt, \PW\PQb$ decay modes~\cite{CMS:2022fck, Benbrik:2025vlq}, but these limits do not directly apply to the $\PQT \to \cPGf\PQt$ decay considered here. A recent ATLAS search has targeted the $\PQT \to \cPGf\PQt$ topology with $\cPGf\to\PGg\PGg$~\cite{ATLAS:2026TtScalar}.

The new scalar boson has four main electroweak bosonic decay channels: $\cPGf \to \PZ\PGg$, $\cPGf \to \PZ\PZ$, $\cPGf \to \PW^+\PW^-$, and $\cPGf \to \PGg\PGg$.  This search targets events in which one of the scalar bosons produced in the decays of the pair-produced \PQT quarks decays to two photons.  Such an approach allows for the retention of substantially more signal events while still maintaining good background rejection.  Although the branching fraction for $\cPGf \to \PGg\PGg$ can be as small as 0.004~\cite{Benbrik:2019zdp} due to the dominance of the $\cPGf \to \Pg\Pg$ decay channel, the diphoton channel offers a clean and identifiable experimental signature.  The scalar boson is assumed to be a pure scalar state. Consequently, CP-violating effects associated with scalar--pseudoscalar mixing are not considered.

\begin{figure}[!ht]
\centering
\includegraphics[width=0.7\columnwidth]{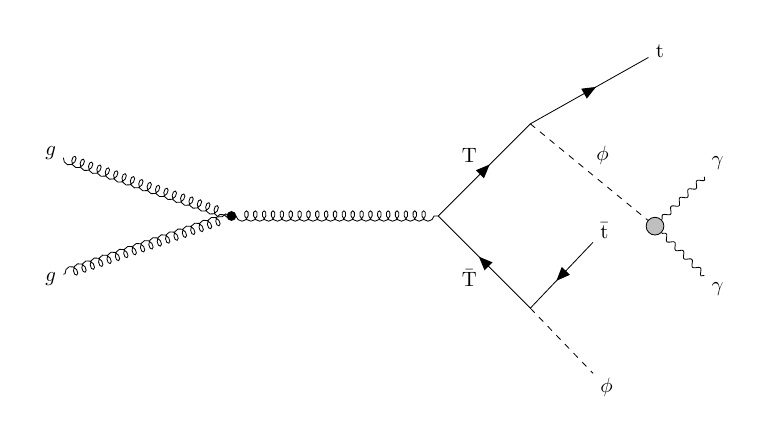}
            \caption{A representative leading order Feynman diagram for the pair production of vector-like top quarks $\PQT$ with subsequent $\PQT \to \cPGf\PQt$ decays.}
            \label{fig:FeynD}
\end{figure}

Figure~\ref{fig:FeynD} presents a leading-order Feynman diagram illustrating the pair production of VLQs, followed by the $\PQT \to \cPGf\PQt$ decays.  Besides the $\Pg\Pg$-initiated final state shown in Fig.~\ref{fig:FeynD}, production can also occur through $\PQq\PAQq$ initial states.  The search focuses on a final state where both top quarks undergo hadronic decay.  For the \PQT masses considered in this analysis, the top quarks are produced with significant Lorentz boosts, leading to their decay products being collimated and reconstructed as single large-radius jets.  Consequently, the experimental signature is characterized by two large-radius jets from the top quarks and two photons resulting from the decay of a \cPGf boson. 

The production of a top quark pair in association with two photons is a rare process in the SM, and an accurate prediction of its cross section requires next-to-leading order (NLO) quantum chromodynamics (QCD) corrections~\cite{Stremmer:2023}. This process, and the backgrounds arising from misidentified photons, are difficult to model reliably in simulation. The SM background in this search is therefore determined from control regions in data.

Searches for the pair production of vector-like top and bottom quarks have previously been performed by the ATLAS and CMS collaborations~\cite{CMS:2022fck,ATLAS:2024xyz,CMS:2024vlq}. These searches have excluded vector-like bottom quark masses up to 1.52\TeV.  In addition, searches for vector-like top quarks, specifically in a singlet representation of isospin with standard decay channels $\PZ\PQt$, $\PH\PQt$, and $\PW\PQb$, have excluded \PQT masses up to 1.48\TeV. In comparison, this analysis constitutes the first exclusive search in the boosted regime.  Furthermore, an advanced neural network-based top quark tagging algorithm, called ParticleNet, is used to identify large-radius jets originating from hadronic decays of Lorentz-boosted top quarks~\cite{Qu:2019gqs}, while a multivariate discriminant is used for photon identification~\cite{CMS:2020uim}.  The diphoton invariant mass corresponds to the reconstructed scalar boson mass, $m^\text{reco}_{\cPGf}$.  The four momenta of the top quark candidate and the diphoton system are combined to obtain the \PQT mass, $m^\text{reco}_{\PQT}$.  The search is performed in the two-dimensional space defined by $m^\text{reco}_{\PQT}$ and $m^\text{reco}_{\cPGf}$.

The remainder of the paper is organized as follows. The CMS detector and particle reconstruction are introduced in Section~\ref{sec:reconstruction}. Data sets and Monte Carlo (MC) simulated samples are described in Section~\ref{sec:signalandbackground}. The event selection is detailed in Section~\ref{sec:objectselection}. The statistical model and background estimation are presented in Section~\ref{sec:statmodel}, followed by the systematic uncertainties in Section~\ref{sec:sysunc}. The results are presented in Section~\ref{sec:results}, followed by the summary in Section~\ref{sec:summary}. Tabulated results are provided in the HEPData record for this analysis~\cite{hepdata}.

\section{The CMS detector and event reconstruction}\label{sec:reconstruction}

The CMS apparatus~\cite{CMS:2008xjf,CMS:2023gfb} is a multipurpose, nearly hermetic detector, designed to trigger on~\cite{CMS:2020cmk,CMS:2016ngn,CMS:2024aqx} and identify electrons, muons, photons, and (charged and neutral) hadrons~\cite{CMS:2020uim,CMS:2018rym,CMS:2014pgm}. Its central feature is a superconducting solenoid of 6\unit{m} diameter, providing an axial magnetic field of 3.8\unit{T}. Within the solenoid volume lie a silicon pixel and strip tracker, a lead tungstate crystal electromagnetic calorimeter (ECAL), and a brass-scintillator sampling hadron calorimeter (HCAL), each composed of a barrel and two endcap sections. Forward calorimeters extend the pseudorapidity, $\eta$, coverage provided by the barrel and endcap detectors. Muons are reconstructed using gas-ionization detectors interleaved with the layers of the steel flux-return yoke outside the solenoid. More detailed descriptions of the CMS detector, together with a definition of the coordinate system used and the relevant kinematic variables, can be found in Refs.~\cite{CMS:2008xjf,CMS:2023gfb}.

Events of interest are selected using a two-tiered trigger system. The first level (L1), composed of custom hardware processors, uses information from the calorimeters and muon detectors to select events at a rate of around 100\unit{kHz} within a fixed latency of 4\mus~\cite{CMS:2021yvr,CMS:2020cmk}. The second level, known as the high-level trigger, consists of a farm of processors running a version of the full event reconstruction software optimized for fast processing, and reduces the event rate to a few kHz before data storage~\cite{CMS:2016ngn,CMS:2024aqx}.

During the 2016 and 2017 data-taking periods, a gradual shift in the timing of the inputs of the ECAL L1 trigger in the $\abs{\eta} > 2.0$ region caused a trigger inefficiency, known as ``prefiring''~\cite{CMS:2021yvr,CMS:2020cmk}. The trigger missed between 10 and 20\% of the events containing an electron (of transverse momentum $\pt \gtrsim 50$\GeV) or a jet (of $\pt \gtrsim 100$\GeV) in the range $2.5 < \abs{\eta} < 3.0$, the exact loss depending on the \pt and $\eta$ of the trigger object. Correction factors were computed from data and applied to the acceptance evaluated by simulation.

The global, also called particle-flow (PF), event reconstruction~\cite{CMS:2017yfk} aims to reconstruct and identify each individual particle in an event by optimally combining all subdetector information. In this process, the identification of the particle type (photon, electron, muon, charged and neutral hadron) plays an important role in the determination of the particle direction and energy. Photons (\eg, coming from \PGpz decays or from electron bremsstrahlung) are identified as ECAL energy clusters not linked to the extrapolation of any charged particle trajectory to the ECAL. Electrons (\eg, those coming from photon conversions in the tracker material or from \PB hadron semileptonic decays) are identified as primary charged particle tracks and potentially associated with many ECAL energy clusters corresponding to each track extrapolation to the ECAL and to possible bremsstrahlung photons emitted along the way through the tracker material. Muons (\eg, from \PB hadron semileptonic decays) are identified as tracks in the silicon tracker consistent with either a track or several hits in the muon system, and associated with calorimeter deposits compatible with the muon hypothesis. Charged hadrons are identified as charged particle tracks neither identified as electrons, nor as muons. Finally, neutral hadrons are identified as HCAL energy clusters not linked to any charged hadron trajectory, or as a combined ECAL and HCAL energy excess with respect to the expected charged hadron energy deposit.

The energies of photons are obtained solely from the ECAL measurements. The energy of each electron is determined from a combination of the track momentum at the primary interaction vertex, the corresponding ECAL cluster energy, and the energy sum of all bremsstrahlung photons associated with the track.  The energies of muons are obtained from the corresponding track momenta. The energies of charged hadrons are determined from combinations of the track momenta and the ECAL and HCAL energies, which are corrected for the response to hadronic showers.  Finally, the energies of neutral hadrons are obtained from the corrected ECAL and HCAL energies.

A multivariate discriminant~\cite{Hoecker:2007bk} is used for photon identification~\cite{CMS:2020uim}. This discriminant combines several input variables to distinguish prompt from nonprompt photons which primarily arise from neutral-meson decays. There are different working points corresponding to photon identification efficiencies of approximately 80\% and 90\% available. This analysis uses the 80\% efficiency working point (WP80)~\cite{CMS:2020uim}. While the discriminant incorporates isolation variables in its training, it does not impose explicit isolation criteria. Electrons misidentified as photons are suppressed by rejecting photon candidates that are matched to an electron track originating from the primary vertex, while retaining candidates consistent with photon conversions.

In the barrel section of the ECAL, an energy resolution of approximately 1\% is achieved for unconverted or late-converting photons with energies of tens of \GeV. The energy resolution for the remaining barrel photons is about 1.3\% up to $\abs{\eta} = 1.0$, changing to about 2.5\% at $\abs{\eta} = 1.4$. Resolution in the endcaps is about 2.5\% for unconverted or late-converting photons, while it ranges between 3\% and 4\% for others~\cite{CMS:2015myp}.  The diphoton mass resolution, as measured in $\PH \to \PGg\PGg$ decays, is typically in the 1--2\% range, depending on the measured photon energies in the ECAL and the topology of the photons in the event~\cite{CMS:2020xrn}.

For each event, hadronic jets are clustered from PF particles using the infrared and collinear safe anti-\kt algorithm~\cite{Cacciari:2008gp,Cacciari:2011ma} with a distance parameter of 0.4.  Jet momentum is determined as the vectorial sum of all particle momenta within the jet.  Simulation studies indicate that this calculation is, on average, accurate within 5 to 10\% of the true momentum over the entire \pt spectrum and detector acceptance. Additional proton-proton interactions within the same or nearby bunch crossings (pileup) can contribute extra tracks and calorimetric energy clusters, which may inflate the apparent jet momentum. To mitigate this effect, tracks identified to be originating from pileup vertices are discarded and an offset correction is applied to correct for any remaining contributions~\cite{Sirunyan:2020foa}. Jet energy corrections are derived from simulation studies, ensuring that the average measured energy of jets aligns with that of particle-level jets. In situ measurements of the momentum balance in dijet, $\text{photon} + \text{jet}$, $\PZ + \text{jet}$, and multijet events are used to determine any residual differences between the jet energy scale in data and in simulation, allowing appropriate corrections to be made~\cite{CMS:2016lmd}. Additional selection criteria are applied to each jet to eliminate those that may be affected by instrumental effects or reconstruction failures~\cite{Sirunyan:2020foa}.

Anti-\kt jets with a distance parameter of 0.8 (AK8 jets) are used to reconstruct boosted top quarks.  As boosted jets are affected by soft, wide-angle radiation, a grooming technique is used to remove peripheral soft components of the jet as well as to provide a more reliable and precise measurement of the jet mass.  The groomed jet mass is calculated after applying to AK8 jets a modified mass-drop algorithm~\cite{Dasgupta:2013ihk,Butterworth:2008iy}, known as the soft drop declustering~\cite{Larkoski:2014wba}, with the soft-drop angular exponent $\beta=0$, and soft threshold $z_\text{cut}=0.1$. The groomed jet mass is calibrated in a top quark pair enriched sample~\cite{Sirunyan:2020foa}.

The ParticleNet algorithm operates on the constituent particles of a jet using a dynamic graph convolutional neural network architecture.  It provides discriminant scores for different jet hypotheses. Specifically, the top quark ParticleNet discriminant can be used to identify top quark candidate jets and to suppress jets that originate from light quarks or gluons. The working point used in this analysis corresponds to a misidentification rate of 0.5\% for light-quark and gluon jets while retaining an efficiency of about 40\% for top quark jets~\cite{Qu:2019gqs}.

\section{Data and simulated samples}\label{sec:signalandbackground}

Signal MC samples are generated at leading order (LO) using \MGvATNLO version 2.6.5~\cite{Alwall:2011uj} with the VLQ pair-production model \textsc{eVLQ}\_v8.1~\cite{Banerjee:2022evx}, assuming a four-flavor scheme (4FS). The samples are produced on a two-dimensional mass grid consisting of 55 mass combinations, covering $100 \leq m_{\cPGf} \leq 300\GeV$ in steps of 50\GeV and $800 \leq m_{\PQT} \leq 1800\GeV$ in steps of 100\GeV.

The \PQT pair-production cross section is determined by the strong interaction and is independent of the $\PQT \to \cPGf\PQt$ coupling strength, $\kappa_{\PQT\cPGf\PQt}$. The left- and right-handed coupling strengths control the $\PQT \to \cPGf\PQt$ partial width and are set to $\kappa_{\PQT\cPGf\PQt}^L = \kappa_{\PQT\cPGf\PQt}^R = 4.0$ in the signal simulation. The total \PQT and \cPGf widths are set to 10\% and 1\% of their respective masses~\cite{Benbrik:2019zdp}. The resulting \PQT width is sufficiently large that finite width effects may be relevant.  In the signal samples, one of the two \cPGf bosons is forced to decay to $\PGg\PGg$, while the second is left undecayed and is therefore detector-stable in the simulation. Since either of the two \cPGf bosons in a physical event can provide the diphoton decay, a factor of two is applied when interpreting the samples for the small $\mathcal{B}(\cPGf\to\PGg\PGg)$ values considered here. The absence of the physical decay products of the second \cPGf boson in the simulation may lead to a slight underestimate of the signal acceptance.

The major SM background processes include top quark pair ($\PQt\PAQt$) production with hadronic or semileptonic top quark decays, $\PQt\PAQt$ production in association with two photons ($\PQt\PAQt\PGg\PGg$), and $\PQt\PAQt$ in association with a photon and additional jets ($\PQt\PAQt\PGg+\text{jets}$). Samples of \PW or \PZ bosons produced in association with one or more jets are also included, although they contribute only a small fraction of the total background. The $\PQt\PAQt$ all hadronic and semileptonic samples are generated at NLO using \POWHEG 2.0~\cite{Frixione:2007vw}. The $\PQt\PAQt\PGg\PGg$ and $\PQt\PAQt\PGg+\text{jets}$ samples are generated at NLO using \MGvATNLO version 2.6.5, while the \PW+jets and \PZ+jets samples are generated at LO using \MGvATNLO version 2.6.5.

The simulated samples use the NNPDF3.1 parton distribution function (PDF) set~\cite{Ball:2017nwa}. Parton showering and hadronization are simulated using \PYTHIA~8.240 with the CP5 event tune~\cite{Sjostrand:2014zea,CMS:2019csb}. To account for differences in the number of additional proton-proton interactions between data and simulation, simulated events are reweighted according to the ratio of the pileup distributions in data and simulation. The detector response for simulated events is modeled using the \GEANTfour-based CMS detector simulation~\cite{Agostinelli:2002hh}.

\section{Event selection}
\label{sec:objectselection}

The triggers used in the baseline event selection comprise all available diphoton triggers and a variety of \HT triggers with thresholds from 800 to 1050\GeV.  Online, \HT is the scalar sum of the $\pt$ of all selected jets in the event.  Offline, \HT is defined as the scalar sum of the $\pt$ of AK8 jets with $\pt >$ 350\GeV and $\abs{\eta}<2.4$.  An offline requirement of $\HT > 900\GeV$ is applied to select a region of phase space where the trigger efficiency depends only weakly on the photon $\pt$, jet $\pt$, and \HT.  The trigger efficiency is measured in an independent single muon data sample as a function of \HT, jet $\pt$, and photon $\pt$. For \HT$< 1500\GeV$ the average efficiency is roughly 80\%.

Events must contain at least two AK8 jets, each being consistent with the decay products of a top quark. Candidate jets must satisfy $\pt > 350\GeV$, $\abs{\eta}<2.4$, and a groomed jet mass of $m^{\text{SD}} > 50\GeV$. Jet pairs are arranged in descending order of jet \pt, and the first pair satisfying $\abs{\Delta\eta} < 1.6$ is selected. Events for which no jet pair satisfies this requirement are rejected. The two selected AK8 jets must be top quark tagged, with ParticleNet discriminant scores exceeding the threshold corresponding to a 0.5\% misidentification rate.

Events are required to contain at least two reconstructed photons corresponding to the decay products of one of the scalar bosons \cPGf, each satisfying $\pt > 25\GeV$ and $\abs{\eta}<2.5$.  Only photons reconstructed in the ECAL barrel ($\abs{\eta}<1.44$) or endcap region ($1.57 < \abs{\eta} < 2.5$) are considered.  Photons in the barrel--endcap transition region are excluded because of the increased amount of inactive material associated with detector services and support structures, which degrades the photon energy measurement and identification performance.  Photon candidates with multivariate discriminant values greater than $-0.75$ are prioritized in the photon baseline selection as follows: if more than two photon candidates are present, the photons which pass the discriminant score of $-0.75$ are selected first.  If only one or no photons pass, the remaining photon(s) are selected according to largest \pt.

After the baseline selection described above, the two photons are used to reconstruct the scalar boson candidate invariant mass, $m^\text{reco}_{\cPGf}$. The \PQT quark candidate is reconstructed by pairing the \cPGf candidate with one of the two AK8 jets to obtain the invariant mass, $m^\text{reco}_{\PQT}$.  Boosted \PQT decay products are characterized as back-to-back in the lab, with \PQT decay products favoring large angular separation and informing the selection strategy used in the search.  The reconstructed jet with the larger angular separation from the reconstructed diphoton system, $\Delta R_{(j,\PGg\PGg)}$, is selected, thereby improving the probability of selecting the correct jet assignment. The efficiency of selecting the correct jet to pair with the scalar boson ranges from 50 to 75\% for most signal mass points. For the two mass points corresponding to the smallest $m^\text{reco}_{\PQT}$ values considered, the efficiency decreases to about 40\%.

\begin{figure*}[!ht]
\centering
\includegraphics[width=0.48\textwidth]{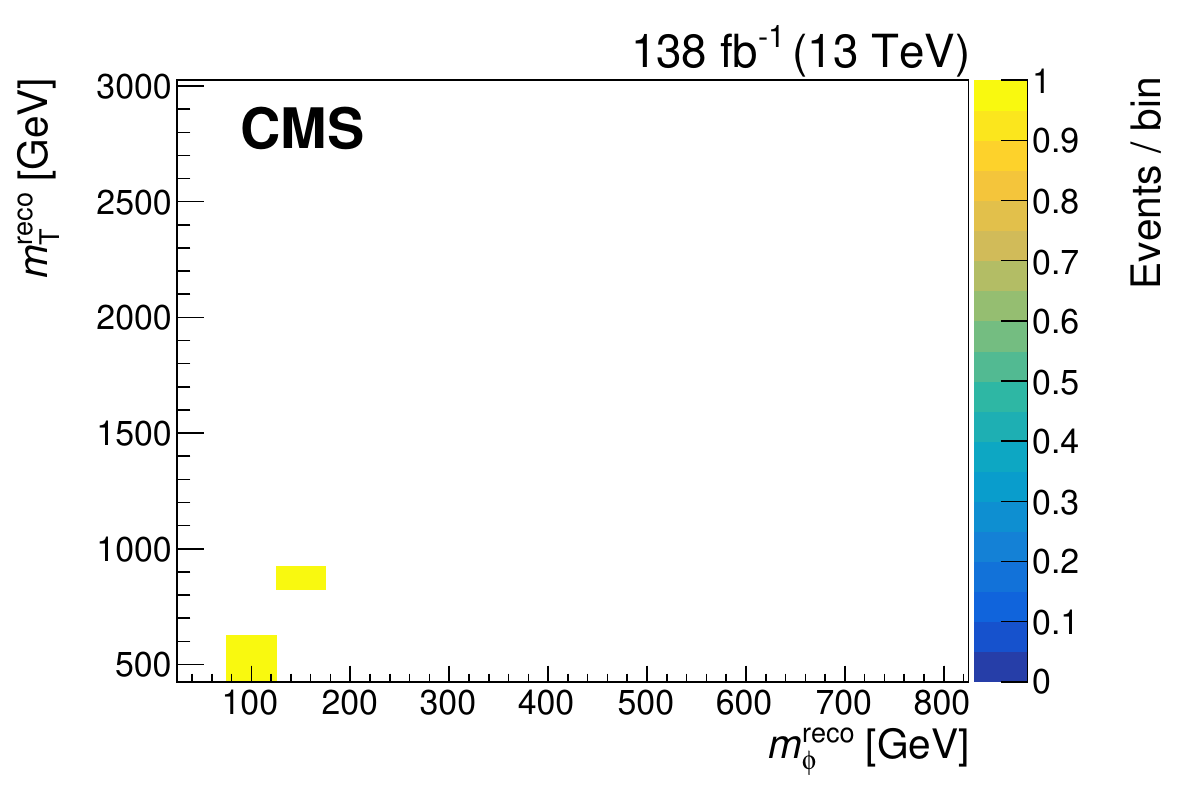} \quad
\includegraphics[width=0.48\textwidth]{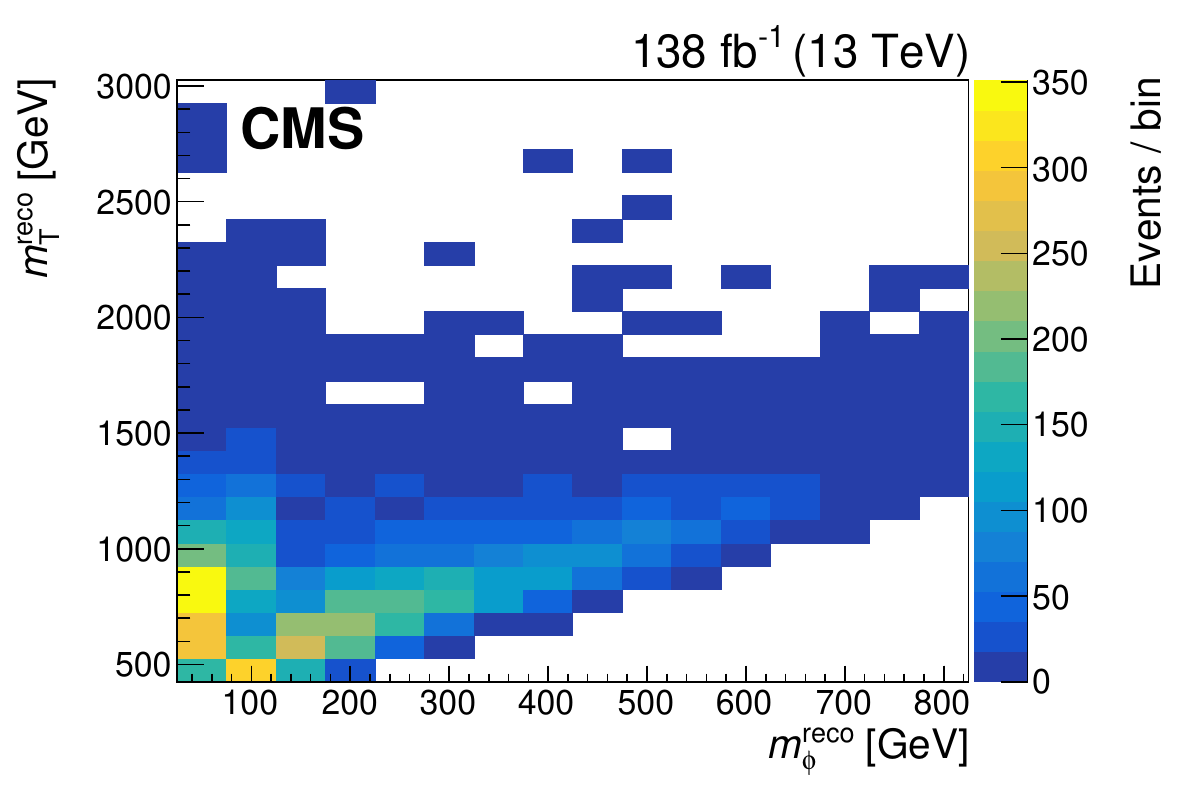} \\
\includegraphics[width=0.48\textwidth]{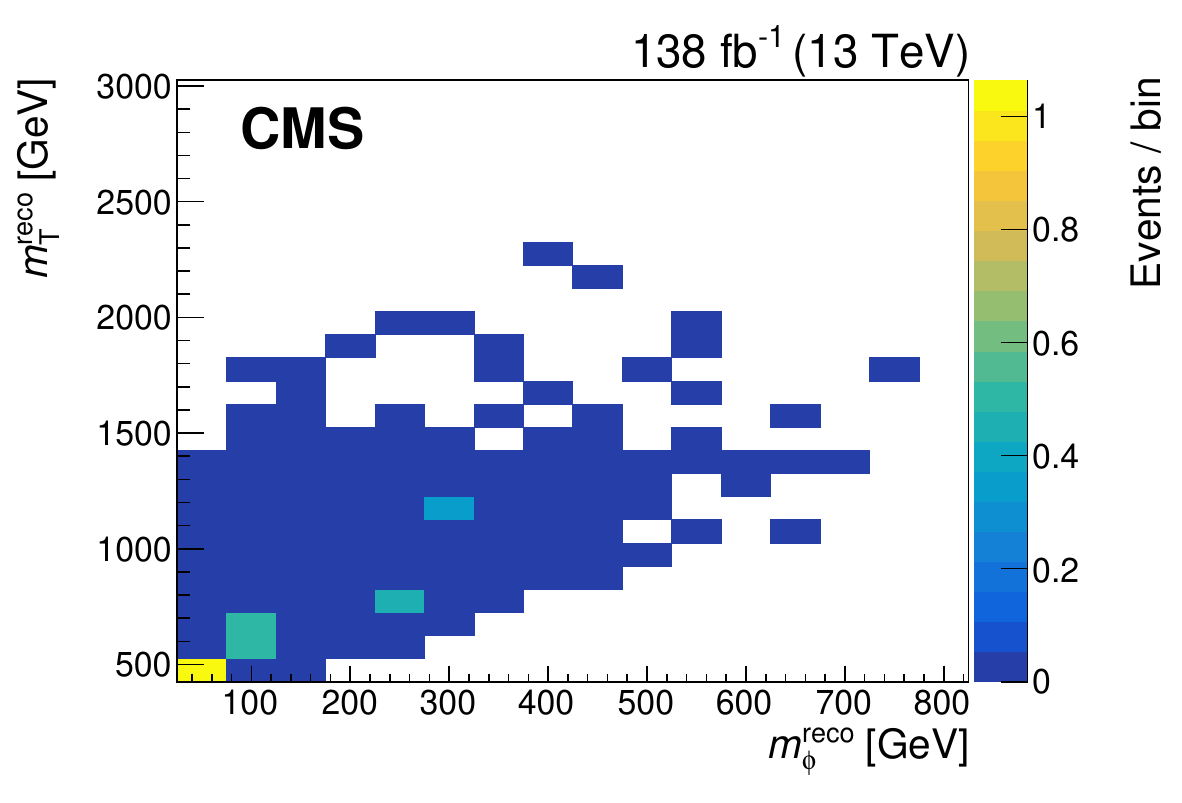} \quad
\includegraphics[width=0.48\textwidth]{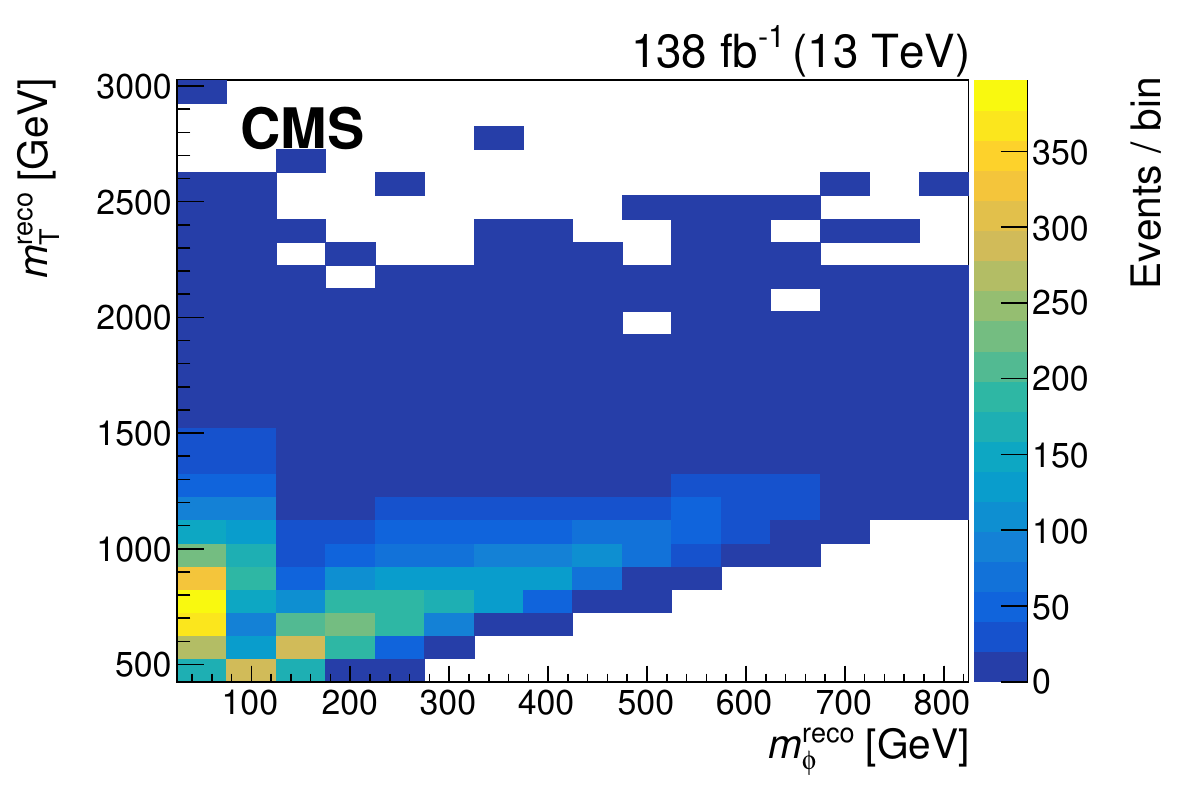}
            \caption{Distribution of reconstructed \PQT candidates in the $m^\text{reco}_{\PQT}$--$m^\text{reco}_{\cPGf}$ plane for data (upper) and simulated $\PQt\PAQt$ events (lower), shown for the SR (left) and CR (right).  The simulated $\PQt\PAQt$ events are normalized to sum to the total data events in the SR and the CR, respectively.}
            \label{fig:DataVsTtbar}
\end{figure*}

Two regions are defined according to whether the photons pass or fail the WP80 identification criterion. The signal region (SR) contains events with two selected photons passing the WP80 identification requirement, while the control region (CR) contains events with both photons failing this requirement. Table~\ref{tab:eventsel} summarizes the baseline selection, SR, and CR definitions. Figure~\ref{fig:DataVsTtbar} shows the distributions in the $m^\text{reco}_{\PQT}$--$m^\text{reco}_{\cPGf}$ plane for data and simulated $\PQt\PAQt$ events in the SR and CR.  For the ease of comparison, the simulated $\PQt\PAQt$ events are normalized to sum to the total data events in the SR and the CR, respectively.  The similarity of the yield distributions in CR data and CR $\PQt\PAQt$ motivates the background prediction strategy.  In both cases, we see a concentration in the yields for small $m_\cPGf^\text{reco}$.  The diagonal concentration of yields is a natural result of the correlation between $m^\text{reco}_{\PQT}$ and $m^\text{reco}_{\cPGf}$.

\begin{table*}[!htbp]
\centering
\caption{The baseline, signal region, and control region event selection.}
\label{tab:eventsel}
\begin{scotch}{ccc}
 & \multicolumn{2}{c}{Baseline selection} \\
\hline
$\HT$ & \multicolumn{2}{c}{$>900\GeV$} \\
Number of AK8 jets & \multicolumn{2}{c}{$\geq 2$ ($\pt > 350\GeV$, $\abs{\eta}<2.4$, $m^{\text{SD}} > 50\GeV$) } \\
Highest $\pt$ AK8-jet pair satisfying $\abs{\Delta\eta}$   & \multicolumn{2}{c}{$<1.6$} \\
Selected jets pass the top quark tag & \multicolumn{2}{c}{$2$} \\
Number of photons & \multicolumn{2}{c}{$\geq 2$ ($\pt > 25\GeV$, $\abs{\eta}<1.44$, $1.57<\abs{\eta}<2.5$)} \\
\\
 & Signal region (SR) & Control region (CR)\\[\cmsTabSkip]
Number of selected WP80 photons & 2 & 0 \\ 
\end{scotch}
\end{table*}

Data and simulated events are organized across five $m_\cPGf$ bins, centered at 100 to 300\GeV in 50\GeV increments and eleven $m_\PQT$ bins, centered at 800 to 1800\GeV in 100\GeV increments, comprising a subset of the two-dimensional mass grid shown in Fig.~\ref{fig:DataVsTtbar}.  Yields corresponding to the simulated signal sample with $m_\cPGf=100\GeV$ and $m_\PQT=1000\GeV$ organized across this two-dimensional array sum up to approximately 72 in the SR and 39 in the CR.

\section{Statistical model and background estimation}\label{sec:statmodel}

Backgrounds in the SR arise primarily from events containing nonprompt photons, such as events with $\pi^0$ decays, or objects misidentified as photons, such as electrons.  The contribution of such backgrounds is estimated using data events in the CR, where the reconstructed photons fail the identification requirement, and are extrapolated to the SR.

After reconstructing the \cPGf and \PQT candidates, a maximum-likelihood fit in the two-dimensional $(m^\text{reco}_{\PQT}, m^\text{reco}_{\cPGf})$ plane is used to predict the background in the SR from events in the CR. The CR background is decomposed into a $\PQt\PAQt$-dominated component and a smaller \PW/\PZ+jets contribution. A transfer function (TF), $R = \re^{-\alpha - \beta\, m^\text{reco}_{\cPGf}}$, relates the $\PQt\PAQt$-dominated component in the CR to that in the SR.  The exponential form is chosen because it guarantees a positive SR prediction and provides stable fits in the low-yield SR. Alternative linear and quadratic parameterizations were found to produce negative predicted yields and/or unstable fits.  Fitting $R$ independently in slices of $m^\text{reco}_{\PQT}$ yields values of $\alpha$ and $\beta$ that are compatible within their uncertainties, supporting the assumption that the aforementioned TF depends on $m^\text{reco}_{\cPGf}$ only.  The extrapolation is further tested in a validation region (VR) described below.

The background includes $\PQt\PAQt\PGg\PGg$, $\PQt\PAQt\PGg$+jets, prompt $\PGg\PGg$, $\PQt\PAQt$ all hadronic and semileptonic contributions, along with contributions from electrons misidentified as photons.  To model this, event yields in the SR and CR are used to predict the number of background events in the SR for a given bin $i$ in the two-dimensional mass plane, $n^{\text{Pred BG}}_{\text{SR}}(i)$, as
\begin{equation}
n^{\text{Pred BG}}_{\text{SR}}(i) =
R\, n^{\PQt\PAQt}_{\text{CR}}(i)
+ n^{\text{V+jets}}_{\text{SR}}(i) .
\label{eq:first}
\end{equation}
Here $n^{\PQt\PAQt}_{\text{CR}}(i)$ denotes the floating yield of the $\PQt\PAQt$-dominated background component in bin $i$ of the CR, and $R$ is the TF defined earlier. The term $n^{\text{V+jets}}_{\text{SR}}(i)$ represents the contribution from $\PW/\PZ+$jets events in bin $i$ of the SR obtained from simulation, also with floating normalization.  Setting $R$ to 1 and replacing $n^{\text{V+jets}}_{\text{SR}}(i)$ with $n^{\text{V+jets}}_{\text{CR}}(i)$ provides the prediction for the number of background events in the CR corresponding to a bin $i$ in the two-dimensional mass plane, $n^{\text{Pred BG}}_{\text{CR}}(i)$.  The performance of the statistical model is thus predicated on how well the shape of the $\PQt\PAQt$ distribution in the CR predicts the shape of data in the CR.

The parameters of the background model are determined by maximizing a likelihood constructed from the event yields observed in the SR and CR, $n^{\mathrm{Data}}_{\mathrm{SR}}$ and $n^{\text{Data}}_{\mathrm{CR}}$, respectively. The negative log-likelihood is written as
\begin{equation}
\begin{aligned}
-\ln L ={}&
\sum_{i}^{N_{\mathrm{CR}}}
\left(
n^{\text{Pred}}_{\mathrm{CR}}(i)
- n^{\text{Data}}_{\mathrm{CR}}(i)
\ln n^{\text{Pred}}_{\mathrm{CR}}(i)
\right)
\\
&+
\sum_{i}^{N_{\mathrm{SR}}}
\left(
n^{\text{Pred}}_{\mathrm{SR}}(i)
- n^{\text{Data}}_{\mathrm{SR}}(i)
\ln n^{\text{Pred}}_{\mathrm{SR}}(i)
\right),
\end{aligned}
\label{eq:nll}
\end{equation}
where $N_{\text{SR}}$ and $N_{\text{CR}}$ denote the total number of bins in the SR and CR, respectively.

The model is fitted simultaneously in the SR and CR using the CMS statistical analysis tool \textsc{Combine}~\cite{CMS:2024onh,Verkerke:2003ir,Moneta:2010pm}. The fit determines the TF parameters $\alpha$ and $\beta$, along with the signal strength parameter $r$.  The $\PQt\PAQt$ normalization in each CR bin is treated as an unconstrained parameter in the fit. A background-only fit is obtained by excluding the signal component from the likelihood.

The validation of the background modeling is performed in the VR, which is introduced solely for this purpose.  To suppress potential signal contamination, events in the VR are required to fail the tight top quark tagging requirement corresponding to the 0.5\% misidentification rate. In addition, one of the two jets must fail the looser 1.0\% misidentification rate top quark tagger while the other satisfies it.  The photon selection defines two VR categories analogous to the SR and CR, VR-SR and VR-CR. The VR-SR requires both photons to pass the WP80 identification requirement, while the VR-CR requires both photons to fail this requirement.

\begin{figure*}[!ht]
\centering
\includegraphics[width=0.48\textwidth]{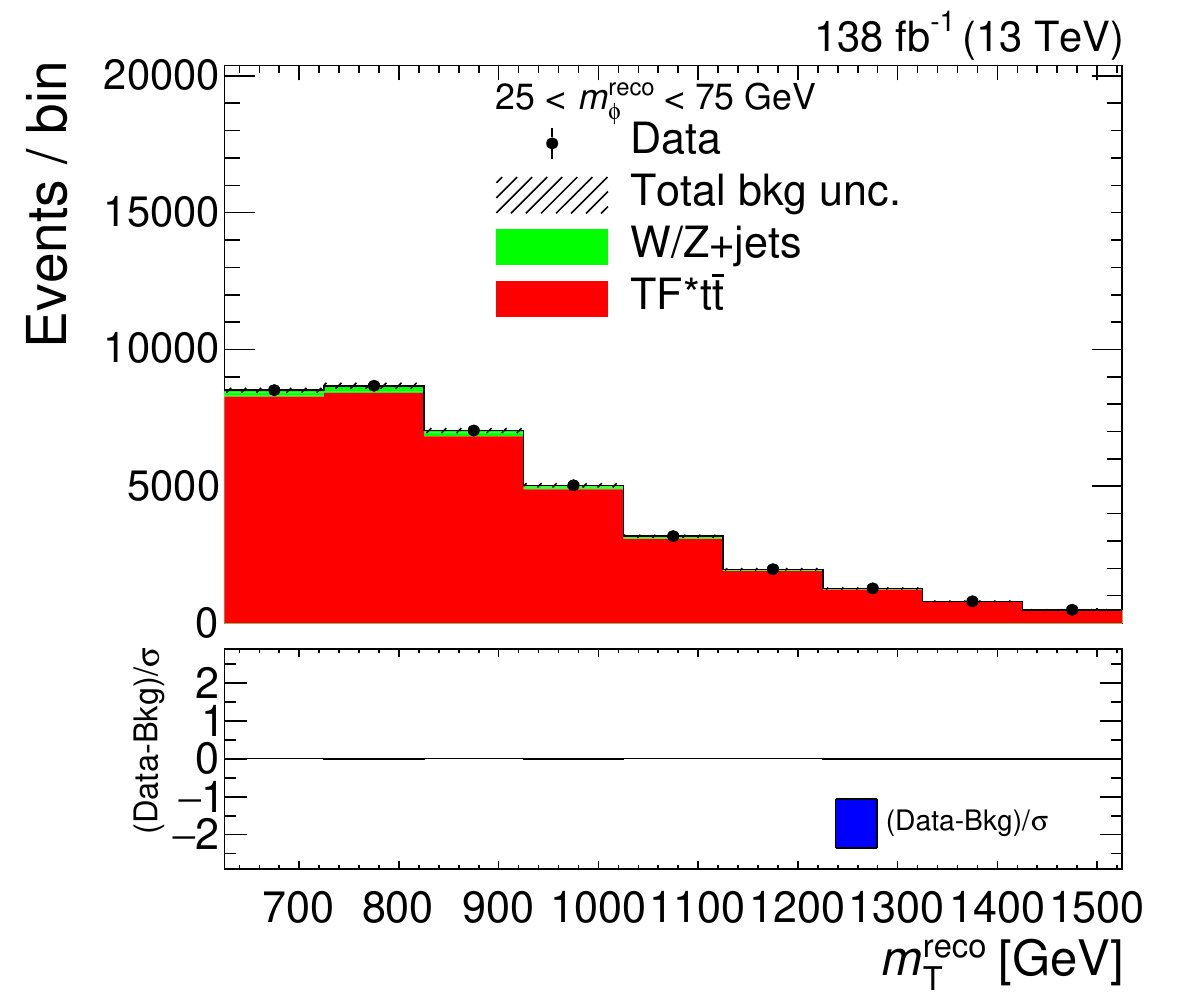}\
\includegraphics[width=0.48\textwidth]{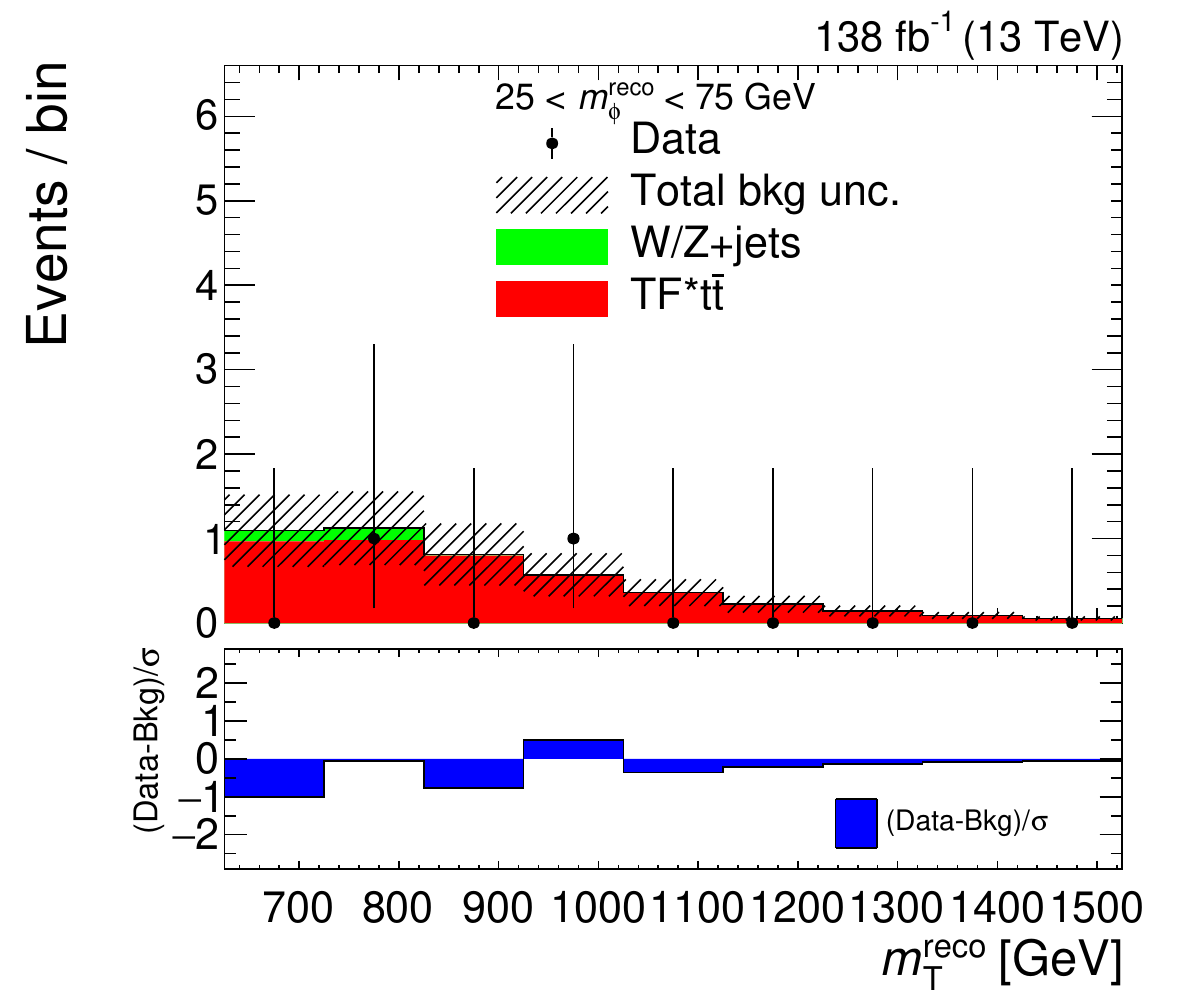}\\
\includegraphics[width=0.48\textwidth]{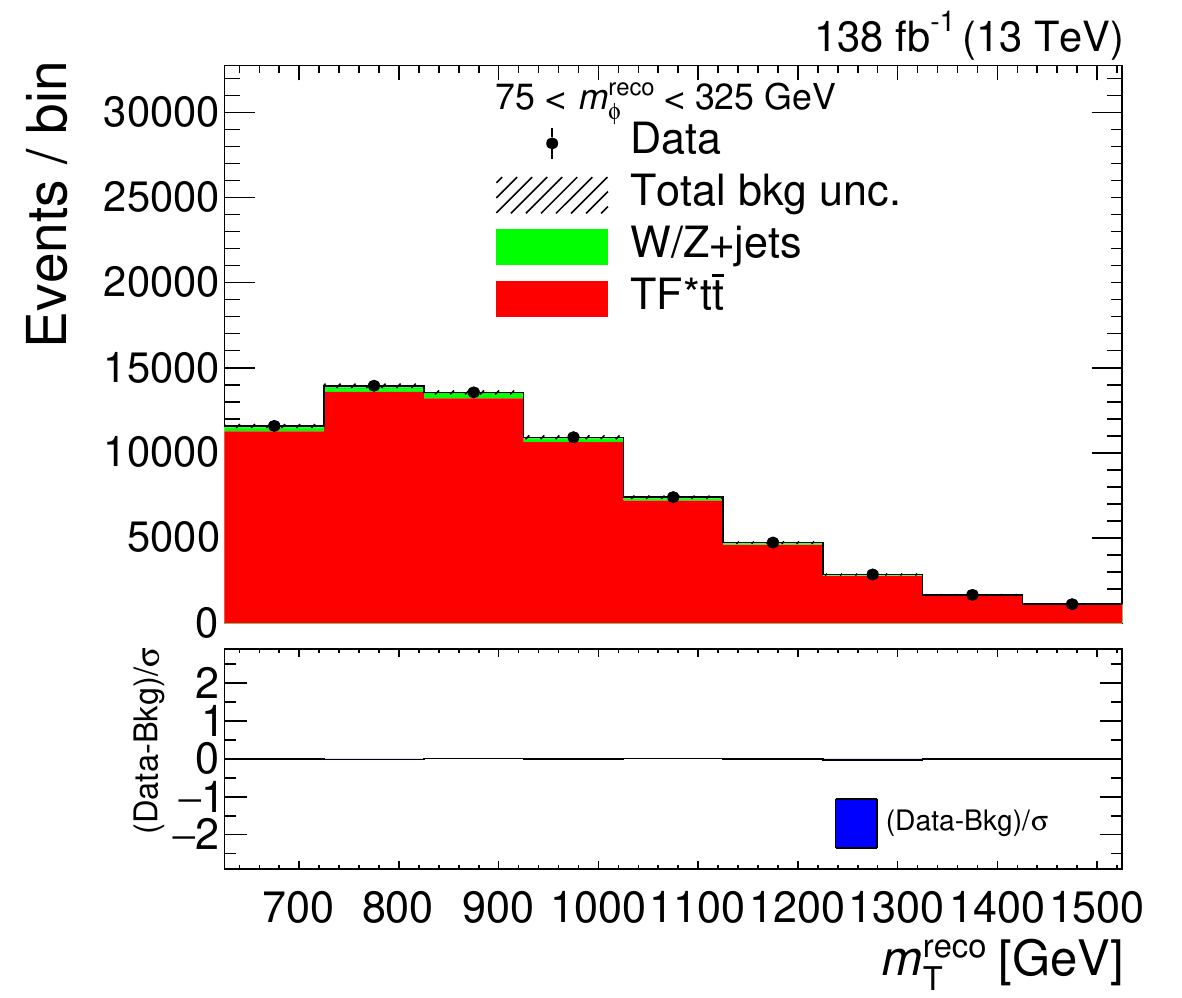}\
\includegraphics[width=0.48\textwidth]{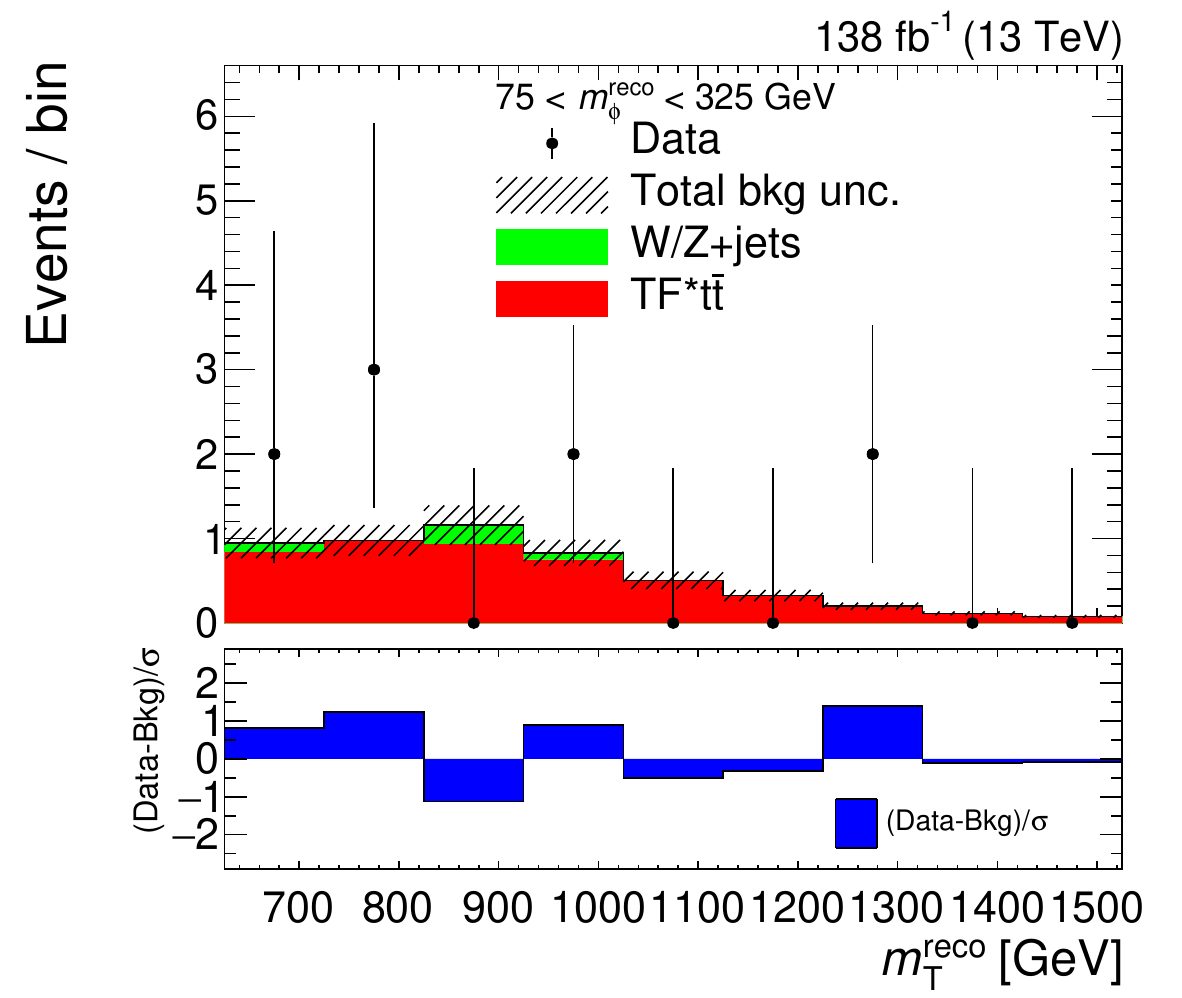}\\
\includegraphics[width=0.48\textwidth]{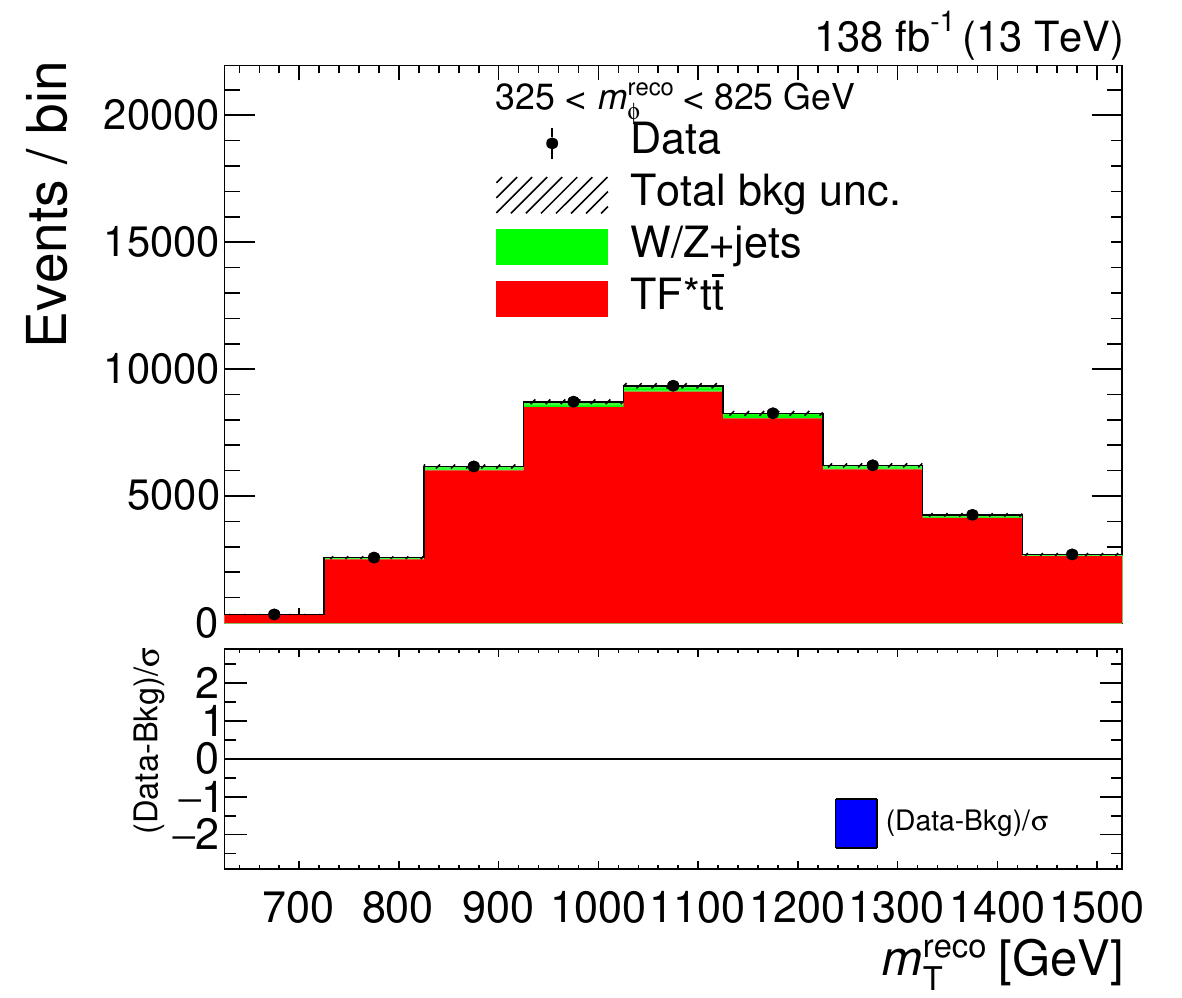}\
\includegraphics[width=0.48\textwidth]{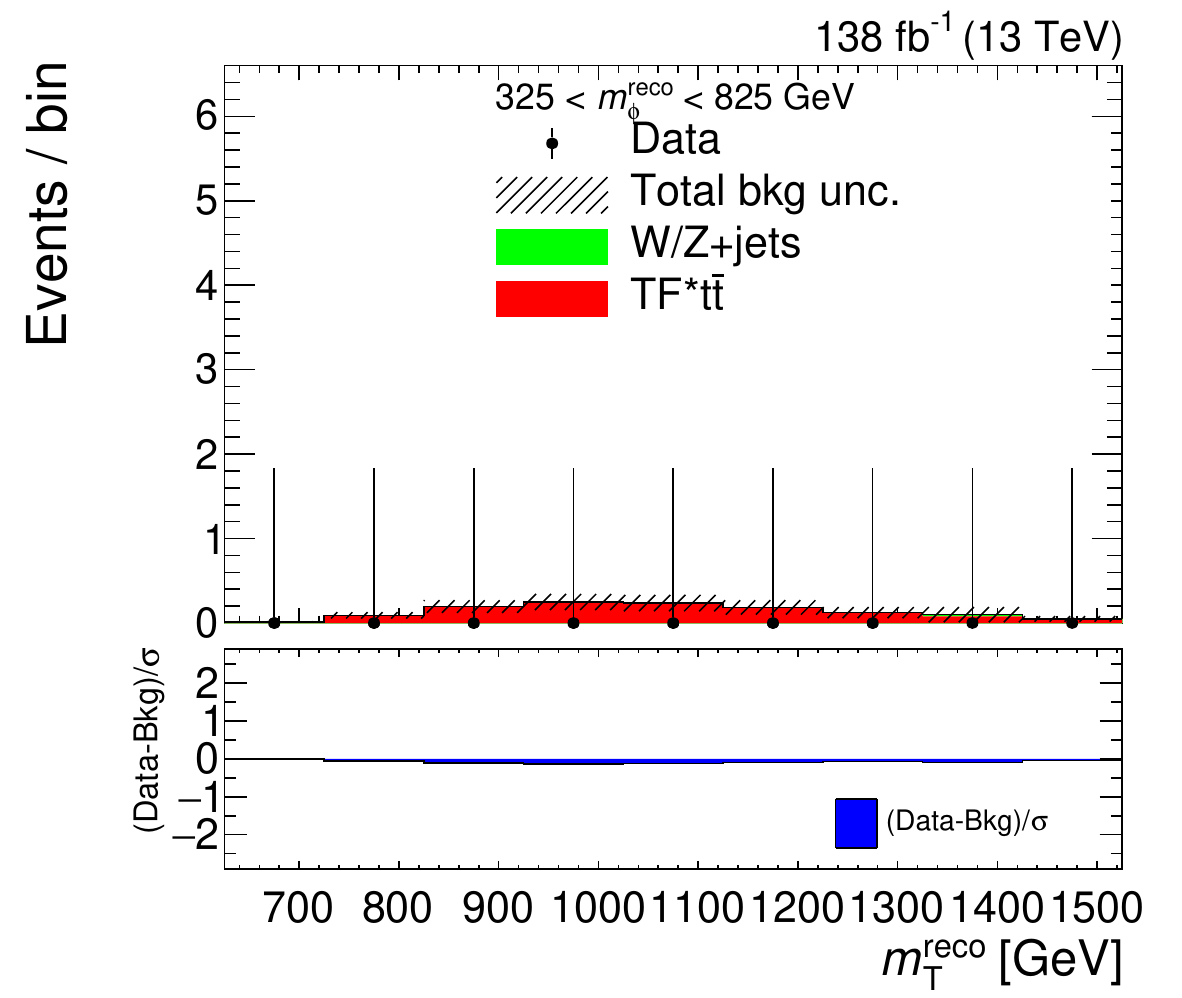}
            \caption{Comparison of data and postfit background predictions from the background-only fit in the VR. The VR-CR (left) and VR-SR (right) regions are shown. The distributions are projected onto $m^\text{reco}_{\PQT}$ in three intervals of $m^\text{reco}_{\cPGf}$: $25 < m^\text{reco}_{\cPGf} < 75\GeV$ (upper), $75 < m^\text{reco}_{\cPGf} < 325\GeV$ (middle), and $325 < m^\text{reco}_{\cPGf} < 825\GeV$ (lower). The $\PQt\PAQt$ component scaled by the transfer function is shown in red and the $\PW/\PZ$+jets contribution in green. The lower panels show the difference between the data and the background normalized by the uncertainty.}
            \label{fig:Fit_VVRT2}
\end{figure*}

\begin{figure*}[!ht]
\centering
\includegraphics[width=0.48\textwidth]{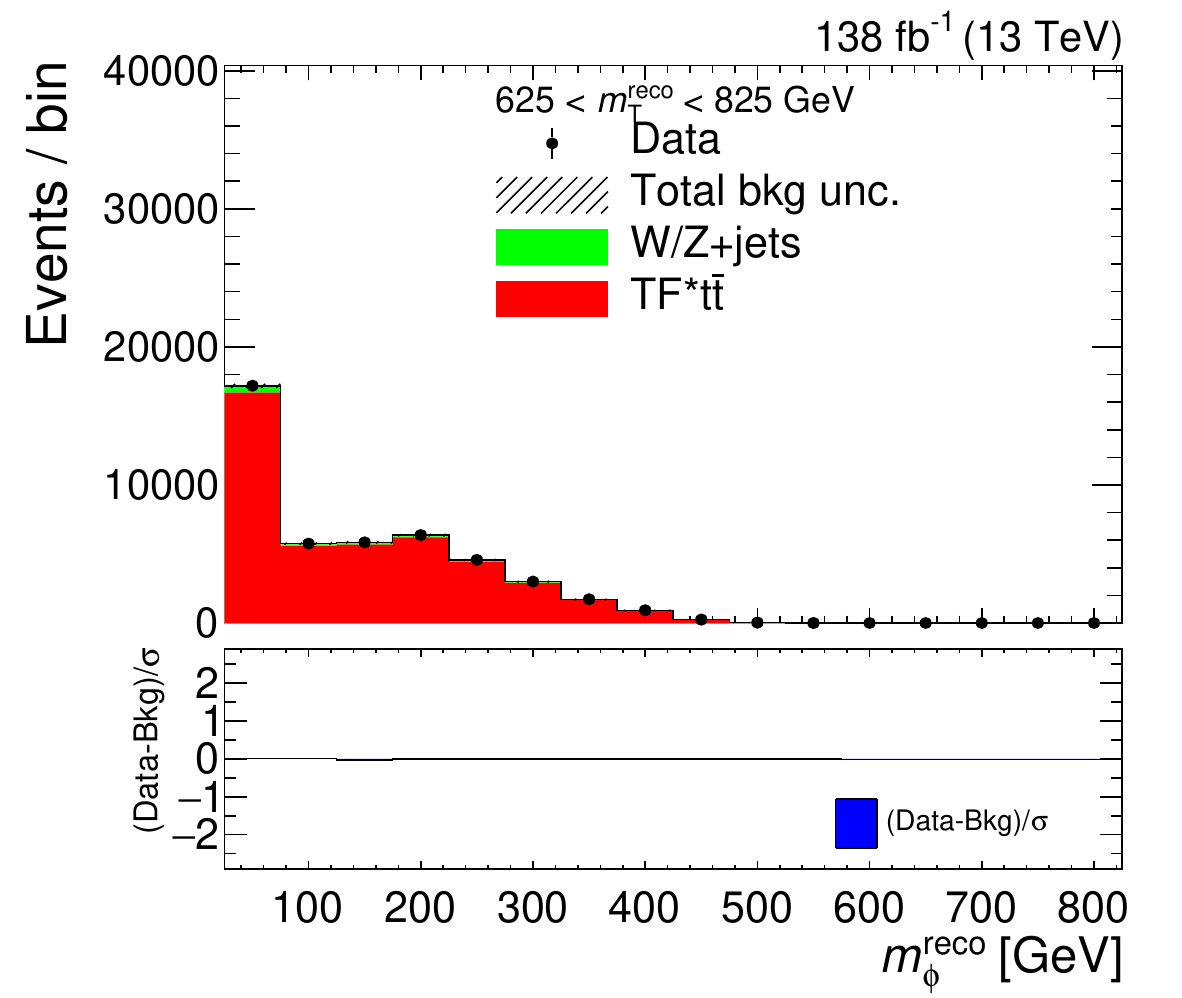}\
\includegraphics[width=0.48\textwidth]{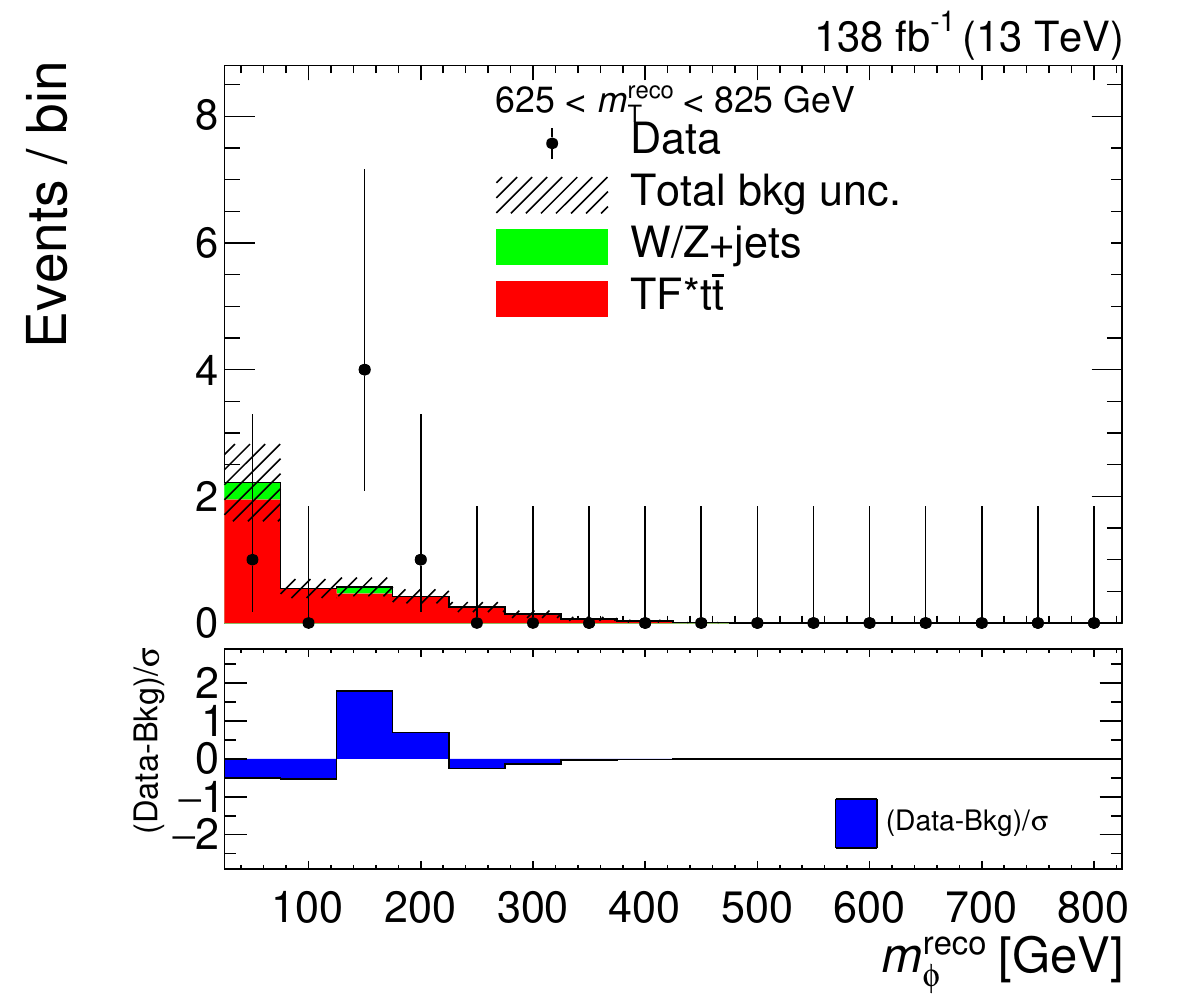}\\
\includegraphics[width=0.48\textwidth]{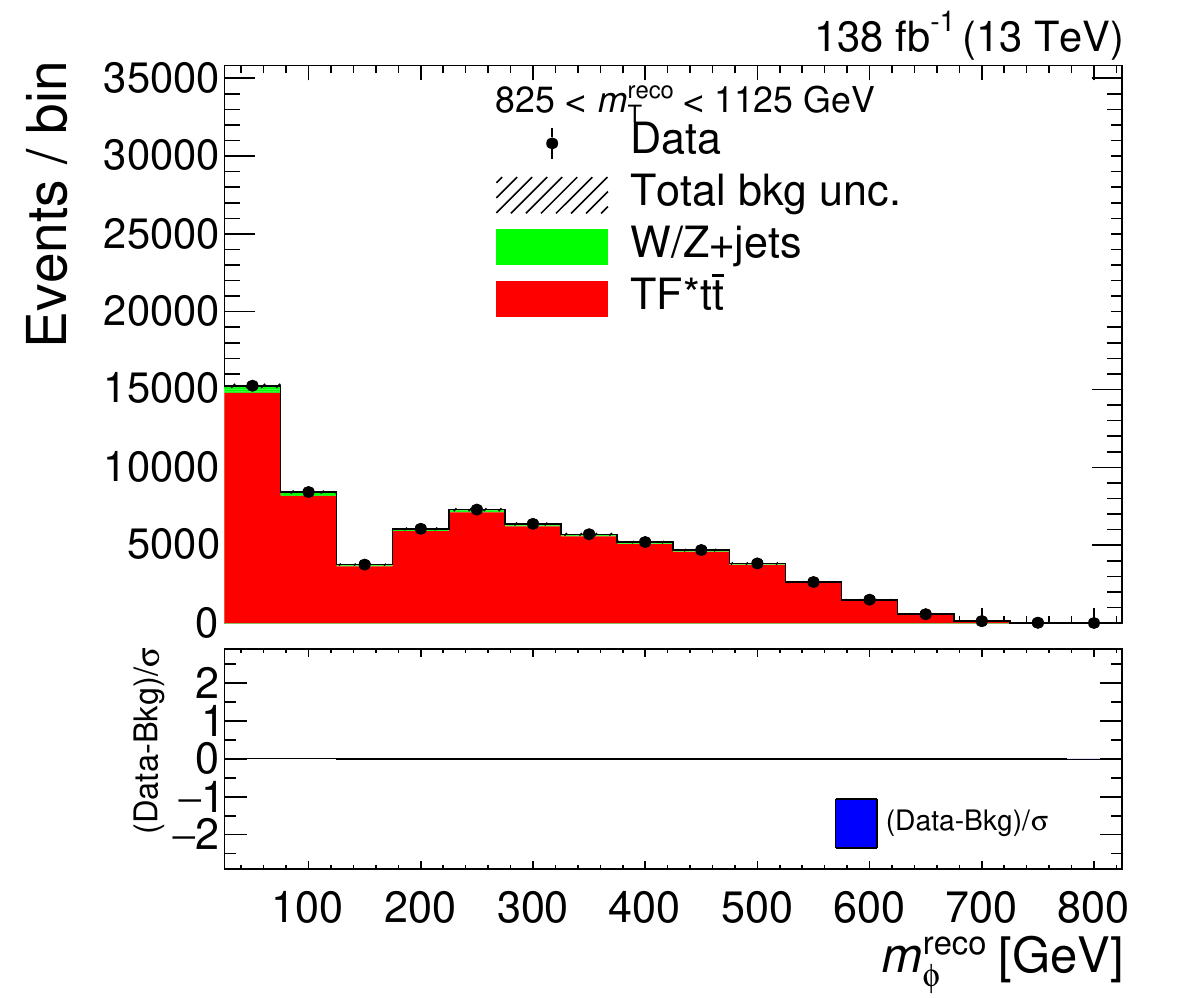}\
\includegraphics[width=0.48\textwidth]{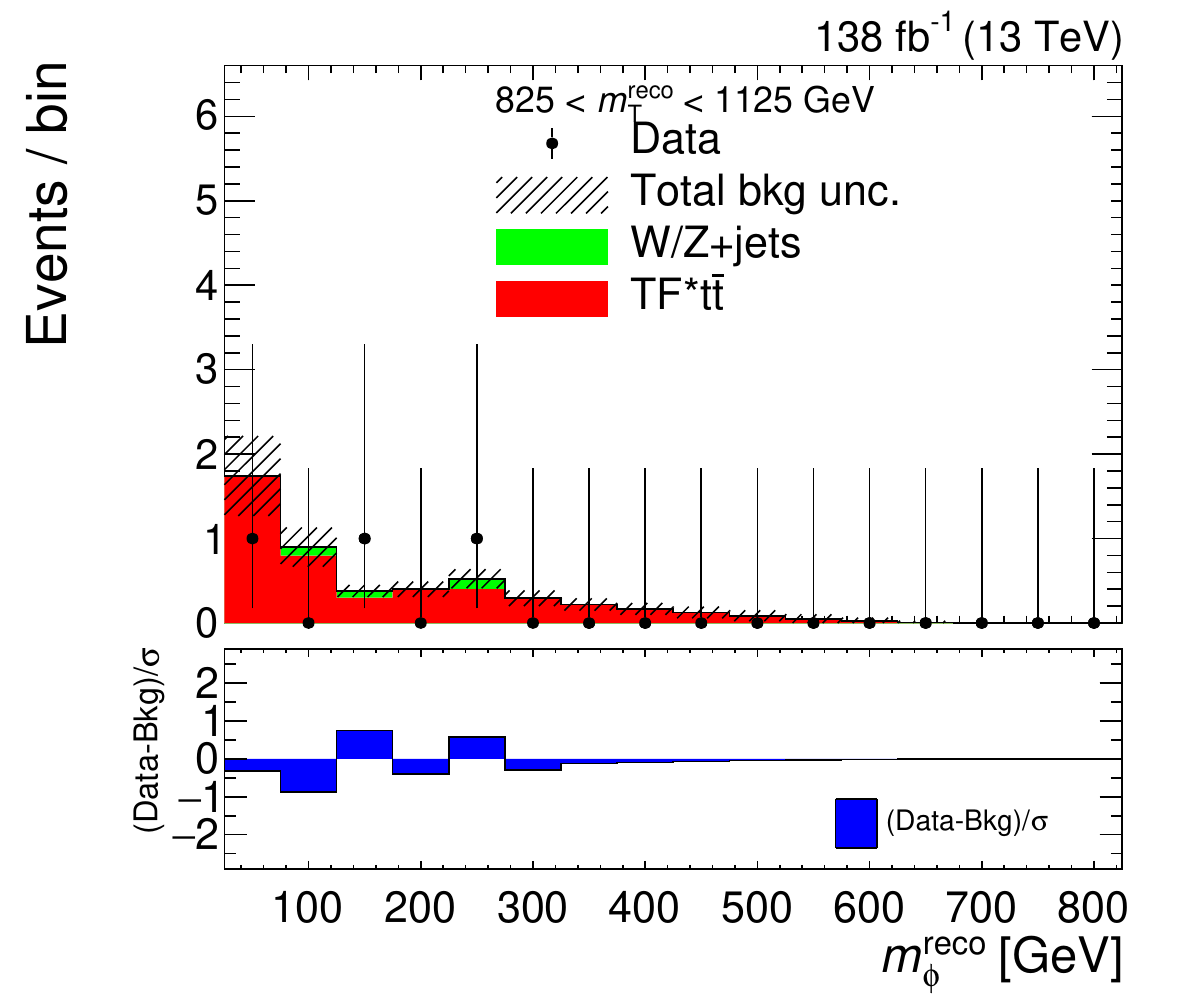}\\
\includegraphics[width=0.48\textwidth]{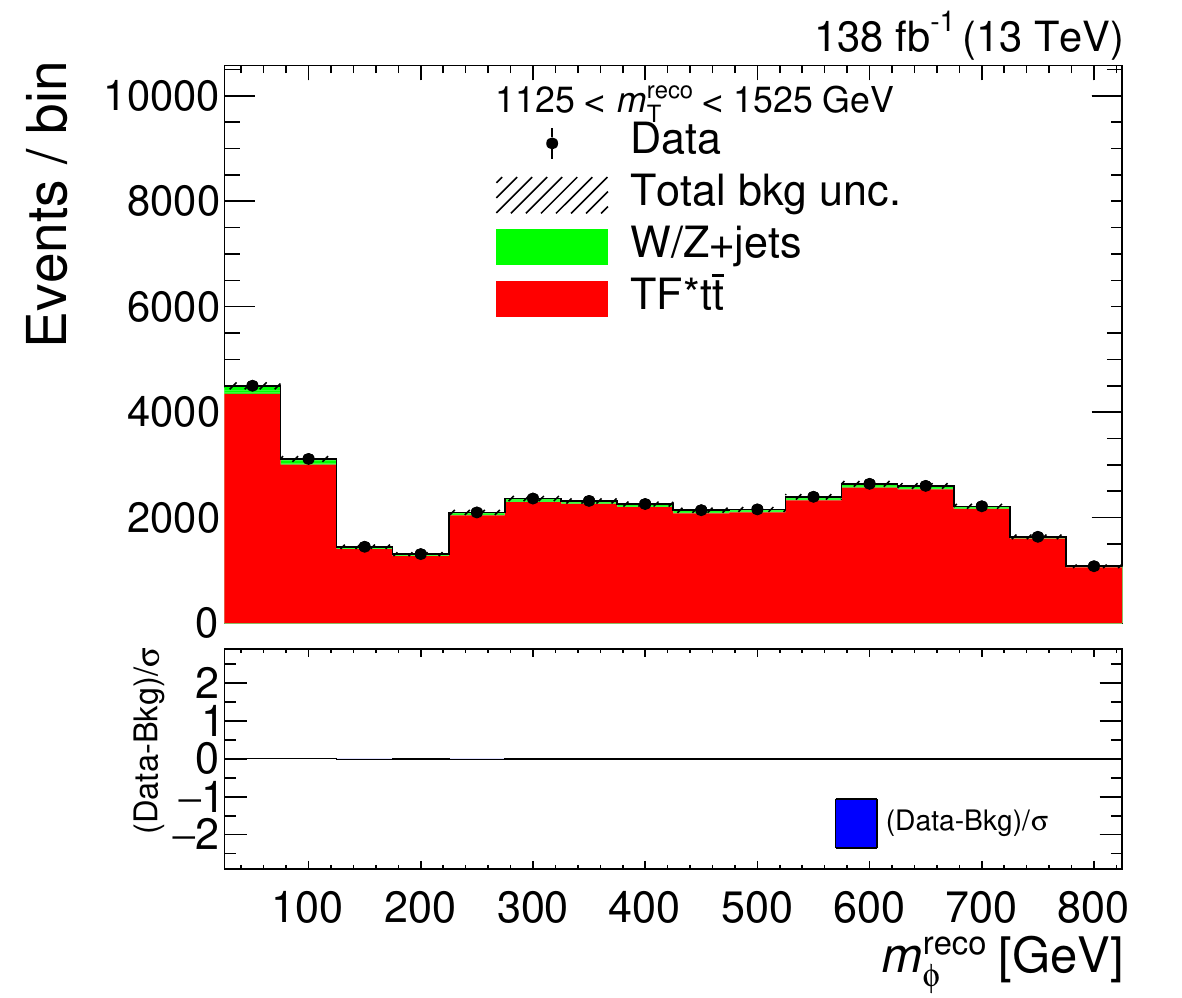}\
\includegraphics[width=0.48\textwidth]{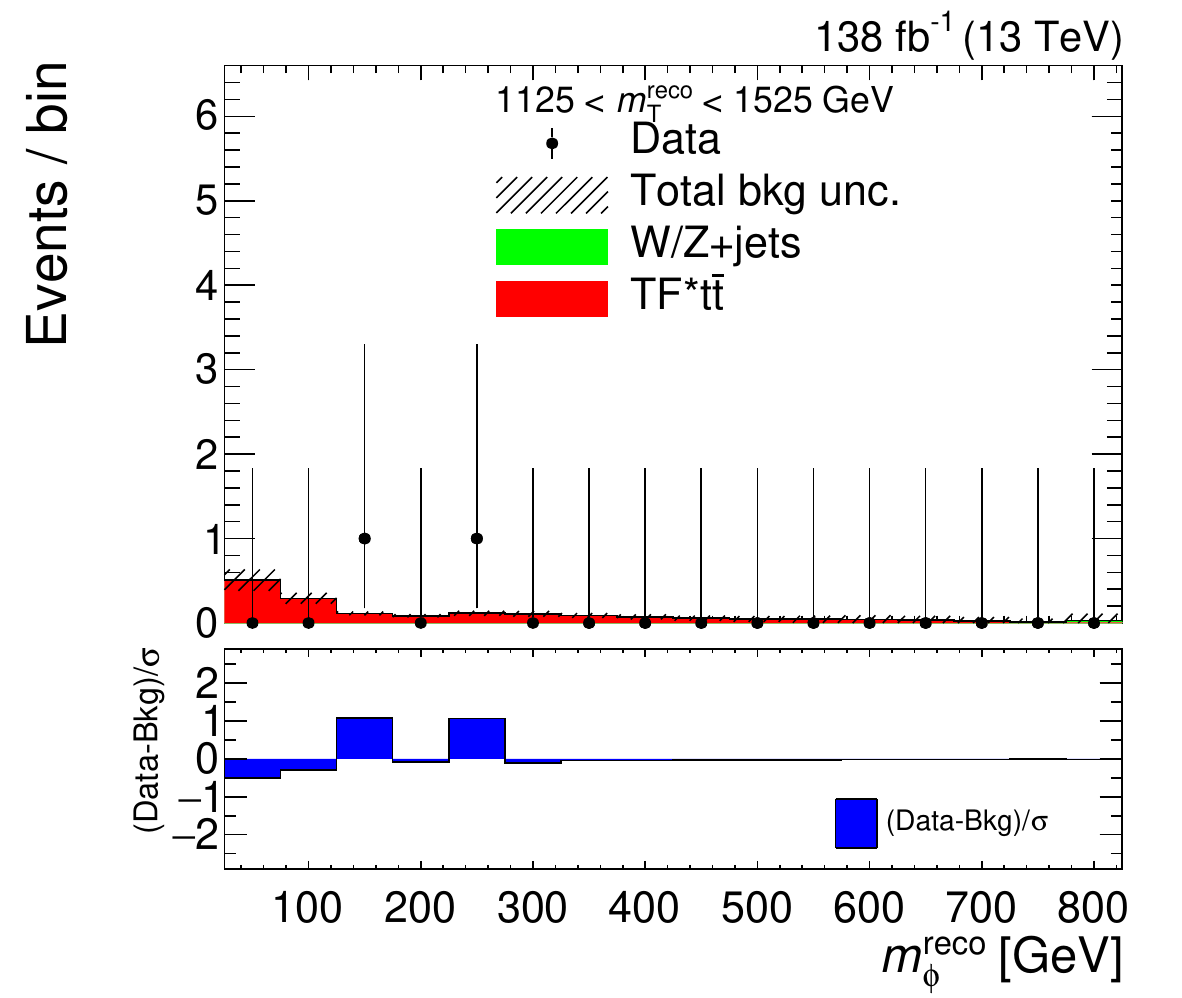}
            \caption{Comparison of data and postfit background predictions from the background-only fit in the VR. The VR-CR (left) and VR-SR (right) regions are shown. The distributions are projected onto $m^\text{reco}_{\cPGf}$ in three intervals of $m^\text{reco}_{\PQT}$: $625 < m^\text{reco}_{\PQT} < 825\GeV$ (upper), $825 < m^\text{reco}_{\PQT} < 1125\GeV$ (middle), and $1125 < m^\text{reco}_{\PQT} < 1525\GeV$ (lower). The $\PQt\PAQt$ component scaled by the transfer function is shown in red and the $\PW/\PZ$+jets contribution in green. The lower panels show the difference between the data and the background normalized by the uncertainty.}
            \label{fig:Fit_VVRS2}
\end{figure*}

Figures~\ref{fig:Fit_VVRT2} and~\ref{fig:Fit_VVRS2} show the postfit distributions obtained from a background-only fit in the VR-CR and VR-SR. Figure~\ref{fig:Fit_VVRT2} shows the distributions projected onto $m^\text{reco}_{\PQT}$, while Fig.~\ref{fig:Fit_VVRS2} shows the projection onto $m^\text{reco}_{\cPGf}$. The lower panels show the difference between the data and the background prediction normalized by the standard deviation.  The normalized differences in the CR region are exactly zero by construction.  The background-only fit in the VR yields a goodness-of-fit $p$-value of 0.14 corresponding to the combined VR-SR and VR-CR fit.  The VR provides a statistically limited test of the extrapolation given the relatively few events in the VR-SR compared to the VR-CR.

\clearpage

\section{Systematic uncertainties}\label{sec:sysunc}

Several sources of systematic uncertainty are considered.  These uncertainties affect both the shape and the normalization of the distributions of the simulated processes. The integrated luminosity uncertainty affecting only the normalization is 1.6\%~\cite{CMS-LUM-17-003, CMS-LUM-20-001}. Statistical uncertainties dominate the analysis, while uncertainties related to the final-state radiation (FSR), luminosity, and pileup are the leading sources of systematic uncertainty, shown in Table~\ref{tab:uncert1} corresponding to the signal sample with $m_{\PQT}=900\GeV$ and $m_{\cPGf}=150\GeV$. All uncertainties are applied to the simulated signal as well as to the $\PQt\PAQt$ and $\PW/\PZ$+jets samples.

\begin{table}[!htbp]
\centering
\caption{Largest postfit relative impacts of the dominant systematic uncertainties on the fitted signal-strength parameter $r$ for mass point ($m_{\PQT}=900\GeV$,$m_{\cPGf}=150\GeV$).}
\label{tab:uncert1}
\begin{scotch}{ll}
Uncertainty & Impact (\% change in $r$) \\
\hline
FSR  & 3.0 \\
Integrated luminosity & 1.6 \\
Pileup & 1.3 \\
PDF & 1.3 \\
ParticleNet & 1.0 \\
\end{scotch}
\end{table}

Pileup, PDF, and trigger uncertainties affect both the shape and normalization of the two-dimensional mass distribution.  The systematic uncertainty associated with pileup is evaluated by varying the assumed total inelastic proton--proton cross section used to derive the pileup distribution, by $\pm$4.6\%~\cite{CMS:2018mlc}. PDF uncertainties are evaluated using the NNPDF3.1 PDF set following the PDF4LHC prescription~\cite{Butterworth:2015oua}.  Uncertainties in the jet energy scale (JES), jet mass scale (JMS), jet energy resolution (JER), and jet mass resolution (JMR), are smaller than the uncertainty due to ParticleNet top quark tagging.  Trigger uncertainties are evaluated separately for each data-taking period and correspond to the variation of the trigger efficiency within the one standard deviation Clopper--Pearson interval~\cite{Clopper:1934}.

Uncertainties in JES, JMS, JER, and JMR are treated as uncorrelated across different data-taking periods. Additional uncertainties include the uncertainty due to the photon SFs, photon energy scale and resolution, found to be negligible compared to the jet-related uncertainties, with impact on the fitted signal-strength parameter $r$ less than 0.1\%, and are therefore not included in this analysis~\cite{CMS:2020uim}. 

Uncertainties associated with QCD scale are evaluated for all simulated samples, including the signal, by varying the renormalization and factorization scales ($\mu_\text{R}$ and $\mu_\text{F}$) independently up and down by factors of 0.5 and 2.0. The envelope of the resulting variations is taken as the uncertainty. Individual uncertainties due to $\mu_\text{R}$ and $\mu_\text{F}$ are also calculated. Since the dominant $\PQt\PAQt$ background contribution is constrained using a maximum-likelihood fit to data, the corresponding QCD scale uncertainty has only a limited impact on the background prediction.

\section{Results}\label{sec:results}

The signal-plus-background fit described in Section~\ref{sec:statmodel} is used to extract the signal strength parameter $r$, determine the TF parameters $\alpha$ and $\beta$, and obtain the postfit background prediction in the SR. The fit is performed simultaneously in the SR and CR, accounting for all systematic uncertainties and their correlations.

\begin{figure*}[!htbp]
\centering
\includegraphics[width=0.48\textwidth]{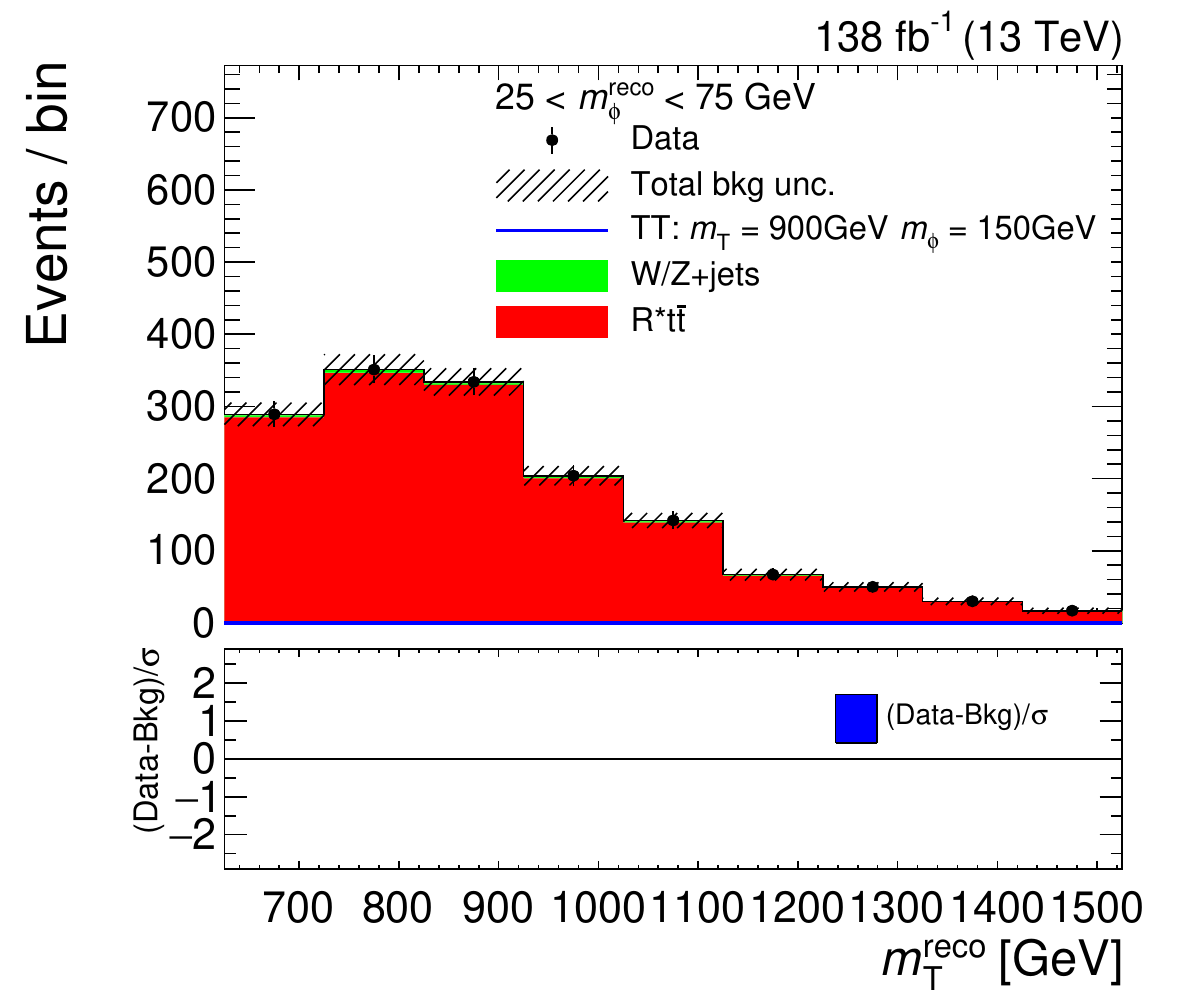}\
\includegraphics[width=0.48\textwidth]{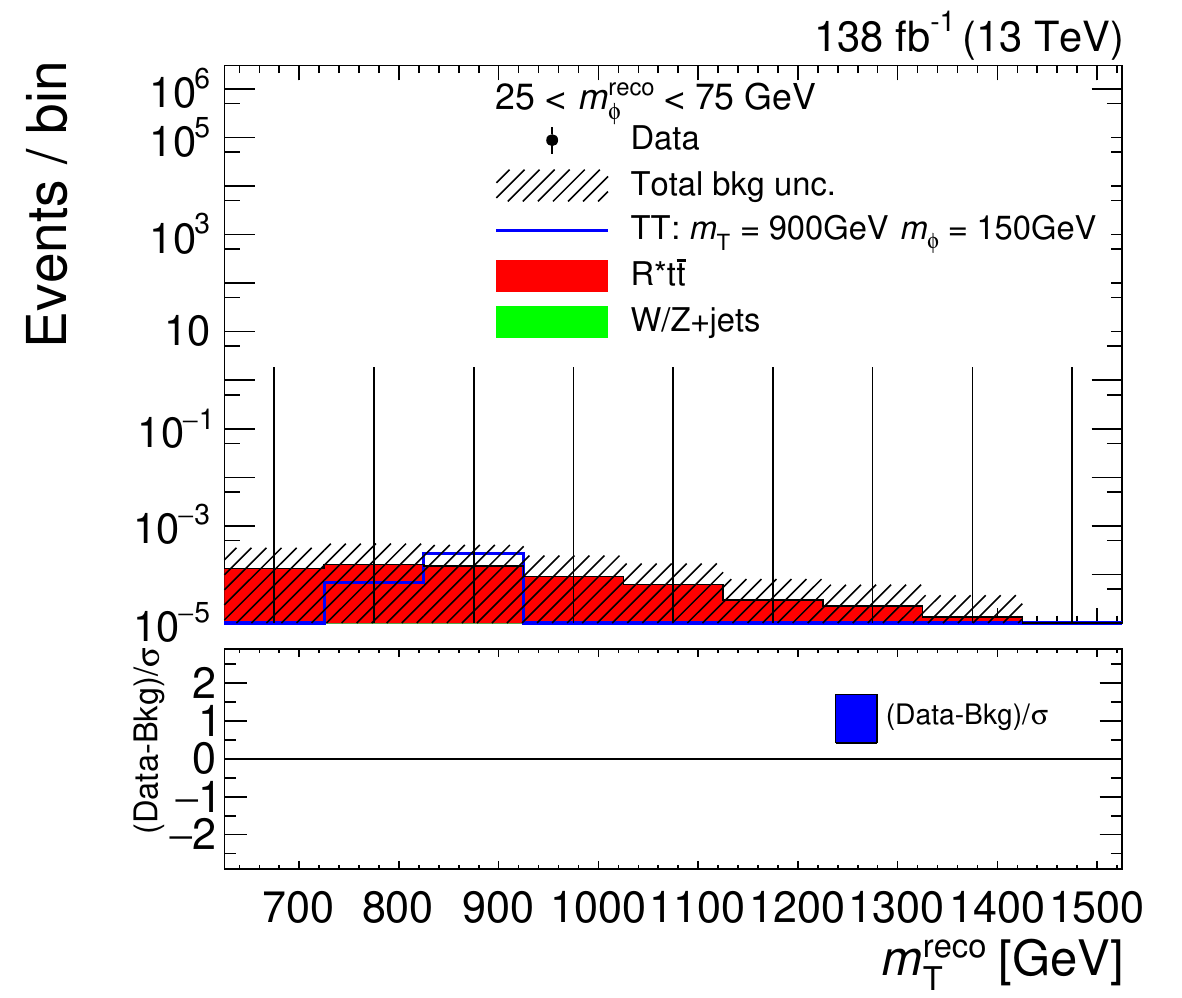}\\
\includegraphics[width=0.48\textwidth]{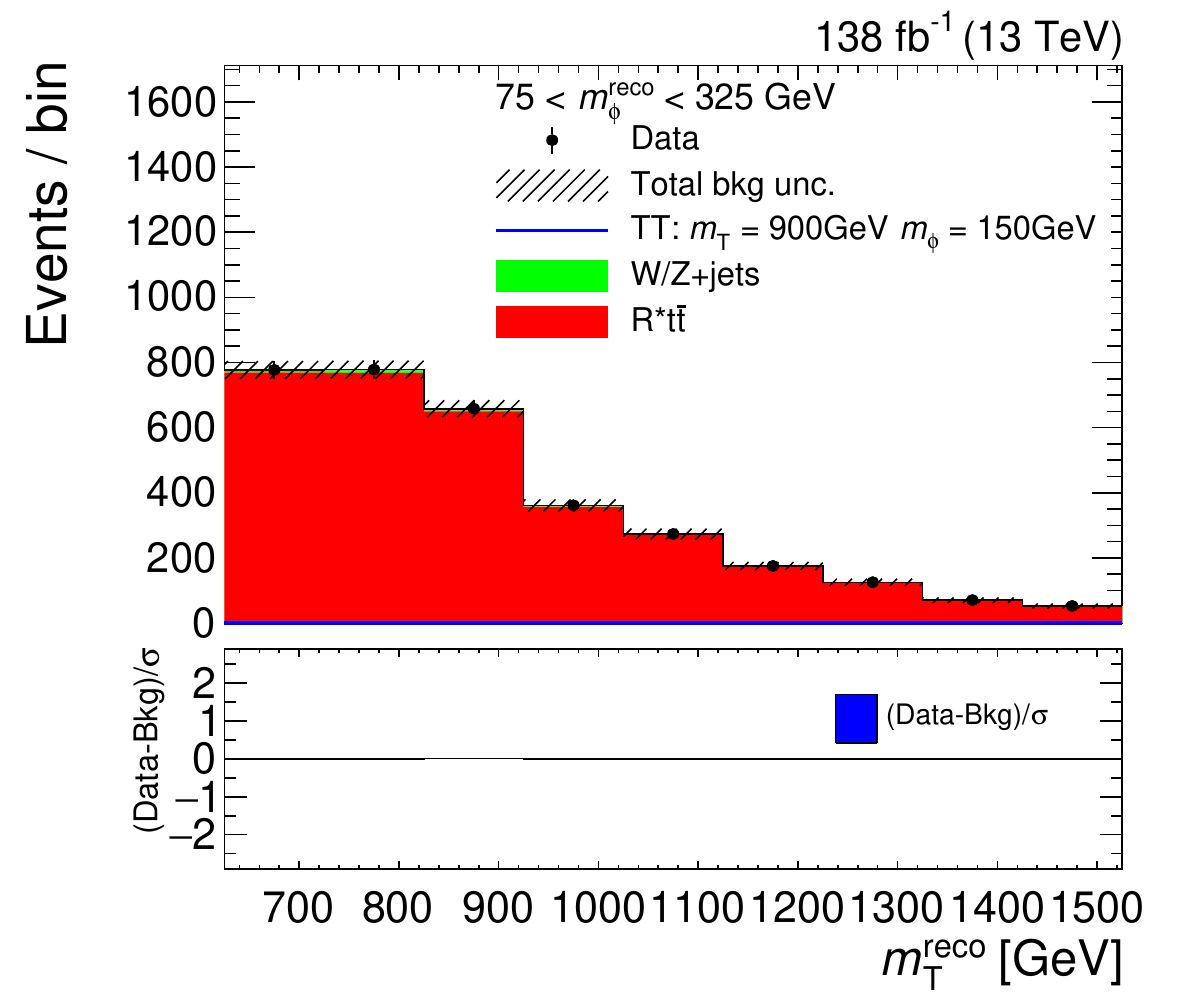}\
\includegraphics[width=0.48\textwidth]{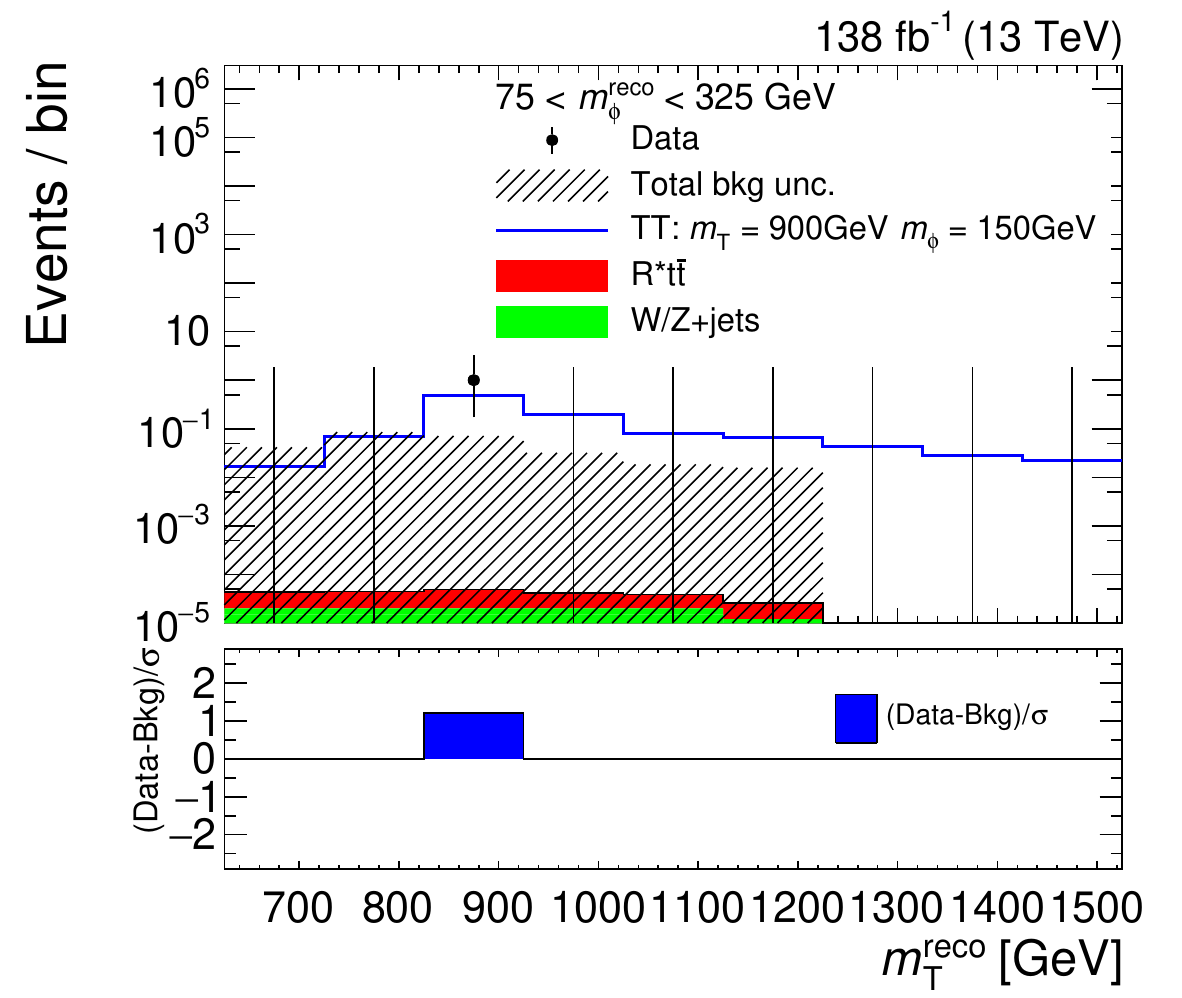}\\
\includegraphics[width=0.48\textwidth]{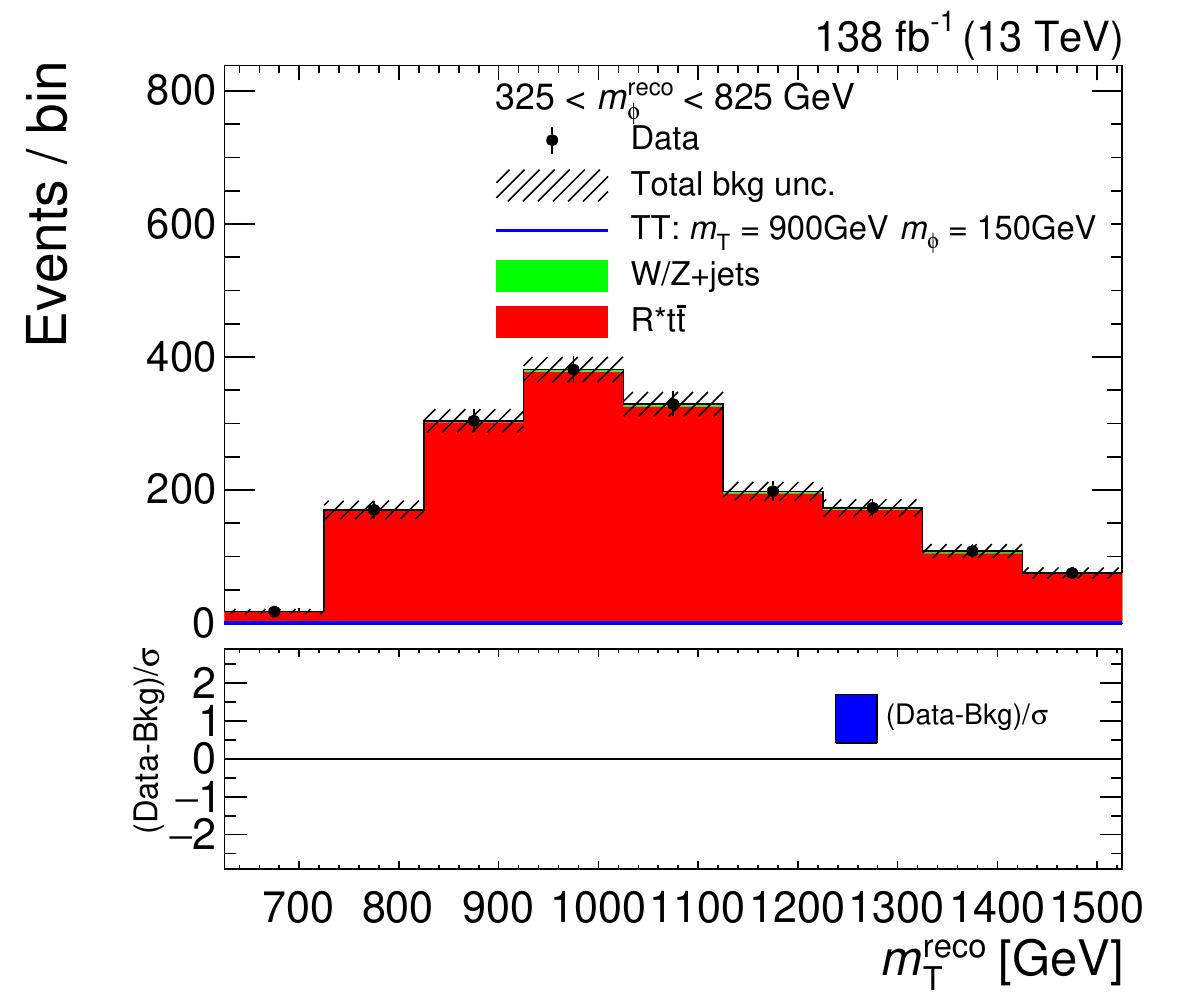}\
\includegraphics[width=0.48\textwidth]{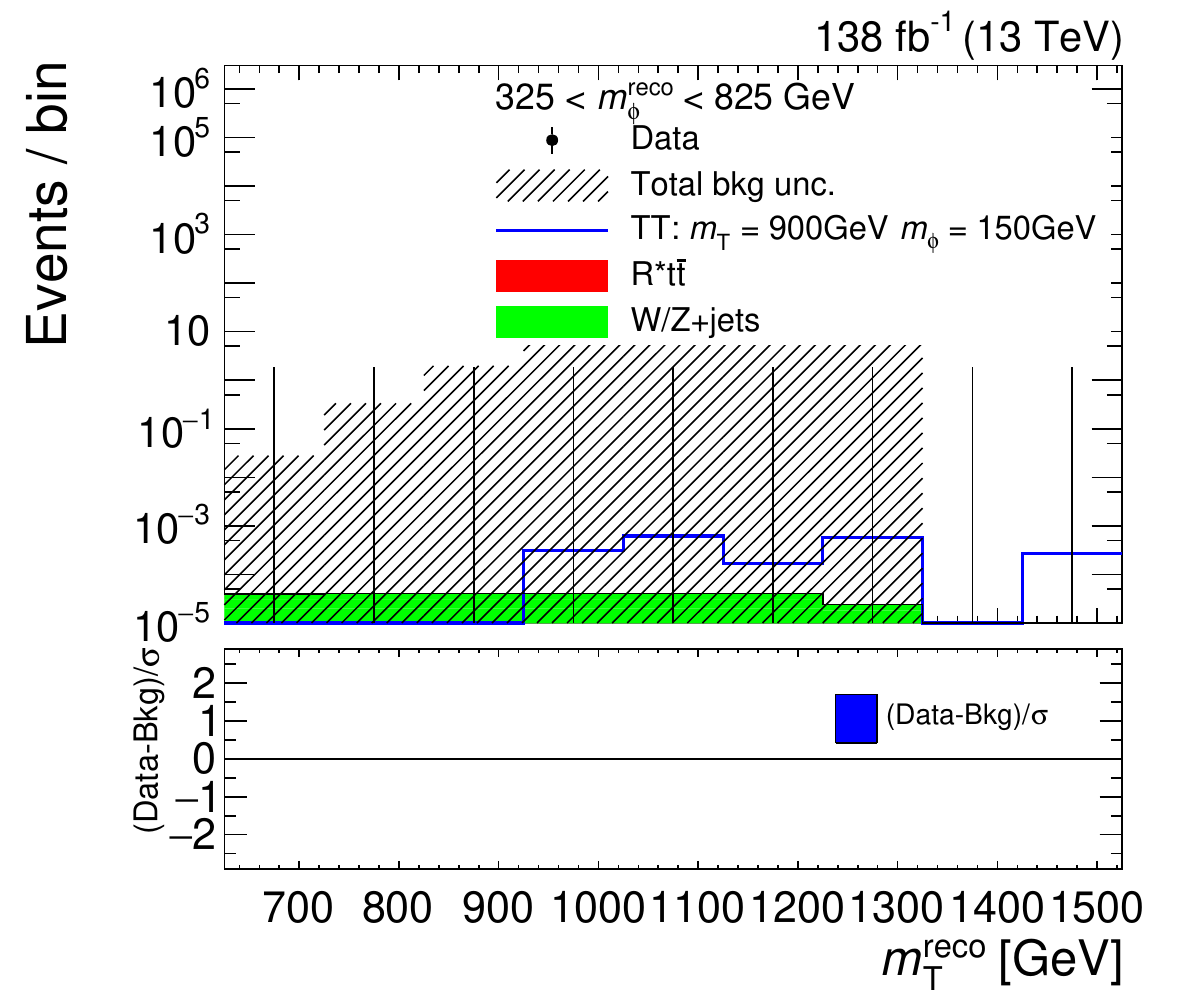}
            \caption{Comparison of the data and postfit background predictions projected onto $m^\text{reco}_{\PQT}$ for the signal hypothesis $m_{\PQT}=900\GeV$ and $m_{\cPGf}=150\GeV$. The CR (left) and SR (right) regions are shown. The distributions are shown in three intervals of $m^\text{reco}_{\cPGf}$: $25 < m^\text{reco}_{\cPGf} < 75\GeV$ (upper), $75 < m^\text{reco}_{\cPGf} < 325\GeV$ (middle), and $325 < m^\text{reco}_{\cPGf} < 825\GeV$ (lower). The transferred $\PQt\PAQt$ contribution is shown in red and the $\PW/\PZ$+jets contribution in green. The signal distribution is overlaid as a blue line. The shaded band represents the uncertainty in the total background predictions at 95\% confidence level.  The lower panels show the difference between the data and the background normalized by the uncertainty.}
            \label{fig:Fit_VRT2}
\end{figure*}

\begin{figure*}[!htbp]
\centering
\includegraphics[width=0.48\textwidth]{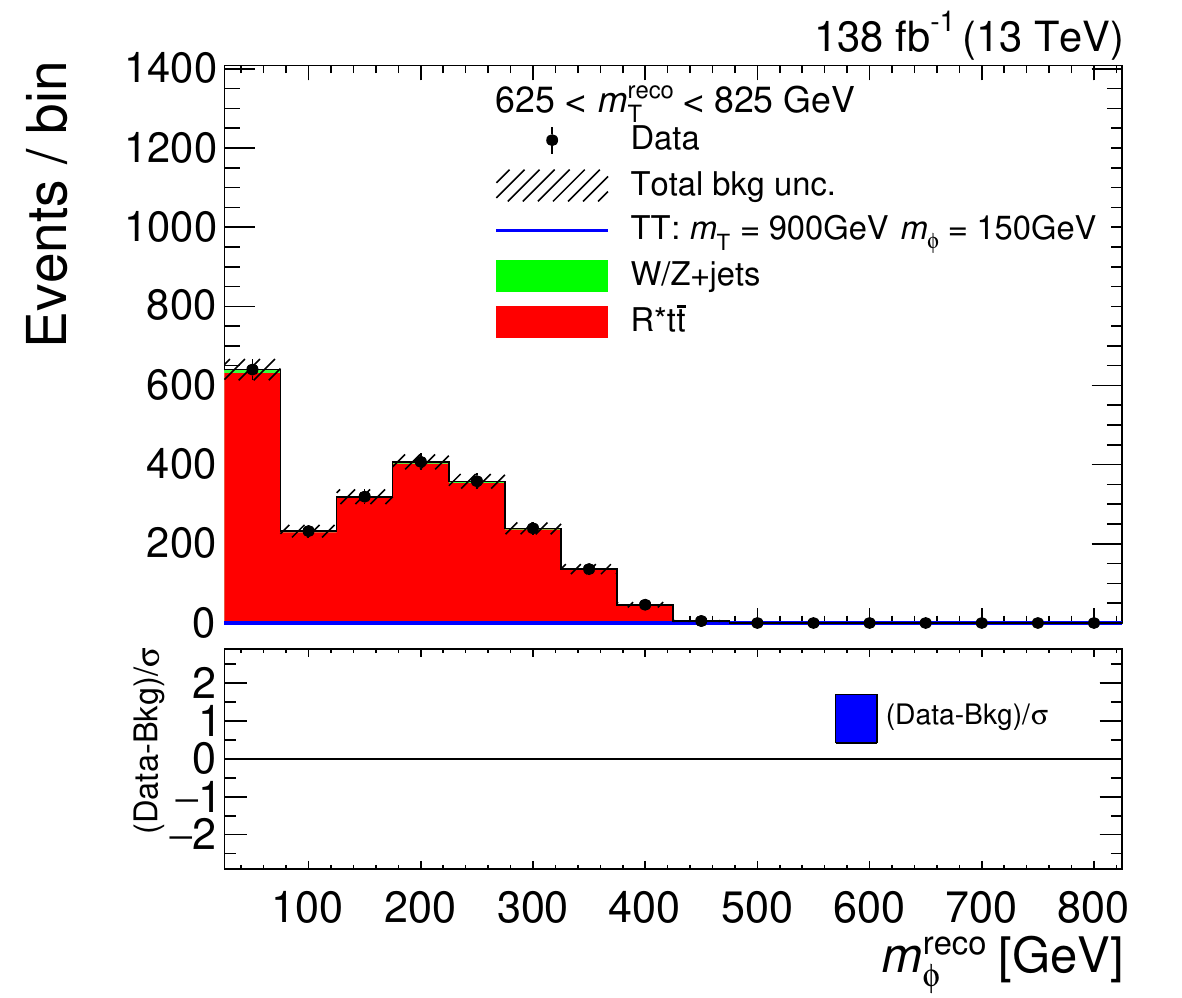}\
\includegraphics[width=0.48\textwidth]{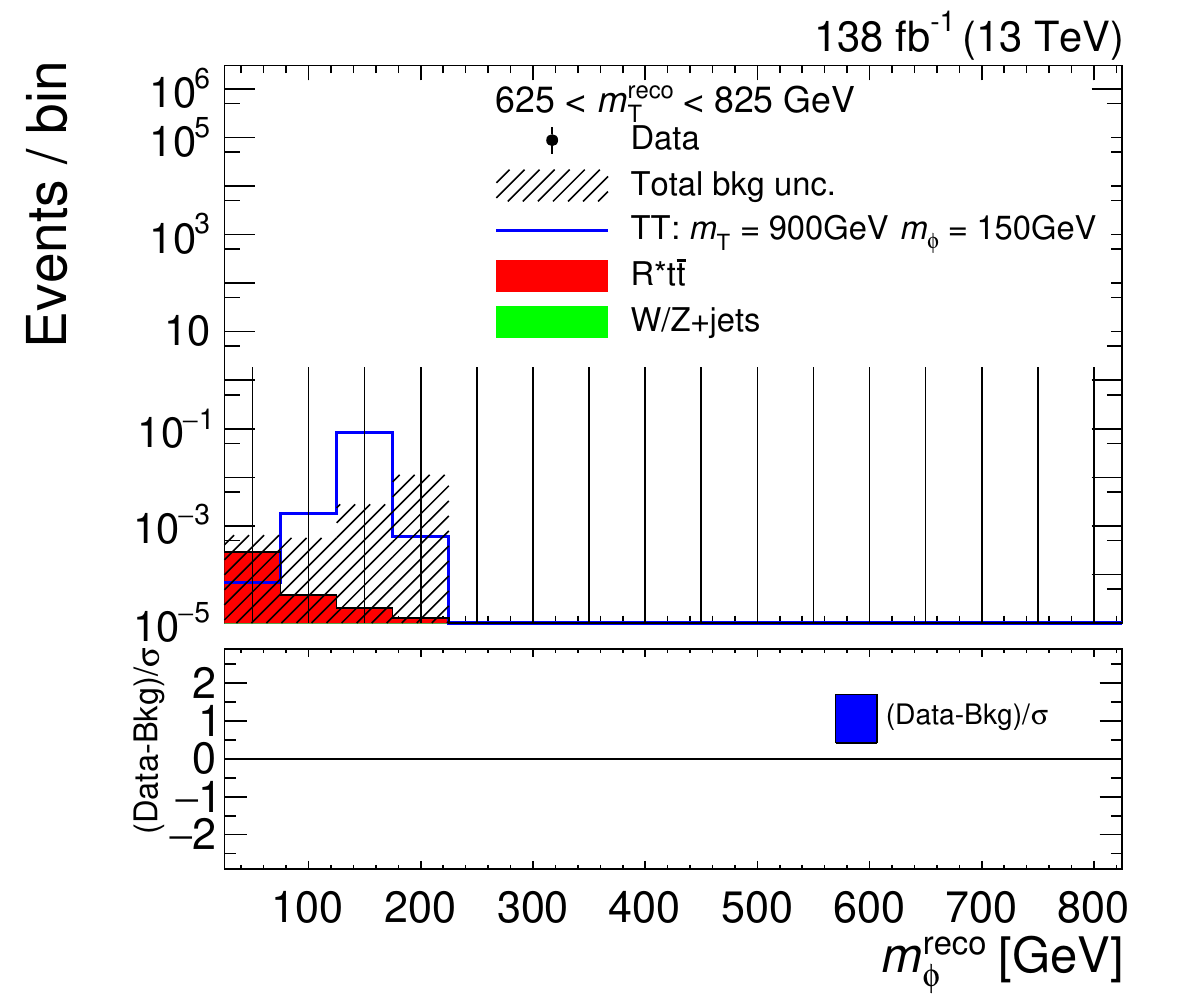}\\
\includegraphics[width=0.48\textwidth]{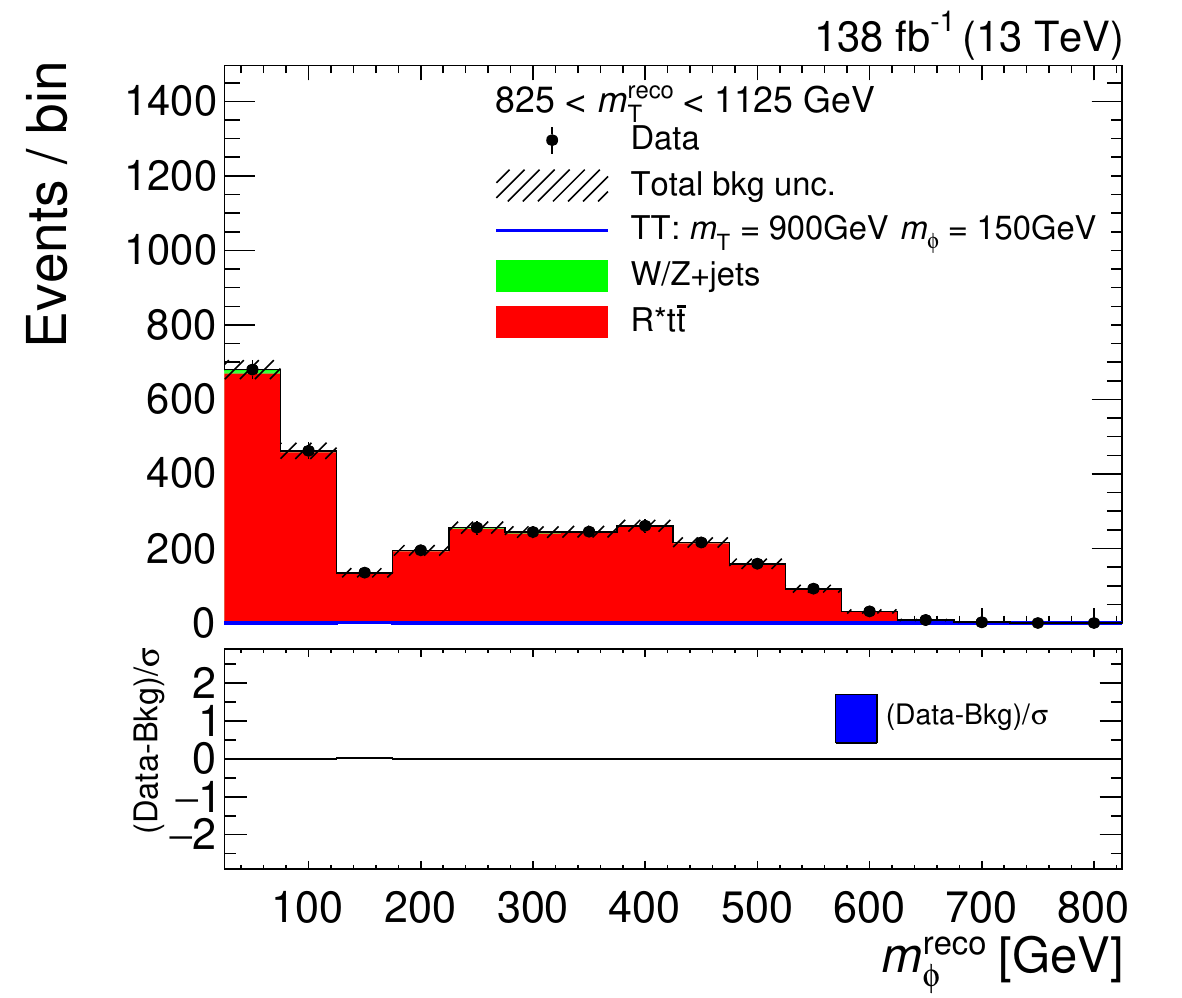}\
\includegraphics[width=0.48\textwidth]{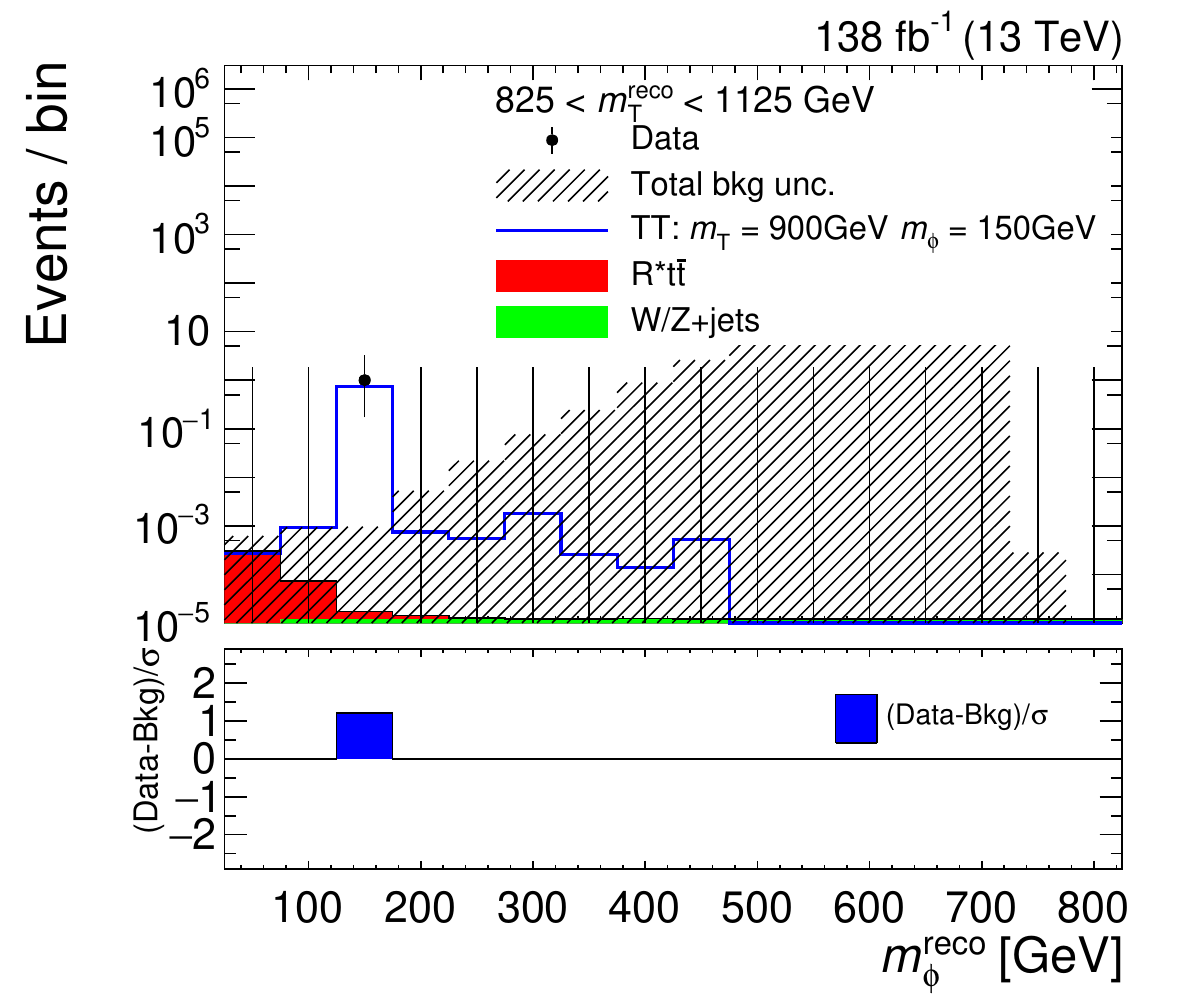}\\
\includegraphics[width=0.48\textwidth]{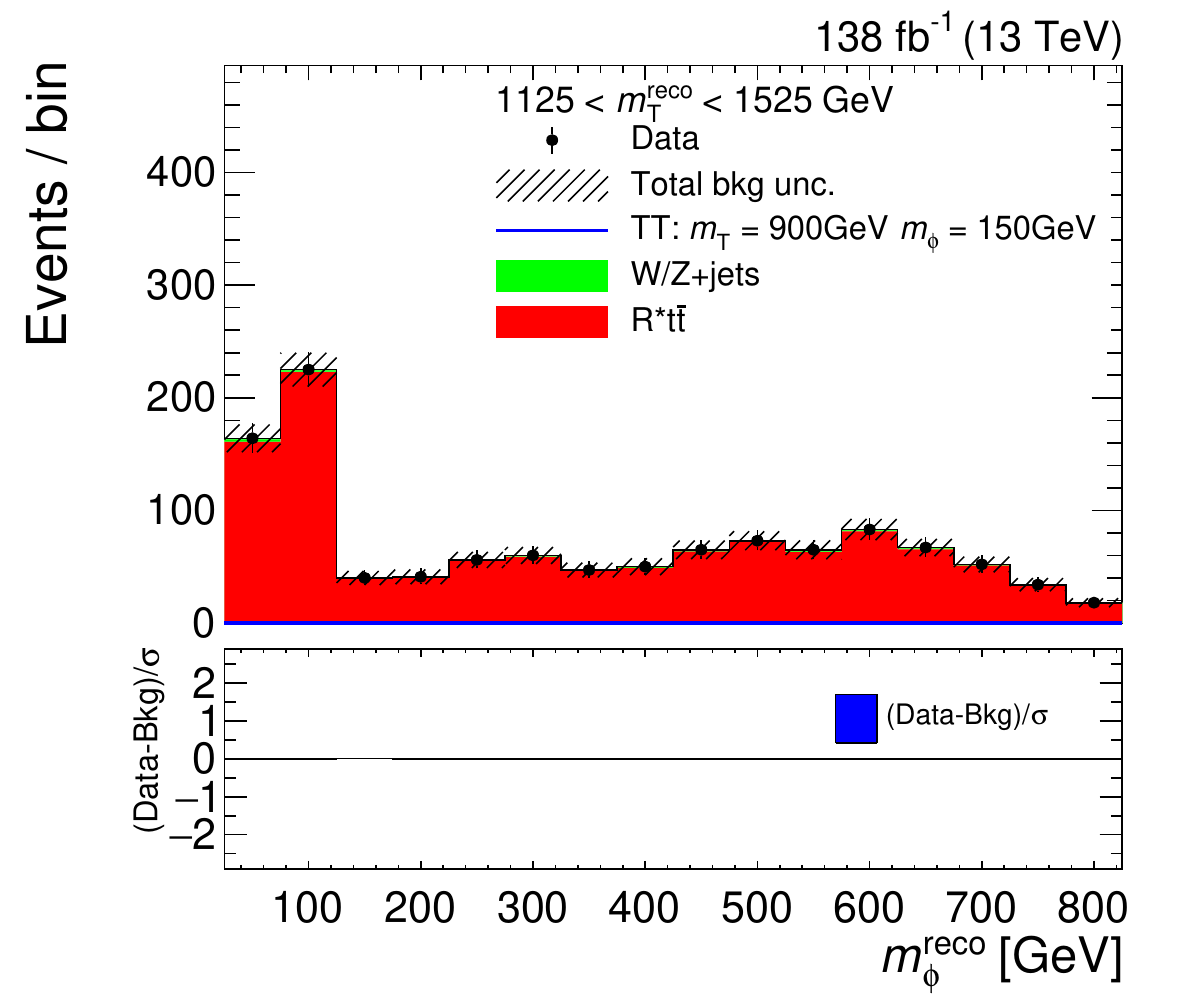}\
\includegraphics[width=0.48\textwidth]{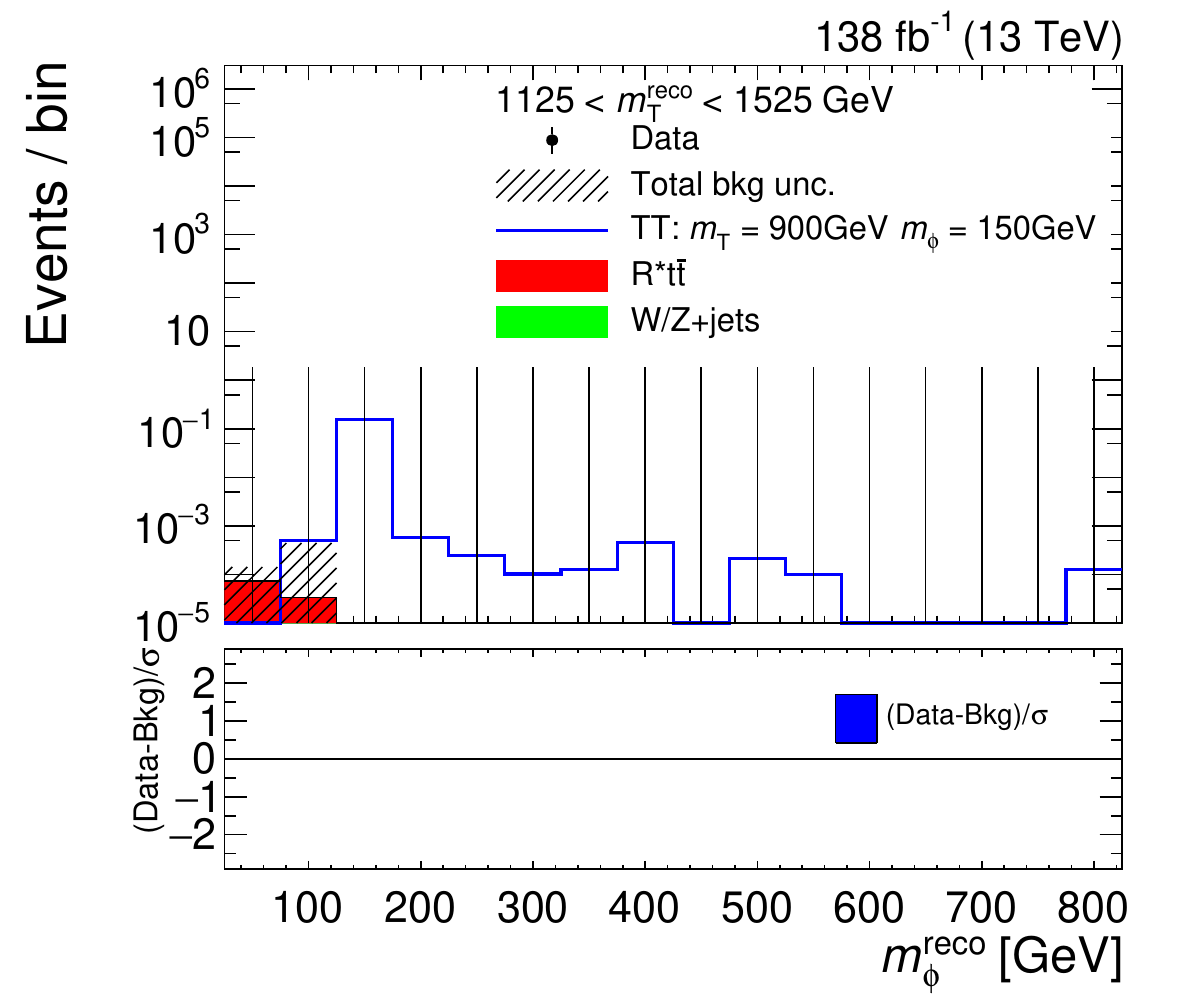}
            \caption{Comparison of the data and postfit background predictions projected onto $m^\text{reco}_{\cPGf}$ for the signal hypothesis $m_{\PQT}=900\GeV$ and $m_{\cPGf}=150\GeV$. The CR (left) and SR (right) regions are shown. The distributions are shown in three intervals of $m^\text{reco}_{\PQT}$: $625 < m^\text{reco}_{\PQT} < 825\GeV$ (upper), $825 < m^\text{reco}_{\PQT} < 1125\GeV$ (middle), and $1125 < m^\text{reco}_{\PQT} < 1525\GeV$ (lower). The transferred $\PQt\PAQt$ contribution is shown in red and the $\PW/\PZ$+jets contribution in green. The signal distribution is overlaid as a blue line. The shaded band represents the uncertainty in the total background predictions at 95\% confidence level.  The lower panels show the difference between the data and the background normalized by the uncertainty.}
            \label{fig:Fit_VRS2}
\end{figure*}

Figures~\ref{fig:Fit_VRT2} and~\ref{fig:Fit_VRS2} show the postfit distributions for the signal hypothesis $m_{\PQT}=900\GeV$ and $m_{\cPGf}=150\GeV$. The distributions are projected onto $m^\text{reco}_{\PQT}$ and $m^\text{reco}_{\cPGf}$, respectively, and are shown for both the SR and CR. The predicted background in the SR is negligible, and one event is observed in the bin corresponding to $m^\text{reco}_\PQT = 900\GeV$ and $m^\text{reco}_\cPGf=150\GeV$.  The background prediction for this bin is $7 \times 10^{-6}$ with an uncertainty of $8 \times 10^{-4}$.  In Fig.~\ref{fig:Fit_VRT2}, this bin contributes to the background prediction in the $m^\text{reco}_{\PQT}=900\GeV$ bin in the middle $m^\text{reco}_{\cPGf}$ interval, $75 < m^\text{reco}_{\cPGf} < 325\GeV$. In Fig.~\ref{fig:Fit_VRS2}, it contributes to the background prediction in the $m^\text{reco}_{\cPGf}=150\GeV$ bin in the middle $m^\text{reco}_{\PQT}$ interval, $825 < m^\text{reco}_{\PQT} < 1125\GeV$.

The data are consistent with the background prediction in both the CR and SR, and no significant deviation is observed~\cite{Read:2002hq,Junk:1999kv}.  The largest deviation is observed for the mass hypothesis $m_{\PQT}=900\GeV$ and $m_{\cPGf}=150\GeV$, corresponding to a local $p$-value of 0.004 and a local significance of 2.7 standard deviations, obtained using the q$_0$ profile likelihood ratio test statistic with 5000 pseudoexperiments~\cite{Cowan:2010js}.  The corresponding global significance, accounting for the look-elsewhere effect across the 55 tested mass hypotheses, is 1.5 standard deviations, obtained from the distribution of the largest local significance among the 55 mass points in each pseudoexperiment.  For all other mass hypotheses, the observed limits are consistent with the expected ones. 

\begin{figure*}[!htbp]
\centering
\includegraphics[width=0.9\textwidth]{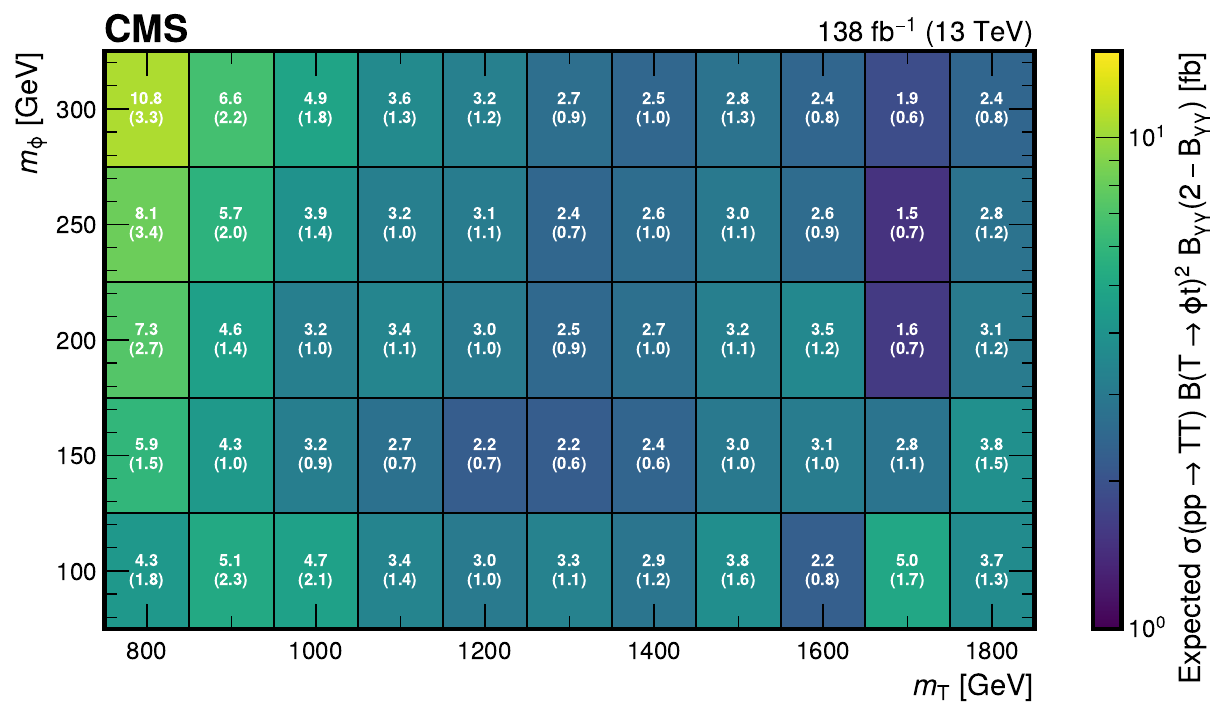}\\
\includegraphics[width=0.9\textwidth]{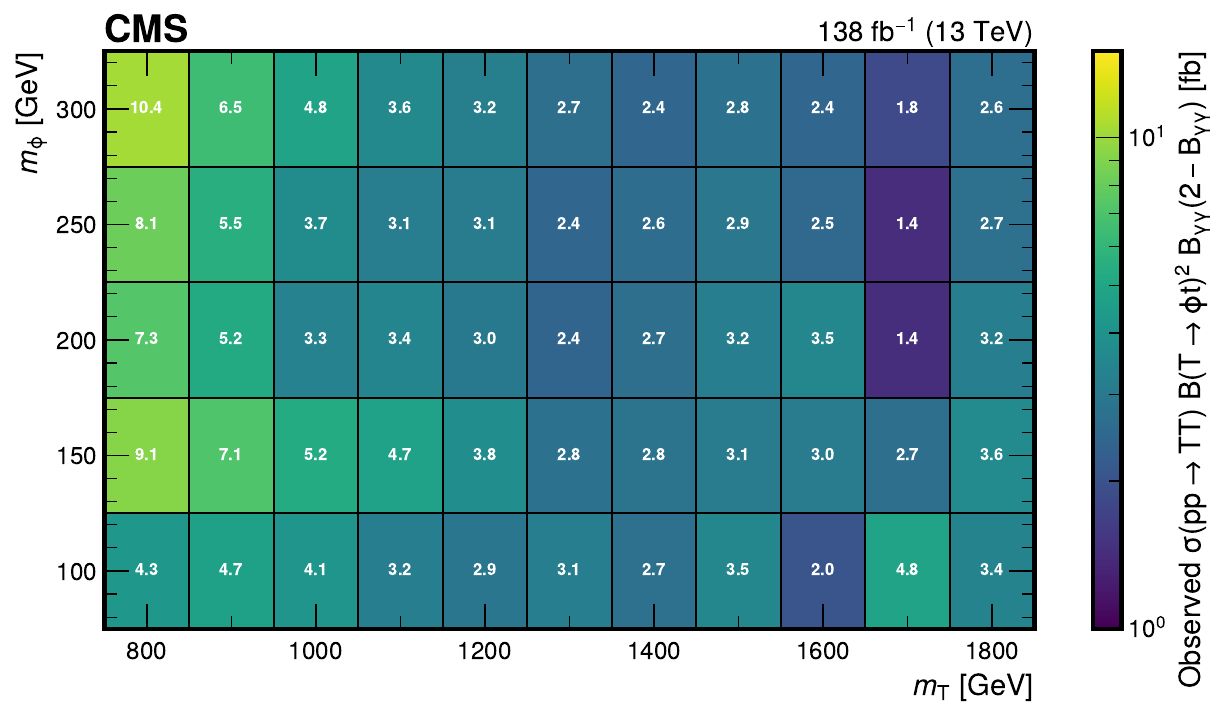}\\
            \caption{Expected (upper) and observed (lower) upper limits on $\sigma_{\PQT\PQT} \ \mathcal{B}(\PQT \to \cPGf \PQt)^2 \ \mathcal{B}_{\PGg\PGg} (2 - \mathcal{B}_{\PGg\PGg})$ in the $(m_{\PQT}, m_{\cPGf})$ plane where $\mathcal{B}_{\PGg\PGg} \equiv \mathcal{B}(\cPGf \to \PGg\PGg)$ is the branching fraction for \cPGf decay to two photons.  The uncertainty for the expected upper limits is shown in parentheses. }
            \label{fig:Flim2D}
\end{figure*}

\begin{figure*}[!htbp]
    \centering
    \includegraphics[width=0.48\textwidth]{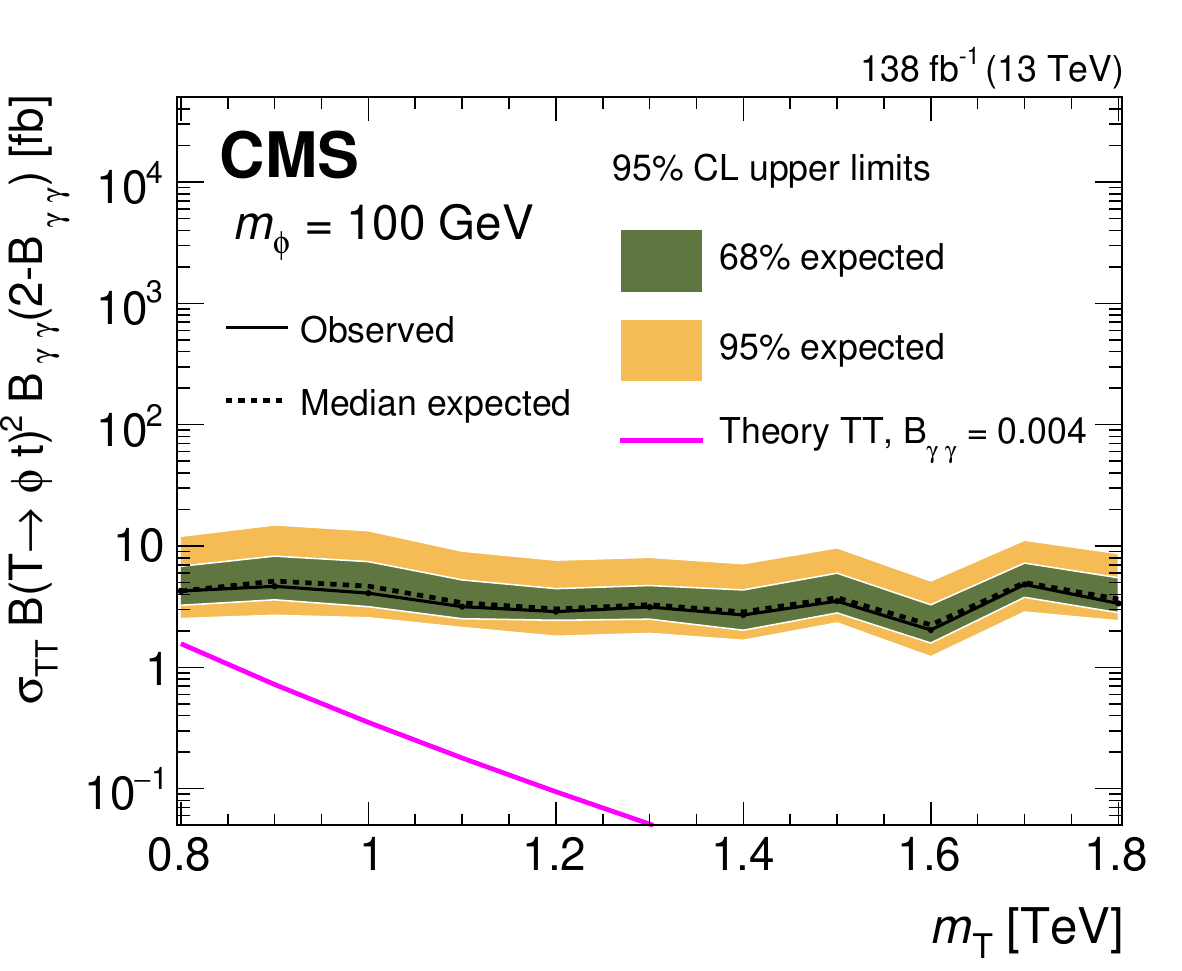}\
    \includegraphics[width=0.48\textwidth]{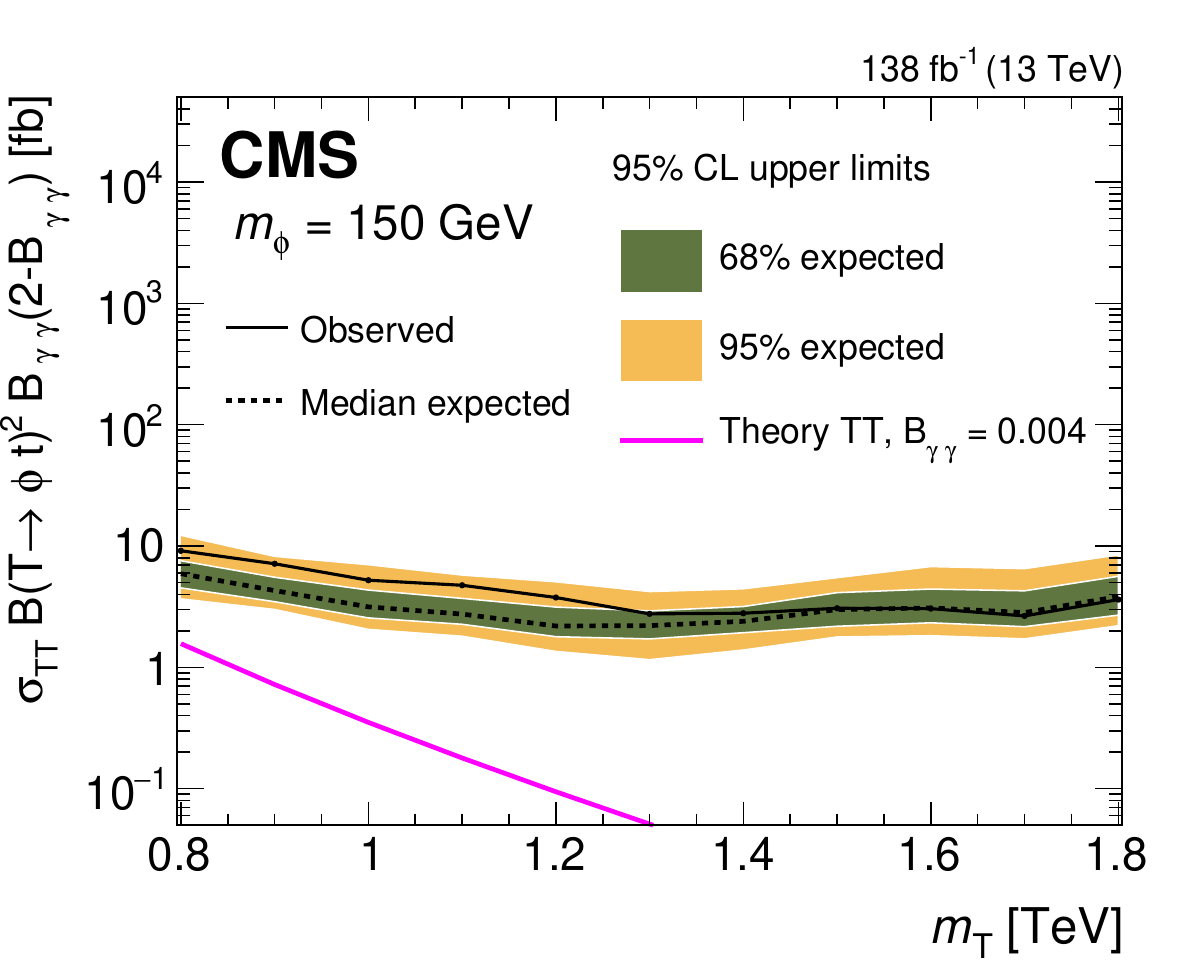}\\
    \includegraphics[width=0.48\textwidth]{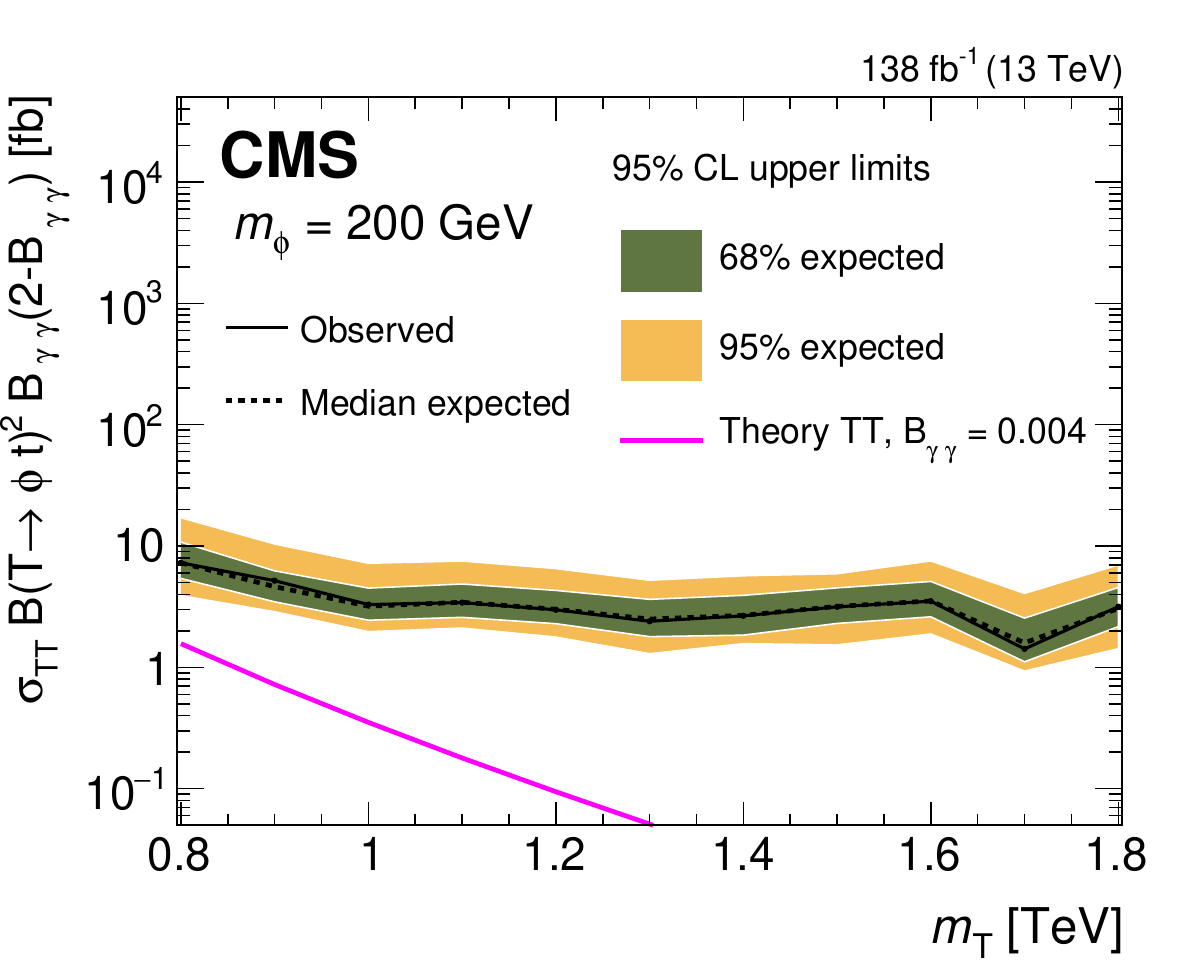}\
    \includegraphics[width=0.48\textwidth]{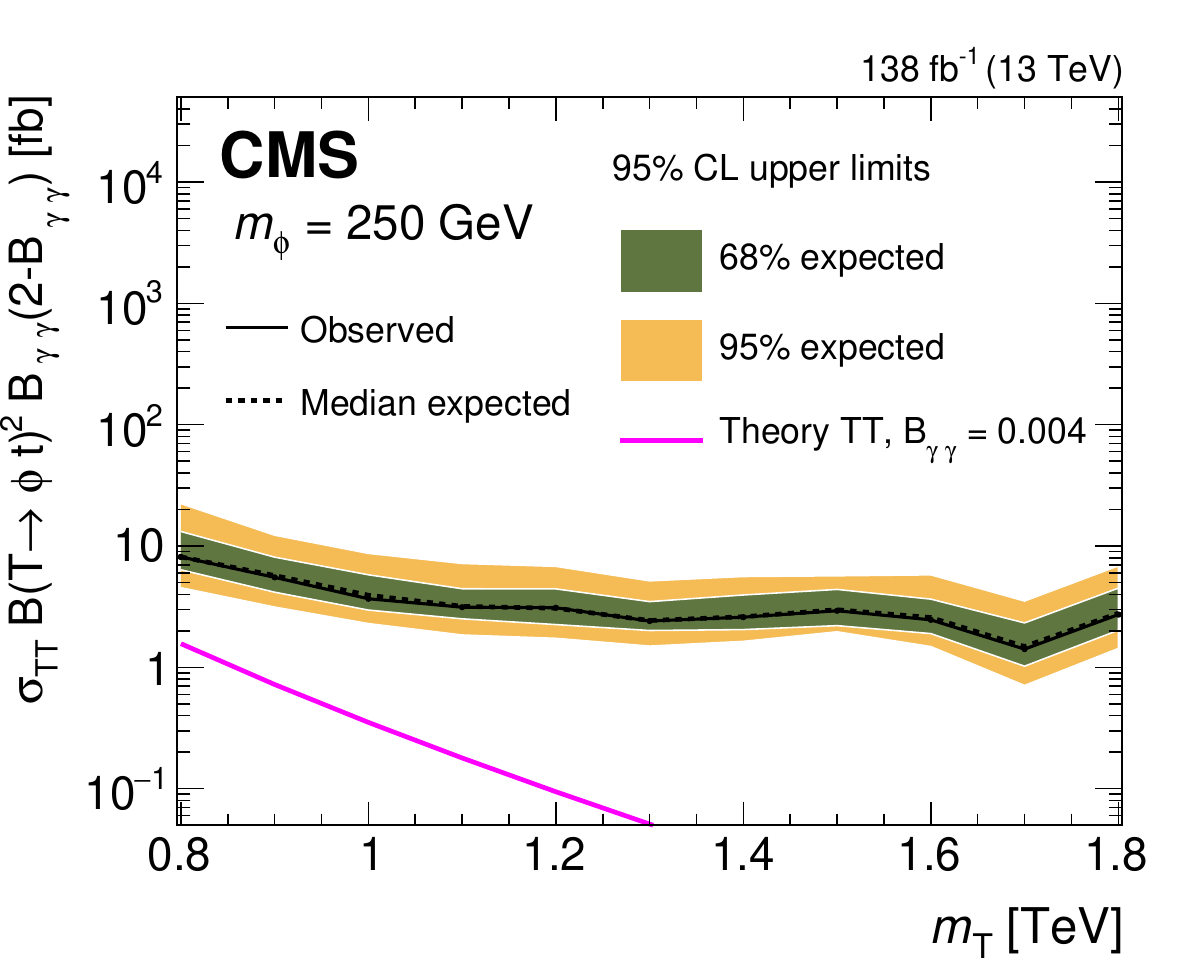}\\
    \includegraphics[width=0.48\textwidth]{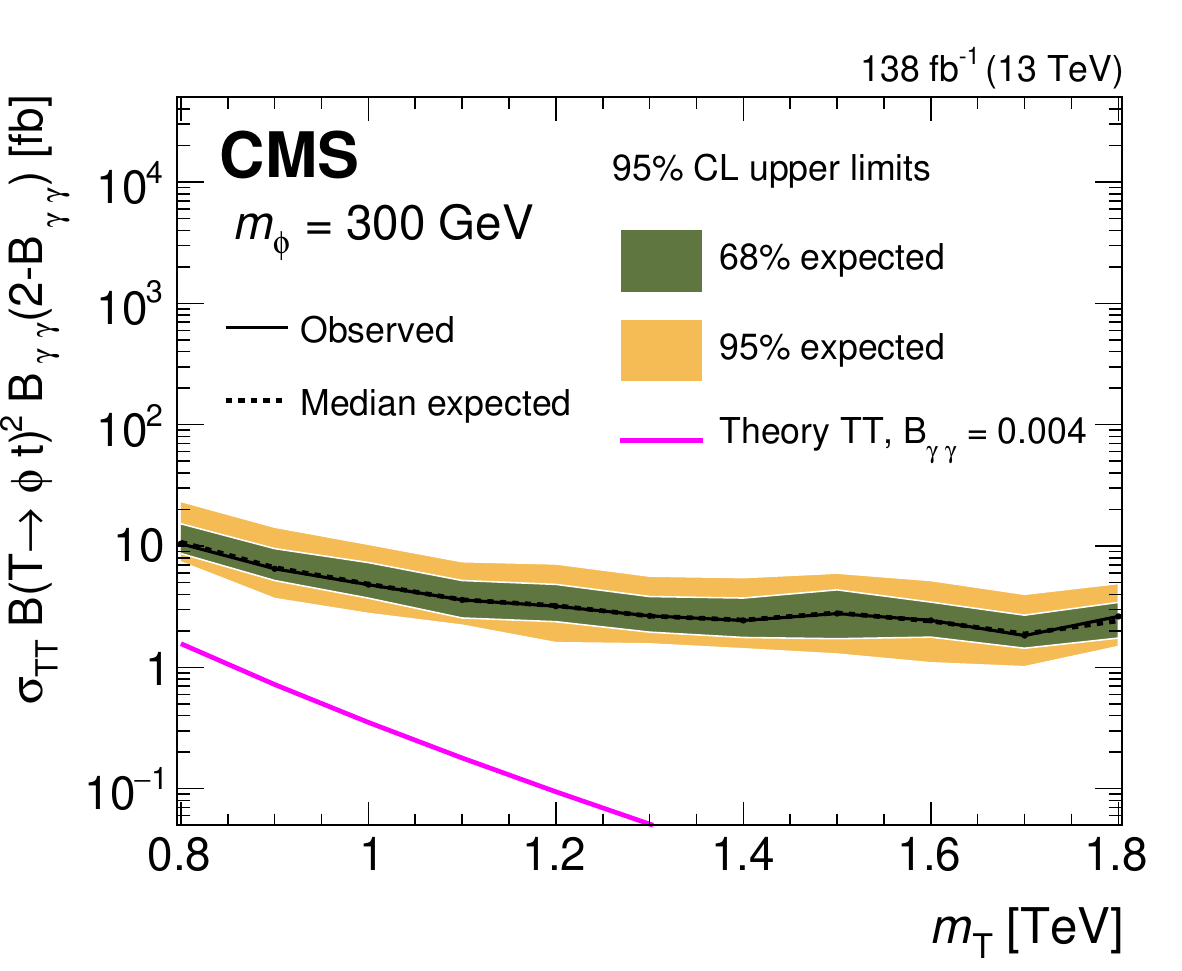}
                \caption{
                    Expected (dashed) and observed (solid) upper limits on $\sigma_{\PQT\PQT} \ \mathcal{B}(\PQT \to \cPGf \PQt)^2 \ \mathcal{B}_{\PGg\PGg} (2 - \mathcal{B}_{\PGg\PGg})$ as a function of $m_{\PQT}$ for different values of $m_{\cPGf}$. The first row shows $m_{\cPGf}=100$ and $150\GeV$ (left to right), the second row shows $m_{\cPGf}=200$ and $250\GeV$ (left to right), and the third row shows $m_{\cPGf}=300\GeV$. The product of the theoretical cross section and the \cPGf  branching fraction is shown for $\mathcal{B}_{\PGg\PGg} = $ 0.004 (magenta).}
                \label{fig:Flim1D150}
\end{figure*}

Upper limits on the product of $\PQT\PAQT$ production cross section and the branching fraction at 95\% confidence level are obtained using the modified frequentist $\CL_s$ method~\cite{Read:2002hq,Junk:1999kv}. The limits are interpreted assuming the branching fraction $\mathcal{B}(\PQt \rightarrow \PQb \PQq \PQq)^2=0.675^2$ for the all-hadronic top quark decay mode~\cite{ParticleDataGroup}. The theoretical cross section $\sigma_{\PQT\PQT}$ is calculated at NNLO with NNLL soft-gluon resummation using \textsc{Top++}~2.0~\cite{Czakon:2011xx,Czakon:2013goa} and ranges in value from $196.0^{+8.0}_{-8.0}$ for $m_{\PQT} = 800$\GeV to $0.39^{+0.04}_{-0.03}$ for $m_{\PQT} = 1800$\GeV.

The expected and observed limits in the $(m_{\PQT}, m_{\cPGf})$ plane are shown in Fig.~\ref{fig:Flim2D}. The corresponding limits as a function of $m_{\PQT}$ for fixed values of $m_{\cPGf}=100$, $150$, $200$, $250$, and $300\GeV$ are shown in Fig.~\ref{fig:Flim1D150}. These are the first limits on \PQT pair production in this final state.

\section{Summary}\label{sec:summary}

A search for the pair production of vector-like top quarks $\PQT$ in a scenario involving an additional scalar boson has been presented. The search uses proton-proton collision data collected by the CMS experiment at the LHC during 2016--2018, at center-of-mass energy 13\TeV, corresponding to an integrated luminosity of 138\fbinv.

The analysis considers $\PQT\PAQT$ production, with each $\PQT$ decaying into a standard model top quark and a scalar boson \cPGf. The top quarks are required to decay hadronically, and one of the scalar bosons is required to decay to a pair of photons. The top quarks are reconstructed as large-radius jets and identified using a ParticleNet discriminant, while a multivariate discriminant is used for photon identification.

The background in the signal region is predicted from the distribution of data in a $\PQt\PAQt$ control region via a transfer function. No significant deviation from the background prediction is observed.  The largest local excess has a significance of 2.7 standard deviations, reduced to 1.5 standard deviations after accounting for the look-elsewhere effect.  This is the first exclusive search for pair-produced $\PQT$ quarks in the boosted regime.

\begin{acknowledgments}
We congratulate our colleagues in the CERN accelerator departments for the excellent performance of the LHC and thank the technical and administrative staffs at CERN and at other CMS institutes for their contributions to the success of the CMS effort. In addition, we gratefully acknowledge the computing centers and personnel of the Worldwide LHC Computing Grid and other centers for delivering so effectively the computing infrastructure essential to our analyses. Finally, we acknowledge the enduring support for the construction and operation of the LHC, the CMS detector, and the supporting computing infrastructure provided by the following funding agencies: SC (Armenia), BMFWF and FWF (Austria); FNRS and FWO (Belgium); CNPq, CAPES, FAPERJ, FAPERGS, and FAPESP (Brazil); MES and BNSF (Bulgaria); CERN; CAS, MoST, and NSFC (China); MINCIENCIAS (Colombia); MSES and CSF (Croatia); RIF (Cyprus); SENESCYT (Ecuador); ERC PRG and PSG, TARISTU24-TK10 and MoER TK202 (Estonia); Academy of Finland, MEC, and HIP (Finland); CEA and CNRS/IN2P3 (France); SRNSF (Georgia); BMFTR, DFG, and HGF (Germany); GSRI (Greece); MATE and NKFIH (Hungary); DAE and DST (India); IPM (Iran); SFI (Ireland); INFN (Italy); MSIT and NRF (Republic of Korea); MES (Latvia); LMTLT (Lithuania); MOE and UM (Malaysia); BUAP, CINVESTAV, CONACYT, LNS, SEP, and UASLP-FAI (Mexico); MOS (Montenegro); MBIE (New Zealand); PAEC (Pakistan); MSHE, NSC, and NAWA (Poland); FCT (Portugal); MESTD (Serbia); MICIU/AEI and PCTI (Spain); MOSTR (Sri Lanka); Swiss Funding Agencies (Switzerland); NSTC (Taiwan); MHESI (Thailand); TUBITAK and TENMAK (T\"{u}rkiye); NASU (Ukraine); STFC (United Kingdom); DOE and NSF (USA).

\begin{sloppypar}
\setlength\emergencystretch{\hsize}
\hyphenation{Rachada-pisek} Individuals have received support from the Marie-Curie programme and the European Research Council and Horizon 2020 Grant, contract Nos.\ 675440, 724704, 752730, 758316, 765710, 824093, 101115353, 101002207, 101001205, and COST Action CA16108 (European Union); the Leventis Foundation; the Alfred P.\ Sloan Foundation; the Alexander von Humboldt Foundation; the Science Committee, project no. 22rl-037 (Armenia); the Fonds pour la Formation \`a la Recherche dans l'Industrie et dans l'Agriculture (FRIA) and Fonds voor Wetenschappelijk Onderzoek contract No. 1228724N (Belgium); the Beijing Municipal Science \& Technology Commission, No. Z191100007219010, the Fundamental Research Funds for the Central Universities, the Ministry of Science and Technology of China under Grant No. 2023YFA1605804, the Natural Science Foundation of China under Grant No. 12535004, and USTC Research Funds of the Double First-Class Initiative No.\ YD2030002017 (China); the Ministry of Education, Youth and Sports (MEYS) of the Czech Republic; the Shota Rustaveli National Science Foundation (Georgia); the Deutsche Forschungsgemeinschaft (DFG), among others, under Germany's Excellence Strategy -- EXC 2121 ``Quantum Universe" -- 390833306, and under project number 400140256 - GRK2497; the Hellenic Foundation for Research and Innovation (HFRI), Project Number 2288 (Greece); the Hungarian Academy of Sciences, the New National Excellence Program - \'UNKP, the NKFIH research grants K 131991, K 138136, K 143460, K 143477, K 147557, K 146913, K 146914, K 147048, TKP2021-NKTA-64, 2026-4.1.1-MISSZI\'{O}K-2026-00005 and 2025-1.1.5-NEMZ\_KI-2025-00004, and MATE KKP and KKPCs Research Excellence and Flagship Research Groups grants (Hungary); the Council of Science and Industrial Research, India; ICSC -- National Research Center for High Performance Computing, Big Data and Quantum Computing, FAIR -- Future Artificial Intelligence Research, and CUP I53D23001070006 (Mission 4 Component 1), funded by the NextGenerationEU program, the Italian Ministry of University and Research (MUR) under Bando PRIN 2022 -- CUP I53C24002390006, PRIN PRIMULA 2022RBYK7T (Italy); the Latvian Council of Science; the Ministry of Science and Higher Education, project no. 2022/WK/14, and the National Science Center, contracts Opus 2021/41/B/ST2/01369, 2021/43/B/ST2/01552, 2023/49/B/ST2/03273, and the NAWA contract BPN/PPO/2021/1/00011 (Poland); the Funda\c{c}\~ao para a Ci\^encia e a Tecnologia (Portugal); the National Priorities Research Program by Qatar National Research Fund; MICIU/AEI/10.13039/501100011033, ERDF/EU, ``European Union NextGenerationEU/PRTR", projects PID2022-142604OB-C21, PID2022-139519OB-C21, PID2023-147706NB-I00, PID2023-148896NB-I00, PID2023-146983NB-I00, PID2023-147115NB-I00, PID2023-148418NB-C41, PID2023-148418NB-C42, PID2023-148418NB-C43, PID2023-148418NB-C44, PID2024-158190NB-C22, RYC2021-033305-I, RYC2024-048719-I, CNS2023-144781, CNS2024-154769 and Plan de Ciencia, Tecnolog{\'i}a e Innovaci{\'o}n de Asturias, Spain; the Chulalongkorn Academic into Its 2nd Century Project Advancement Project, the National Science, Research and Innovation Fund program IND\_FF\_68\_369\_2300\_097, and the Program Management Unit for Human Resources \& Institutional Development, Research and Innovation, grant B39G680009 (Thailand); the Eric \& Wendy Schmidt Fund for Strategic Innovation through the CERN Next Generation Triggers project under grant agreement number SIF-2023-004; the Kavli Foundation; the Nvidia Corporation; the SuperMicro Corporation; the Welch Foundation, contract C-1845; and the Weston Havens Foundation (USA).
\end{sloppypar}
\end{acknowledgments}\section*{Data availability} Release and preservation of data used by the CMS Collaboration as the basis for publications is guided by the  \href{https://doi.org/10.7483/OPENDATA.CMS.1BNU.8V1W}{CMS data preservation, re-use and open access policy}.

\bibliography{auto_generated} 

\cleardoublepage \appendix\section{The CMS Collaboration \label{app:collab}}\begin{sloppypar}\hyphenpenalty=5000\widowpenalty=500\clubpenalty=5000\input{B2G-24-016-public-authorlist.tex}\end{sloppypar}
\end{document}

%% file: B2G-24-016-public-authorlist.tex
\cmsinstitute{Yerevan Physics Institute, Yerevan, Armenia}
{\tolerance=6000
A.~Gevorgyan\cmsorcid{0000-0003-2751-9489}, A.~Hayrapetyan, V.~Makarenko\cmsorcid{0000-0002-8406-8605}, A.~Tumasyan\cmsAuthorMark{1}\cmsorcid{0009-0000-0684-6742}
\par}
\cmsinstitute{Institut f\"{u}r Hochenergiephysik, Vienna, Austria}
{\tolerance=6000
P.S.~Hussain\cmsorcid{0000-0002-4825-5278}, M.~Sonawane\cmsorcid{0000-0003-0510-7010}
\par}
\cmsinstitute{Marietta Blau Institute for Particle Physics, Vienna, Austria}
{\tolerance=6000
W.~Adam\cmsorcid{0000-0001-9099-4341}, L.~Benato\cmsorcid{0000-0001-5135-7489}, T.~Bergauer\cmsorcid{0000-0002-5786-0293}, M.~Dragicevic\cmsorcid{0000-0003-1967-6783}, S.~Gundacker\cmsorcid{0000-0003-2087-3266}, A.K.~Guven\cmsorcid{0009-0004-5670-5138}, M.~Jeitler\cmsAuthorMark{2}\cmsorcid{0000-0002-5141-9560}, N.~Krammer\cmsorcid{0000-0002-0548-0985}, A.~Li\cmsorcid{0000-0002-4547-116X}, D.~Liko\cmsorcid{0000-0002-3380-473X}, M.~Matthewman, A.~Pfeiffer\cmsorcid{0000-0001-5328-448X}, J.~Schieck\cmsAuthorMark{2}\cmsorcid{0000-0002-1058-8093}, R.~Sch\"{o}fbeck\cmsAuthorMark{2}\cmsorcid{0000-0002-2332-8784}, M.~Shooshtari\cmsorcid{0009-0004-8882-4887}, N.~Van~Den~Bossche\cmsorcid{0000-0003-2973-4991}, W.~Waltenberger\cmsorcid{0000-0002-6215-7228}, C.E.~Wulz\cmsAuthorMark{2}\cmsorcid{0000-0001-9226-5812}
% ...
\par}
$^{1}$Also at Yerevan State University, Yerevan, Armenia\\
$^{2}$Also at Technische Universit\"{a}t Wien, Vienna, Austria\\
% ...


%% file: auto_generated.bib
@ARTICLE{ATLAS:2012yve,
	AUTHOR=	"{ATLAS Collaboration}",
	TITLE=	"{Observation of a new particle in the search for the Standard Model Higgs boson with the ATLAS detector at the LHC}",
	EPRINT=	"1207.7214",
	ARCHIVEPREFIX=	"arXiv",
	PRIMARYCLASS=	"hep-ex",
	REPORTNUMBER=	"CERN-PH-EP-2012-218",
	DOI=	"10.1016/j.physletb.2012.08.020",
	JOURNAL=	"Phys. Lett. B",
	VOLUME=	"716",
	PAGES=	"1",
	YEAR=	"2012",
}

@ARTICLE{CMS:2012qbp,
	AUTHOR=	"{CMS Collaboration}",
	TITLE=	"Observation of a New Boson at a Mass of 125 {GeV} with the {CMS} Experiment at the {LHC}",
	EPRINT=	"1207.7235",
	ARCHIVEPREFIX=	"arXiv",
	PRIMARYCLASS=	"hep-ex",
	REPORTNUMBER=	"CMS-HIG-12-028, CERN-PH-EP-2012-220",
	DOI=	"10.1016/j.physletb.2012.08.021",
	JOURNAL=	"Phys. Lett. B",
	VOLUME=	"716",
	PAGES=	"30",
	YEAR=	"2012",
}

@ARTICLE{CMS:2022dwd,
	AUTHOR=	"{CMS Collaboration}",
	TITLE=	"A portrait of the {Higgs} boson by the {CMS} experiment ten years after the discovery",
	EPRINT=	"2207.00043",
	ARCHIVEPREFIX=	"arXiv",
	PRIMARYCLASS=	"hep-ex",
	REPORTNUMBER=	"CMS-HIG-22-001, CERN-EP-2022-039",
	DOI=	"10.1038/s41586-022-04892-x",
	JOURNAL=	"Nature",
	VOLUME=	"607",
	PAGES=	"60",
	YEAR=	"2022",
	ERRATUM=	"10.1038/s41586-023-06164-8",
}

@ARTICLE{Barbieri:1987fn,
	AUTHOR=	"R. Barbieri and G. F. Giudice",
	TITLE=	"Upper Bounds on Supersymmetric Particle Masses",
	JOURNAL=	"Nucl. Phys. B",
	VOLUME=	"306",
	PAGES=	"63",
	YEAR=	"1988",
	DOI=	"10.1016/0550-3213(88)90171-X",
}

@ARTICLE{Peskin:2025hp,
	AUTHOR=	"Peskin, M. E.",
	TITLE=	"{What is the hierarchy problem?}",
	JOURNAL=	"Nucl. Phys. B",
	VOLUME=	"1018",
	YEAR=	"2025",
	PAGES=	"116971",
	DOI=	"10.1016/j.nuclphysb.2025.116971",
}

@ARTICLE{Guo:2018uxx,
	AUTHOR=	"Guo, Bin and Liu, Yu-Xiao and Yang, Ke and Wei, Shao-Wen",
	TITLE=	"{Hierarchy problem and new warped extra dimension}",
	EPRINT=	"1805.03976",
	ARCHIVEPREFIX=	"arXiv",
	PRIMARYCLASS=	"hep-th",
	DOI=	"10.1103/PhysRevD.98.085022",
	JOURNAL=	"Phys. Rev. D",
	VOLUME=	"98",
	PAGES=	"085022",
	YEAR=	"2018",
}

@ARTICLE{Giudice:2007fh,
	AUTHOR=	"Giudice, G. F. and Grojean, C. and Pomarol, A. and Rattazzi, R.",
	TITLE=	"The Strongly-Interacting Light {Higgs}",
	EPRINT=	"hep-ph/0703164",
	ARCHIVEPREFIX=	"arXiv",
	REPORTNUMBER=	"CERN-PH-TH-2007-47",
	DOI=	"10.1088/1126-6708/2007/06/045",
	JOURNAL=	"JHEP",
	VOLUME=	"06",
	PAGES=	"045",
	YEAR=	"2007",
}

@ARTICLE{Lodone:2008yy,
	AUTHOR=	"Lodone, Paolo",
	TITLE=	"{Vector-like quarks in a 'composite' Higgs model}",
	EPRINT=	"0806.1472",
	ARCHIVEPREFIX=	"arXiv",
	PRIMARYCLASS=	"hep-ph",
	DOI=	"10.1088/1126-6708/2008/12/029",
	JOURNAL=	"JHEP",
	VOLUME=	"12",
	PAGES=	"029",
	YEAR=	"2008",
}

@ARTICLE{Benbrik:2019zdp,
	AUTHOR=	"Benbrik, Rachid and others",
	TITLE=	"Signatures of vector-like top partners decaying into new neutral scalar or pseudoscalar bosons",
	EPRINT=	"1907.05929",
	ARCHIVEPREFIX=	"arXiv",
	PRIMARYCLASS=	"hep-ph",
	DOI=	"10.1007/JHEP05(2020)028",
	JOURNAL=	"JHEP",
	VOLUME=	"05",
	PAGES=	"028",
	YEAR=	"2020",
}

@ARTICLE{CMS:2022fck,
	AUTHOR=	"{CMS Collaboration}",
	TITLE=	"Search for pair production of vector-like quarks in leptonic final states in proton-proton collisions at $ \sqrt{s} = 13$ {TeV}",
	EPRINT=	"2209.07327",
	ARCHIVEPREFIX=	"arXiv",
	PRIMARYCLASS=	"hep-ex",
	REPORTNUMBER=	"CMS-B2G-20-011, CERN-EP-2022-175",
	DOI=	"10.1007/JHEP07(2023)020",
	JOURNAL=	"JHEP",
	VOLUME=	"07",
	PAGES=	"020",
	YEAR=	"2023",
}

@ARTICLE{Benbrik:2025vlq,
	AUTHOR=	"Benbrik, R. and others",
	TITLE=	"{Vector-like quarks at the LHC: A unified perspective from ATLAS and CMS exclusion limits}",
	JOURNAL=	"JHEP",
	VOLUME=	"03",
	YEAR=	"2025",
	PAGES=	"020",
	DOI=	"10.1007/JHEP03(2025)020",
	EPRINT=	"2412.01761",
	ARCHIVEPREFIX=	"arXiv",
	PRIMARYCLASS=	"hep-ph",
}

@ARTICLE{ATLAS:2026TtScalar,
	AUTHOR=	"{ATLAS Collaboration}",
	TITLE=	"{Search for pair-production of vector-like $T$ quarks decaying into a top quark and a spin-0 particle in the diphoton final state in proton-proton collisions at $\sqrt{s}=13$ TeV with the ATLAS detector}",
	EPRINT=	"2607.28381",
	ARCHIVEPREFIX=	"arXiv",
	PRIMARYCLASS=	"hep-ex",
	YEAR=	"2026",
}

@UNPUBLISHED{Stremmer:2023,
	AUTHOR=	"Daniel Stremmer and Ma{\l}gorzata Worek",
	TITLE=	"Associated production of a top-quark pair with two isolated photons at the {LHC} through {NLO} in {QCD}",
	DOI=	"10.1007/JHEP08(2023)179",
	EPRINT=	"2306.16968",
	ARCHIVEPREFIX=	"arXiv",
	PRIMARYCLASS=	"hep-ph",
	JOURNAL=	"JHEP",
	VOLUME=	"08",
	PAGES=	"179",
	YEAR=	"2023",
}

@ARTICLE{ATLAS:2024xyz,
	AUTHOR=	"{ATLAS Collaboration}",
	TITLE=	"{Exploration at the high-energy frontier: ATLAS Run 2 searches investigating the exotic jungle beyond the Standard Model}",
	JOURNAL=	"Phys. Rept.",
	VOLUME=	"1116",
	YEAR=	"2025",
	PAGES=	"301",
	DOI=	"10.1016/j.physrep.2024.10.001",
	EPRINT=	"2403.09292",
	ARCHIVEPREFIX=	"arXiv",
	PRIMARYCLASS=	"hep-ex",
	REPORTNUMBER=	"CERN-EP-2024-075",
}

@ARTICLE{CMS:2024vlq,
	AUTHOR=	"{CMS Collaboration}",
	TITLE=	"{Review of searches for vector-like quarks, vector-like leptons, and heavy neutral leptons in proton-proton collisions at {\ensuremath{\sqrt{s}}} = 13 TeV at the CMS experiment}",
	JOURNAL=	"Phys. Rept.",
	VOLUME=	"1115",
	YEAR=	"2025",
	PAGES=	"570",
	DOI=	"10.1016/j.physrep.2024.09.012",
	EPRINT=	"2405.17605",
	ARCHIVEPREFIX=	"arXiv",
	PRIMARYCLASS=	"hep-ex",
	REPORTNUMBER=	"CMS-EXO-23-006, CERN-EP-2024-095",
}

@ARTICLE{Qu:2019gqs,
	AUTHOR=	"Qu, Huilin and Gouskos, Loukas",
	TITLE=	"{ParticleNet: Jet tagging via particle clouds}",
	JOURNAL=	"Phys. Rev. D",
	VOLUME=	"101",
	YEAR=	"2020",
	PAGES=	"056019",
	DOI=	"10.1103/PhysRevD.101.056019",
	EPRINT=	"1902.08570",
	ARCHIVEPREFIX=	"arXiv",
	PRIMARYCLASS=	"hep-ph",
}

@ARTICLE{CMS:2020uim,
	AUTHOR=	"{CMS Collaboration}",
	TITLE=	"Electron and photon reconstruction and identification with the {CMS} experiment at the {CERN} {LHC}",
	EPRINT=	"2012.06888",
	JOURNAL=	"JINST",
	VOLUME=	"16",
	PAGES=	"P05014",
	YEAR=	"2021",
	ARCHIVEPREFIX=	"arXiv",
	PRIMARYCLASS=	"hep-ex",
	REPORTNUMBER=	"CMS-EGM-17-001, CERN-EP-2020-219",
	DOI=	"10.1088/1748-0221/16/05/P05014",
}

@MISC{hepdata,
	HOWPUBLISHED=	"{HEPD}ata record for this analysis",
	DOI=	"10.17182/hepdata.168412",
	YEAR=	"2026",
}

@ARTICLE{CMS:2008xjf,
	AUTHOR=	"{CMS Collaboration}",
	TITLE=	"The {CMS} experiment at the {CERN} {LHC}",
	DOI=	"10.1088/1748-0221/3/08/S08004",
	JOURNAL=	"JINST",
	VOLUME=	"3",
	PAGES=	"S08004",
	YEAR=	"2008",
}

@ARTICLE{CMS:2023gfb,
	AUTHOR=	"{CMS Collaboration}",
	TITLE=	"Development of the {CMS} detector for the {CERN} {LHC} {Run} 3",
	DOI=	"10.1088/1748-0221/19/05/P05064",
	JOURNAL=	"JINST",
	VOLUME=	"19",
	PAGES=	"P05064",
	YEAR=	"2024",
}

@ARTICLE{CMS:2020cmk,
	AUTHOR=	"{CMS Collaboration}",
	TITLE=	"{Performance of the CMS Level-1 trigger in proton-proton collisions at $\sqrt{s} = 13$\,TeV}",
	JOURNAL=	"JINST",
	VOLUME=	"15",
	PAGES=	"P10017",
	YEAR=	"2020",
	DOI=	"10.1088/1748-0221/15/10/P10017",
	EPRINT=	"2006.10165",
	ARCHIVEPREFIX=	"arXiv",
	PRIMARYCLASS=	"hep-ex",
	REPORTNUMBER=	"CMS-TRG-17-001, CERN-EP-2020-065",
}

@ARTICLE{CMS:2016ngn,
	AUTHOR=	"{CMS Collaboration}",
	TITLE=	"{The CMS trigger system}",
	JOURNAL=	"JINST",
	VOLUME=	"12",
	PAGES=	"P01020",
	DOI=	"10.1088/1748-0221/12/01/P01020",
	YEAR=	"2017",
	EPRINT=	"1609.02366",
	ARCHIVEPREFIX=	"arXiv",
	PRIMARYCLASS=	"physics.ins-det",
	REPORTNUMBER=	"CMS-TRG-12-001, CERN-EP-2016-160",
	SLACCITATION=	"%%CITATION = ARXIV:1609.02366;%%",
}

@ARTICLE{CMS:2024aqx,
	AUTHOR=	"{CMS Collaboration}",
	TITLE=	"Performance of the {CMS} high-level trigger during {LHC} {Run} 2",
	EPRINT=	"2410.17038",
	ARCHIVEPREFIX=	"arXiv",
	PRIMARYCLASS=	"physics.ins-det",
	REPORTNUMBER=	"CMS-TRG-19-001, CERN-EP-2024-259",
	DOI=	"10.1088/1748-0221/19/11/P11021",
	JOURNAL=	"JINST",
	VOLUME=	"19",
	PAGES=	"P11021",
	YEAR=	"2024",
}

@ARTICLE{CMS:2018rym,
	AUTHOR=	"{CMS Collaboration}",
	TITLE=	"{Performance of the CMS muon detector and muon reconstruction with proton-proton collisions at $\sqrt{s}=$ 13 TeV}",
	EPRINT=	"1804.04528",
	ARCHIVEPREFIX=	"arXiv",
	PRIMARYCLASS=	"physics.ins-det",
	REPORTNUMBER=	"CMS-MUO-16-001, CERN-EP-2018-058",
	DOI=	"10.1088/1748-0221/13/06/P06015",
	JOURNAL=	"JINST",
	VOLUME=	"13",
	PAGES=	"P06015",
	YEAR=	"2018",
}

@ARTICLE{CMS:2014pgm,
	AUTHOR=	"{CMS Collaboration}",
	TITLE=	"{Description and performance of track and primary-vertex reconstruction with the CMS tracker}",
	EPRINT=	"1405.6569",
	ARCHIVEPREFIX=	"arXiv",
	PRIMARYCLASS=	"physics.ins-det",
	REPORTNUMBER=	"CMS-TRK-11-001, CERN-PH-EP-2014-070",
	DOI=	"10.1088/1748-0221/9/10/P10009",
	JOURNAL=	"JINST",
	VOLUME=	"9",
	PAGES=	"P10009",
	YEAR=	"2014",
}

@ARTICLE{CMS:2021yvr,
	AUTHOR=	"{CMS Collaboration}",
	TITLE=	"{Search for resonant and nonresonant new phenomena in high-mass dilepton final states at $\sqrt{s}=13$ TeV}",
	JOURNAL=	"JHEP",
	VOLUME=	"07",
	PAGES=	"208",
	YEAR=	"2021",
	DOI=	"10.1007/JHEP07(2021)208",
	EPRINT=	"2103.02708",
	ARCHIVEPREFIX=	"arXiv",
	PRIMARYCLASS=	"hep-ex",
}

@ARTICLE{CMS:2017yfk,
	AUTHOR=	"{CMS Collaboration}",
	TITLE=	"Particle-flow reconstruction and global event description with the {CMS} detector",
	JOURNAL=	"JINST",
	VOLUME=	"12",
	YEAR=	"2017",
	PAGES=	"P10003",
	DOI=	"10.1088/1748-0221/12/10/P10003",
	EPRINT=	"1706.04965",
	ARCHIVEPREFIX=	"arXiv",
	PRIMARYCLASS=	"physics.ins-det",
	REPORTNUMBER=	"CMS-PRF-14-001, CERN-EP-2017-110",
	SLACCITATION=	"%%CITATION = ARXIV:1706.04965;%%",
}

@ARTICLE{Hoecker:2007bk,
	AUTHOR=	"Hoecker, A. and Speckmayer, P. and Stelzer, J. and Therhaag, J. and von Toerne, E. and Voss, H.",
	TITLE=	"{TMVA: Toolkit for multivariate data analysis}",
	JOURNAL=	"PoS",
	VOLUME=	"ACAT",
	YEAR=	"2007",
	PAGES=	"040",
	EPRINT=	"physics/0703039",
	ARCHIVEPREFIX=	"arXiv",
}

@ARTICLE{CMS:2015myp,
	AUTHOR=	"{CMS Collaboration}",
	TITLE=	"Performance of Photon Reconstruction and Identification with the {CMS} Detector in Proton-Proton Collisions at $\sqrt{s} = 8$ {TeV}",
	EPRINT=	"1502.02702",
	ARCHIVEPREFIX=	"arXiv",
	PRIMARYCLASS=	"physics.ins-det",
	REPORTNUMBER=	"CMS-EGM-14-001, CERN-PH-EP-2015-006",
	DOI=	"10.1088/1748-0221/10/08/P08010",
	JOURNAL=	"JINST",
	VOLUME=	"10",
	PAGES=	"P08010",
	YEAR=	"2015",
}

@ARTICLE{CMS:2020xrn,
	AUTHOR=	"{CMS Collaboration}",
	TITLE=	"A measurement of the {Higgs} boson mass in the diphoton decay channel",
	EPRINT=	"2002.06398",
	ARCHIVEPREFIX=	"arXiv",
	PRIMARYCLASS=	"hep-ex",
	REPORTNUMBER=	"CMS-HIG-19-004, CERN-EP-2020-004",
	DOI=	"10.1016/j.physletb.2020.135425",
	JOURNAL=	"Phys. Lett. B",
	VOLUME=	"805",
	PAGES=	"135425",
	YEAR=	"2020",
}

@ARTICLE{Cacciari:2008gp,
	AUTHOR=	"Cacciari, Matteo and Salam, Gavin P. and Soyez, Gregory",
	TITLE=	"The anti-\kt jet clustering algorithm",
	JOURNAL=	"JHEP",
	VOLUME=	"04",
	YEAR=	"2008",
	PAGES=	"063",
	DOI=	"10.1088/1126-6708/2008/04/063",
	EPRINT=	"0802.1189",
	ARCHIVEPREFIX=	"arXiv",
	PRIMARYCLASS=	"hep-ex",
}

@ARTICLE{Cacciari:2011ma,
	AUTHOR=	"Cacciari, Matteo and Salam, Gavin P. and Soyez, Gregory",
	TITLE=	"{FastJet user manual}",
	JOURNAL=	"Eur. Phys. J. C",
	VOLUME=	"72",
	PAGES=	"1896",
	DOI=	"10.1140/epjc/s10052-012-1896-2",
	YEAR=	"2012",
	EPRINT=	"1111.6097",
	ARCHIVEPREFIX=	"arXiv",
	PRIMARYCLASS=	"hep-ph",
	REPORTNUMBER=	"CERN-PH-TH-2011-297",
	SLACCITATION=	"%%CITATION = ARXIV:1111.6097;%%",
}

@ARTICLE{Sirunyan:2020foa,
	AUTHOR=	"{CMS Collaboration}",
	TITLE=	"{Pileup mitigation at {CMS} in 13 {TeV} data}",
	JOURNAL=	"JINST",
	VOLUME=	"15",
	PAGES=	"P09018",
	YEAR=	"2020",
	EPRINT=	"2003.00503",
	ARCHIVEPREFIX=	"arXiv",
	PRIMARYCLASS=	"hep-ex",
	REPORTNUMBER=	"CMS-JME-18-001, CERN-EP-2020-017",
	DOI=	"10.1088/1748-0221/15/09/P09018",
}

@ARTICLE{CMS:2016lmd,
	AUTHOR=	"{CMS Collaboration}",
	TITLE=	"{Jet energy scale and resolution in the CMS experiment in pp collisions at 8 TeV}",
	JOURNAL=	"JINST",
	VOLUME=	"12",
	YEAR=	"2017",
	PAGES=	"P02014",
	DOI=	"10.1088/1748-0221/12/02/P02014",
	EPRINT=	"1607.03663",
	ARCHIVEPREFIX=	"arXiv",
	PRIMARYCLASS=	"hep-ex",
	REPORTNUMBER=	"CMS-JME-13-004, CERN-PH-EP-2015-305",
	SLACCITATION=	"%%CITATION = ARXIV:1607.03663;%%",
}

@ARTICLE{Dasgupta:2013ihk,
	AUTHOR=	"Dasgupta, Mrinal and Fregoso, Alessandro and Marzani, Simone and Salam, Gavin P.",
	TITLE=	"Towards an understanding of jet substructure",
	JOURNAL=	"JHEP",
	VOLUME=	"09",
	YEAR=	"2013",
	PAGES=	"029",
	DOI=	"10.1007/JHEP09(2013)029",
	EPRINT=	"1307.0007",
	ARCHIVEPREFIX=	"arXiv",
	PRIMARYCLASS=	"hep-ph",
	REPORTNUMBER=	"CERN-PH-TH-2013-145, DCPT-13-86, IPPP-13-43, LPN13-036, MAN-HEP-2013-12",
	SLACCITATION=	"%%CITATION = ARXIV:1307.0007;%%",
}

@ARTICLE{Butterworth:2008iy,
	AUTHOR=	"Butterworth, Jonathan M. and Davison, Adam R. and Rubin, Mathieu and Salam, Gavin P.",
	TITLE=	"Jet substructure as a new {Higgs} search channel at the {LHC}",
	JOURNAL=	"Phys. Rev. Lett.",
	VOLUME=	"100",
	YEAR=	"2008",
	PAGES=	"242001",
	DOI=	"10.1103/PhysRevLett.100.242001",
	EPRINT=	"0802.2470",
	ARCHIVEPREFIX=	"arXiv",
	PRIMARYCLASS=	"hep-ph",
	SLACCITATION=	"%%CITATION = ARXIV:0802.2470;%%",
}

@ARTICLE{Larkoski:2014wba,
	AUTHOR=	"Larkoski, Andrew J. and Marzani, Simone and Soyez, Gregory and Thaler, Jesse",
	TITLE=	"Soft drop",
	JOURNAL=	"JHEP",
	VOLUME=	"05",
	YEAR=	"2014",
	PAGES=	"146",
	DOI=	"10.1007/JHEP05(2014)146",
	EPRINT=	"1402.2657",
	ARCHIVEPREFIX=	"arXiv",
	PRIMARYCLASS=	"hep-ph",
	REPORTNUMBER=	"MIT-CTP-4531, DCPT-14-24, IPPP-14-12",
	SLACCITATION=	"%%CITATION = ARXIV:1402.2657;%%",
}

@ARTICLE{Alwall:2011uj,
	AUTHOR=	"Alwall, Johan and Frederix, Rikkert and Frixione, Stefano and Hirschi, Valentin and Maltoni, Fabio and Mattelaer, Olivier and Shao, Hua-Sheng and Stelzer, Tim and Torrielli, Paolo and Zaro, Marco",
	TITLE=	"The automated computation of tree-level and next-to-leading order differential cross sections, and their matching to parton shower simulations",
	JOURNAL=	"JHEP",
	VOLUME=	"07",
	PAGES=	"079",
	YEAR=	"2014",
	DOI=	"10.1007/JHEP07(2014)079",
	EPRINT=	"1405.0301",
	ARCHIVEPREFIX=	"arXiv",
	PRIMARYCLASS=	"hep-ph",
	REPORTNUMBER=	"CERN-PH-TH-2014-064, CP3-14-18, LPN14-066, MCNET-14-09, ZU-TH-14-14",
}

@UNPUBLISHED{Banerjee:2022evx,
	AUTHOR=	"Banerjee, Avik and Buarque Franzosi, Diogo and Cacciapaglia, Giacomo and Deandrea, Aldo and Ferretti, Gabriele and Flacke, Thomas and Fuks, Benjamin and Kunkel, Manuel and Panizzi, Luca and Porod, Werner and Schwarze, Leonard",
	TITLE=	"Phenomenological aspects of composite {Higgs} scenarios: exotic scalars and vector-like quarks",
	EPRINT=	"2203.07270",
	ARCHIVEPREFIX=	"arXiv",
	PRIMARYCLASS=	"hep-ph",
	YEAR=	"2022",
	REPORTNUMBER=	"KIAS-A22001",
	NOTE=	"Snowmass 2021 White Paper",
}

@ARTICLE{Frixione:2007vw,
	AUTHOR=	"Frixione, Stefano and Nason, Paolo and Oleari, Carlo",
	TITLE=	"{Matching NLO QCD computations with parton shower simulations: the POWHEG method}",
	EPRINT=	"0709.2092",
	ARCHIVEPREFIX=	"arXiv",
	PRIMARYCLASS=	"hep-ph",
	REPORTNUMBER=	"BICOCCA-FT-07-9, GEF-TH-21-2007",
	DOI=	"10.1088/1126-6708/2007/11/070",
	JOURNAL=	"JHEP",
	VOLUME=	"11",
	PAGES=	"070",
	YEAR=	"2007",
}

@ARTICLE{Ball:2017nwa,
	AUTHOR=	"NNPDF Collaboration",
	TITLE=	"{Parton distributions from high-precision collider data}",
	JOURNAL=	"Eur. Phys. J. C",
	VOLUME=	"77",
	YEAR=	"2017",
	PAGES=	"663",
	DOI=	"10.1140/epjc/s10052-017-5199-5",
	EPRINT=	"1706.00428",
	ARCHIVEPREFIX=	"arXiv",
	PRIMARYCLASS=	"hep-ph",
}

@ARTICLE{Sjostrand:2014zea,
	AUTHOR=	"Sjostrand, T. and others",
	TITLE=	"{An introduction to PYTHIA 8.2}",
	JOURNAL=	"Comput. Phys. Commun.",
	VOLUME=	"191",
	YEAR=	"2015",
	PAGES=	"159",
	DOI=	"10.1016/j.cpc.2015.01.024",
	EPRINT=	"1410.3012",
	ARCHIVEPREFIX=	"arXiv",
	PRIMARYCLASS=	"hep-ph",
}

@ARTICLE{CMS:2019csb,
	AUTHOR=	"{CMS Collaboration}",
	TITLE=	"Extraction and validation of a new set of {CMS} {PYTHIA8} tunes from underlying-event measurements",
	EPRINT=	"1903.12179",
	ARCHIVEPREFIX=	"arXiv",
	PRIMARYCLASS=	"hep-ex",
	REPORTNUMBER=	"CMS-GEN-17-001, CERN-EP-2019-007",
	DOI=	"10.1140/epjc/s10052-019-7499-4",
	JOURNAL=	"Eur. Phys. J. C",
	VOLUME=	"80",
	PAGES=	"4",
	YEAR=	"2020",
}

@ARTICLE{Agostinelli:2002hh,
	AUTHOR=	"Agostinelli, S. and others",
	COLLABORATION=	"GEANT4",
	TITLE=	"{\GEANTfour}---a simulation toolkit",
	JOURNAL=	"Nucl. Instrum. Meth. A",
	VOLUME=	"506",
	YEAR=	"2003",
	PAGES=	"250",
	DOI=	"10.1016/S0168-9002(03)01368-8",
	SLACCITATION=	"%%CITATION = NUIMA,A506,250;%%",
}

@ARTICLE{CMS:2024onh,
	AUTHOR=	"{CMS Collaboration}",
	TITLE=	"The {CMS} statistical analysis and combination tool: {COMBINE}",
	EPRINT=	"2404.06614",
	ARCHIVEPREFIX=	"arXiv",
	PRIMARYCLASS=	"physics.data-an",
	REPORTNUMBER=	"CMS-CAT-23-001, CERN-EP-2024-078",
	DOI=	"10.1007/s41781-024-00121-4",
	JOURNAL=	"Comput. Softw. Big Sci.",
	VOLUME=	"8",
	PAGES=	"19",
	YEAR=	"2024",
}

@INPROCEEDINGS{Verkerke:2003ir,
	AUTHOR=	"Verkerke, W. and Kirkby, D.",
	TITLE=	"{The RooFit toolkit for data modeling}",
	BOOKTITLE=	"Proc. 13th Int. Conf. on Computing in High Energy and Nuclear Physics (CHEP 2003), La Jolla, CA, March 24--28, 2003",
	EPRINT=	"physics/0306116",
	ARCHIVEPREFIX=	"arXiv",
	PRIMARYCLASS=	"physics.data-an",
	YEAR=	"2003",
	NOTE=	"[eConf C0303241 (2003) MOLT007]",
}

@INPROCEEDINGS{Moneta:2010pm,
	AUTHOR=	"Moneta, L. and Belasco, K. and Cranmer, K. S. and Kreiss, S. and Lazzaro, A. and Piparo, D. and Schott, G. and Verkerke, W. and Wolf, M.",
	TITLE=	"{The RooStats project}",
	BOOKTITLE=	"Proc. 13th Int. Workshop on Advanced Computing and Analysis Techniques in Physics Research (ACAT 2010), Jaipur, India, February 22--27, 2010",
	EPRINT=	"1009.1003",
	ARCHIVEPREFIX=	"arXiv",
	PRIMARYCLASS=	"physics.data-an",
	DOI=	"10.22323/1.093.0057",
	YEAR=	"2010",
	NOTE=	"[PoS (ACAT2010) 057]",
}

@ARTICLE{CMS-LUM-17-003,
	AUTHOR=	"{CMS Collaboration}",
	TITLE=	"Precision luminosity measurement in proton-proton collisions at $\sqrt{s} = 13$ {TeV} in 2015 and 2016 at {CMS}",
	EPRINT=	"2104.01927",
	ARCHIVEPREFIX=	"arXiv",
	PRIMARYCLASS=	"hep-ex",
	REPORTNUMBER=	"CMS-LUM-17-003, CERN-EP-2021-033",
	DOI=	"10.1140/epjc/s10052-021-09538-2",
	JOURNAL=	"Eur. Phys. J. C",
	VOLUME=	"81",
	PAGES=	"800",
	YEAR=	"2021",
}

@UNPUBLISHED{CMS-LUM-20-001,
	AUTHOR=	"{CMS Collaboration}",
	TITLE=	"Precision luminosity measurement in proton-proton collisions at a center-of-mass energy of 13 {TeV} with the {CMS} detector at the {Large Hadron Collider}",
	EPRINT=	"2606.26832",
	ARCHIVEPREFIX=	"arXiv",
	PRIMARYCLASS=	"hep-ex",
	REPORTNUMBER=	"CMS-LUM-20-001, CERN-EP-2026-134",
	YEAR=	"2026",
	NOTE=	"Submitted to {Phys.Rev.X}",
}

@ARTICLE{CMS:2018mlc,
	AUTHOR=	"{CMS Collaboration}",
	TITLE=	"Measurement of the inelastic proton-proton cross section at $ \sqrt{s}=13 $ {TeV}",
	EPRINT=	"1802.02613",
	ARCHIVEPREFIX=	"arXiv",
	PRIMARYCLASS=	"hep-ex",
	REPORTNUMBER=	"CMS-FSQ-15-005, CERN-EP-2018-004",
	DOI=	"10.1007/JHEP07(2018)161",
	JOURNAL=	"JHEP",
	VOLUME=	"07",
	PAGES=	"161",
	YEAR=	"2018",
}

@ARTICLE{Butterworth:2015oua,
	AUTHOR=	"Butterworth, Jon and others",
	TITLE=	"{PDF4LHC} recommendations for {LHC} {Run II}",
	JOURNAL=	"J. Phys. G",
	VOLUME=	"43",
	YEAR=	"2016",
	PAGES=	"023001",
	DOI=	"10.1088/0954-3899/43/2/023001",
	EPRINT=	"1510.03865",
	ARCHIVEPREFIX=	"arXiv",
	PRIMARYCLASS=	"hep-ph",
	REPORTNUMBER=	"OUTP-15-17P, SMU-HEP-15-12, TIF-UNIMI-2015-14, LCTS-2015-27, CERN-PH-TH-2015-249",
	SLACCITATION=	"%%CITATION = ARXIV:1510.03865;%%",
}

@ARTICLE{Clopper:1934,
	AUTHOR=	"Clopper, C. and Pearson, E.",
	TITLE=	"{The use of confidence or fiducial limits illustrated in the case of the binomial}",
	JOURNAL=	"Biometrika",
	VOLUME=	"26",
	PAGES=	"404",
	YEAR=	"1934",
	DOI=	"10.1093/biomet/26.4.404",
}

@ARTICLE{Read:2002hq,
	AUTHOR=	"Read, A. L.",
	TITLE=	"{Presentation of search results: The CL$_{\text{s}}$ technique}",
	JOURNAL=	"J. Phys. G",
	VOLUME=	"28",
	YEAR=	"2002",
	PAGES=	"2693",
	DOI=	"10.1088/0954-3899/28/10/313",
}

@ARTICLE{Junk:1999kv,
	AUTHOR=	"Junk, T.",
	TITLE=	"{Confidence level computation for combining searches with small statistics}",
	JOURNAL=	"Nucl. Instrum. Meth. A",
	VOLUME=	"434",
	YEAR=	"1999",
	PAGES=	"435",
	DOI=	"10.1016/S0168-9002(99)00498-2",
	EPRINT=	"hep-ex/9902006",
	ARCHIVEPREFIX=	"arXiv",
	PRIMARYCLASS=	"hep-ex",
	REPORTNUMBER=	"CARLETON-OPAL-PHYS-99-01, CERN-EP-99-041",
}

@ARTICLE{Cowan:2010js,
	AUTHOR=	"Cowan, G. and Cranmer, K. and Gross, E. and Vitells, O.",
	TITLE=	"{Asymptotic formulae for likelihood-based tests of new physics}",
	EPRINT=	"1007.1727",
	ARCHIVEPREFIX=	"arXiv",
	PRIMARYCLASS=	"physics.data-an",
	DOI=	"10.1140/epjc/s10052-011-1554-0",
	JOURNAL=	"Eur. Phys. J. C",
	VOLUME=	"71",
	PAGES=	"1554",
	YEAR=	"2011",
	NOTE=	"{Erratum: Eur. Phys. J. C 73 (2013) 2501, doi:10.1140/epjc/s10052-013-2501-z}",
}

@ARTICLE{ParticleDataGroup,
	AUTHOR=	"Navas, S. and others",
	COLLABORATION=	"Particle Data Group",
	TITLE=	"Review of particle physics",
	DOI=	"10.1103/PhysRevD.110.030001",
	JOURNAL=	"Phys. Rev. D",
	VOLUME=	"110",
	PAGES=	"030001",
	YEAR=	"2024",
}

@ARTICLE{Czakon:2011xx,
	AUTHOR=	"Czakon, Michal and Mitov, Alexander",
	TITLE=	"{Top++: A program for the calculation of the top-pair cross-section at hadron colliders}",
	EPRINT=	"1112.5675",
	ARCHIVEPREFIX=	"arXiv",
	PRIMARYCLASS=	"hep-ph",
	REPORTNUMBER=	"CERN-PH-TH-2011-303, TTK-11-58",
	DOI=	"10.1016/j.cpc.2014.06.021",
	JOURNAL=	"Comput. Phys. Commun.",
	VOLUME=	"185",
	PAGES=	"2930",
	YEAR=	"2014",
}

@ARTICLE{Czakon:2013goa,
	AUTHOR=	"Czakon, M. and Fiedler, P. and Mitov, A.",
	TITLE=	"{Total top-quark pair-production cross section at hadron colliders}",
	JOURNAL=	"Phys. Rev. Lett.",
	VOLUME=	"110",
	YEAR=	"2013",
	PAGES=	"252004",
	DOI=	"10.1103/PhysRevLett.110.252004",
	EPRINT=	"1303.6254",
	ARCHIVEPREFIX=	"arXiv",
	PRIMARYCLASS=	"hep-ph",
}
